\documentclass{article}
\usepackage{iclr2027_arxiv,times}
\iclrfinalcopy

\usepackage[T1]{fontenc}
\usepackage[utf8]{inputenc}
\usepackage{microtype}
\usepackage{amsmath,amssymb,amsfonts,bm}
\usepackage{booktabs}
\usepackage{xcolor}
\usepackage{graphicx}
\usepackage{tikz}
\usetikzlibrary{positioning,arrows.meta,calc,fit}
\usepackage[labelfont=bf]{caption}
\usepackage{hyperref}
\usepackage[capitalize,noabbrev]{cleveref}
\usepackage[para]{footmisc}
\usepackage{etoc}
\crefname{appendix}{appendix}{appendices}
\Crefname{appendix}{Appendix}{Appendices}

\graphicspath{{figures/}}

\usepackage{amsmath,amsfonts,bm}

\def\eqref#1{equation~\ref{#1}}

\def\1{\bm{1}}

\DeclareMathAlphabet{\mathsfit}{\encodingdefault}{\sfdefault}{m}{sl}
\SetMathAlphabet{\mathsfit}{bold}{\encodingdefault}{\sfdefault}{bx}{n}

\newcommand{\Ls}{\mathcal{L}}
\newcommand{\R}{\mathbb{R}}

\newcommand{\softmax}{\mathrm{softmax}}

\DeclareMathOperator*{\argmax}{arg\,max}

\newcommand{\lrminzDfour}{3.5}

\newcommand{\lrminzDfive}{3.4}
\newcommand{\lrnarms}{10}
\newcommand{\lrsteps}{3000}
\newcommand{\lrwindow}{250}

\newcommand{\betameandiagOne}{0.013}

\newcommand{\betamodefracOne}{0.999}

\newcommand{\betauniqueOne}{1.000}
\newcommand{\betacoverageOne}{0.179}

\newcommand{\betaobjcorrnatOne}{-0.169}
\newcommand{\betaobjcorrstabOne}{-0.057}
\newcommand{\betaprefoffnatOne}{-0.019}
\newcommand{\betaprefoffstabOne}{-0.024}

\newcommand{\betameandiagEight}{0.110}

\newcommand{\betatrendEight}{-10.9\%}
\newcommand{\betadeltastdEight}{0.86}

\newcommand{\betameandiagThirtyTwo}{0.204}

\newcommand{\betactrlstabThirtyTwo}{0.103}
\newcommand{\betamodefracThirtyTwo}{0.935}

\newcommand{\betauniqueThirtyTwo}{0.997}
\newcommand{\betacoverageThirtyTwo}{0.207}

\newcommand{\betaobjcorrnatThirtyTwo}{-0.112}
\newcommand{\betaobjcorrstabThirtyTwo}{-0.107}
\newcommand{\betaprefoffnatThirtyTwo}{-0.366}
\newcommand{\betaprefoffstabThirtyTwo}{-0.272}
\newcommand{\betatrendThirtyTwo}{-11.5\%}
\newcommand{\betadeltastdThirtyTwo}{4.72}

\newcommand{\betansamplesperpref}{512}
\newcommand{\betaidentity}{70\%}

\newcommand{\betawindowfrac}{40\%}

\newcommand{\betalocalspreaddrop}{9.2\%}
\newcommand{\betansamples}{655{,}360}
\newcommand{\betanprefs}{64}
\newcommand{\betantargets}{20}

\newcommand{\pfparetotargets}{45}

\newcommand{\curationResCut}{2.3}
\newcommand{\curationAfterRes}{873}

\newcommand{\curationBlacklist}{17}

\newcommand{\curationParserDropped}{72}

\newcommand{\curationUniverse}{784}

\newcommand{\curationResLowest}{1.2}
\newcommand{\curationResMedian}{2.0}
\newcommand{\curationResHighest}{2.3}
\newcommand{\curationProtein}{469}
\newcommand{\curationPeptide}{315}
\newcommand{\curationNModels}{24}
\newcommand{\curationShiftChothia}{558}
\newcommand{\curationShiftImgt}{695}

\newcommand{\splitShortDropped}{31}

\newcommand{\splitExclude}{161}
\newcommand{\splitTestLeaks}{48}
\newcommand{\splitValidLeaks}{27}
\newcommand{\splitRabdAvail}{55}
\newcommand{\splitRabdTotal}{60}
\newcommand{\splitRabdCollapsed}{0}
\newcommand{\splitValidHthrees}{5, 7, 9, 9, 10, 11, 12, 14, 16, 18}

\newcommand{\scafIdentityPfifty}{0.856}
\newcommand{\scafIdentityPseventyfive}{0.900}
\newcommand{\scafIdentityPninety}{0.934}
\newcommand{\scafIdentityPninetyfive}{0.981}
\newcommand{\scafIdentityPninetynine}{1.000}

\newcommand{\pcaNobjectives}{14}

\newcommand{\pcaCumDfour}{89.1}
\newcommand{\pcaCumDfive}{93.7}

\newcommand{\pcaGainDfourToDfive}{4.6}

\newcommand{\tradeoffMostOpposed}{ProteinMPNN-antibody vs.\ trainset novelty}
\newcommand{\tradeoffMostOpposedRho}{$-$0.58}

\newcommand{\tradeoffMostAligned}{ProteinMPNN vs.\ ProteinMPNN-soluble}
\newcommand{\tradeoffMostAlignedRho}{0.97}

\newcommand{\sfourexpandedFourteenDFourRosettaTopOne}{-8.88}

\newcommand{\sfourexpandedFourteenDFiveFoldxTopOne}{-5.50}
\newcommand{\sfourexpandedFourteenDFiveRosettaTopOne}{-9.73}

\newcommand{\sfourgflownetFoldxTopOne}{-4.78}

\newcommand{\sfourproteinmpnnvanillaFoldxTopOne}{-6.20}
\newcommand{\sfourproteinmpnnvanillaRosettaTopOne}{-14.52}

\newcommand{\sfouresmifFoldxTopOne}{-6.09}
\newcommand{\sfouresmifRosettaTopOne}{-13.07}

\newcommand{\acKeepPercent}{25\%}
\newcommand{\acTopK}{250}

\newcommand{\ParetoPooledFourTargets}{45}

\newcommand{\cpNArms}{7}
\newcommand{\cpNTargets}{45}
\newcommand{\cpNDesigns}{78{,}711}

\newcommand{\bbArLoopsPersec}{97}

\newcommand{\bbMedianSpan}{10}
\newcommand{\bbMlmLoopsPersec}{9.1}
\newcommand{\bbOasDevLogNll}{0.2381}

\newcommand{\bbOasDevNll}{0.4902}

\newcommand{\bbOasGap}{-0.012}

\newcommand{\bbOasTrainNll}{0.5022}

\newcommand{\bbScopDevLogNll}{1.8970}
\newcommand{\bbScopDevLogPct}{1.8}

\newcommand{\bbScopDevNll}{1.8637}

\newcommand{\bbScopGap}{+0.737}

\newcommand{\bbScopTrainNll}{1.1271}

\newcommand{\bbSpeedup}{10.7}
\newcommand{\bbThroughputN}{55}
\newcommand{\bbUniprotDevLogNll}{1.4296}
\newcommand{\bbUniprotDevLogPct}{1.6}

\newcommand{\bbUniprotDevNll}{1.4074}

\newcommand{\condAmplifyFimPriorPoolRho}{+0.01}
\newcommand{\condAmplifyPriorPoolRho}{-0.04}
\newcommand{\condDFiveAgree}{+0.60}
\newcommand{\condDFiveBestAxis}{slot1}
\newcommand{\condDFiveBestSel}{0.729}
\newcommand{\condDFiveNoveltyMass}{0.387}

\newcommand{\condDFourAgree}{+1.00}
\newcommand{\condDFourBestAxis}{slot2}
\newcommand{\condDFourBestSel}{0.827}
\newcommand{\condDFourNoveltyMass}{0.022}
\newcommand{\condDFourPoolRho}{+0.75}

\newcommand{\condFixedAgree}{+0.60}
\newcommand{\condFixedBestAxis}{mhc2}
\newcommand{\condFixedBestSel}{1.476}

\newcommand{\condFixedPoolRho}{+0.63}

\newcommand{\bx}{\mathbf{x}}
\newcommand{\bw}{\mathbf{w}}

\newcommand{\Epref}{\mathbf{E}_{\mathrm{pref}}}
\newcommand{\Etask}{\mathbf{E}_{\mathrm{task}}}

\newcommand{\bz}{\mathbf{z}}
\newcommand{\br}{\mathbf{r}}

\newcommand{\bA}{\mathbf{A}}
\newcommand{\bTheta}{\boldsymbol{\Theta}}
\newcommand{\balpha}{\boldsymbol{\alpha}}
\newcommand{\cA}{\mathcal{A}}
\newcommand{\cH}{\mathcal{H}}

\newcommand{\cV}{\mathcal{V}}

\newcommand{\cX}{\mathcal{X}}

\title{Multi-Objective Amortized CDRH3 Sequence Sampling Diversifies Promising Antibody Candidates}

\author{%
  \textbf{Joseph D. Viviano}$^{1,*}$ \quad
  \textbf{Almer van der Sloot}$^{1,2,\dagger}$ \quad
  \textbf{Quentin Fournier}$^{1,2,3,\ddagger}$ \\
  \textbf{Robert M. Vernon}$^{3}$ \quad
  \textbf{Christopher James Langmead}$^{4,\S}$ \quad
  \textbf{Yoshua Bengio}$^{1,2,5}$ \\[0.3em]
  {\small\mdseries
    $^{1}$Mila -- Qu\'ebec AI Institute \quad
    $^{2}$Universit\'e de Montr\'eal \quad
    $^{3}$Amgen \quad
    $^{4}$Carnegie Mellon University \quad
    $^{5}$LawZero} \\[0.2em]
  {\footnotesize\mdseries\ttfamily
    $^{*}$joseph@viviano.ca \quad
    $^{\dagger}$almer.van-der-sloot@mila.quebec} \\
  {\footnotesize\mdseries\ttfamily
    $^{\ddagger}$quentin.fournier@mila.quebec \quad
    $^{\S}$cjl@cs.cmu.edu}
}

\begin{document}

\maketitle
\etocdepthtag.toc{mainmatter}

\begin{abstract}
    Antibody therapeutics have transformed treatment across diverse therapeutic areas and have become central to modern medicine. While recent machine learning approaches aim to accelerate antibody discovery through structure-based design and target-affinity optimization, successful candidates must also satisfy broad developability constraints, including structural compatibility, low immunogenicity, and favorable biophysical properties, while remaining sufficiently diverse to provide sufficient coverage of design space to mitigate later-stage development failures. We introduce a preference-conditioned multi-objective generative flow network (GFN) fine-tuning pipeline for fixed-framework CDRH3 loop redesign that fine-tunes a biological sequence model against many inexpensive proxies covering backbone compatibility, sequence likelihood, immunogenicity risk, novelty, and developability liabilities. It combines up to fourteen predictive rewards in a single policy trained to sample CDRH3 loops across 623 SAbDab antibody structures, leveraging a reward reduction strategy to preserve Pareto dominance under this learned map. We also introduce a novel autoregressive span-in-filling variant of AMPLIFY-350M that serves both as a scaffold-conditioned CDRH3 generation prior and as a direct conditional likelihood scorer. We evaluated reward scores, diversity, and design quality on loops sampled on 45 held-out antibody complexes to assess structural consistency, confidence, aggregation propensity, stability, and antigen binding. Compared with the inverse-folding methods ProteinMPNN and ESM-IF, our approach produces roughly $10\times$ more distinct acceptable candidates per design at 50\% sequence identity, while holding $10\%$ of the 14-objective Pareto front pooled across methods, trading per-sample quality for greater exploration of novel possible structures. The results suggest that by relying on a more diverse data distribution and reward distribution, we are able to generate a wider array of promising candidates. All code and data will be made publicly available upon publication.
\end{abstract}

\section{Introduction}
\label{sec:intro}
\vspace{-2mm}

Antibody therapeutics are a major drug modality across therapeutic areas, with more than 200 antibodies and antibody-like biologics now approved ~\citep{Chan2025-zs}. Current antibody discovery still relies heavily on experimental screening of large libraries, followed by extensive engineering and optimization~\citep{Laustsen2021-oa}. Yet such libraries sample only a small fraction of the vast antibody sequence space; a typical 13 residue loop alone allows $20^{13} \approx 8.2 \times 10^{16}$ possible sequences. 
Computational and machine learning-driven approaches increasingly aim to direct this search in high dimensional space~\citep{Laustsen2021-oa,Bennett2026-xw,Shanehsazzadeh2023.01.08.523187,Bio2025-zd,Agarwal2025-bw}, focusing primarily on affinity, structural plausibility, or sequence likelihood, with developability usually addressed later in development. However, binding rank-based methods can perform at parity with or worse than random mutational screens \citep{erasmus2026blinded}, motivating a broader exploration of sequence design space. Moreover, antibody design is inherently multi-objective: candidates must balance structural compatibility with low immunogenicity and favorable biophysical and developability properties~\citep{Khetan2022-ev,Xu2019-tj,Zhang2023-lb, raybould2019five}. These objectives can conflict \citep{adams2001high} and maintaining diversity provides alternatives when otherwise promising candidates fail later in development. This is particularly important, as validation studies repeatedly show that once in-silico structural or binding scores exceed an acceptable quality threshold, further score improvements do not necessarily translate into a better candidate ranking \citep{Chu2026.06.13.732058, Cotet2025.04.17.648362}, motivating the consideration of a broader suite of design criteria when selecting candidates to progress to the lab for further therapeutic development \citep{Overath2025.08.14.670059}.


This paper approaches fixed-framework redesign of antibody heavy-chain complementarity-determining region 3 (CDRH3) loops as a sequence generation problem in the context of a fixed structural scaffold. CDRH3 is a major determinant of antigen binding and specificity due to its central position in the binding interface and high sequence diversity~\citep{xu2000diversity,McConnell2025-pb}\footnote{Antibody variable domains are organized around conserved framework regions, which provide the structural scaffold for six hypervariable complementarity-determining regions (CDRs) responsible for antigen recognition and binding. Here, we focus on CDRH3, which typically makes a major contribution to antigen binding.}, Each design replaces CDRH3 within the fixed framework of a known antibody-antigen crystal complex, avoiding the need for antibody-antigen pose search required in many \emph{de novo} design settings. We adopt Multi Objective Generative Flow Networks (MOGFNs; \citet{jain2023mogfn}) to train a (multi-) preference conditioned CDRH3 sequence sampler across 623 experimentally-derived antibody-antigen complexes, enabling the sampling of diverse, high-reward designs representing different trade-offs along the Pareto front. 
We use this approach to fine-tune a new span-in-filling variant of AMPLIFY-350M \citep{fournier2024amplify}, conditioned on both scaffold and preference. 
Our oracle-agnostic framework supports additional or alternative oracles without changing the generator architecture, and we demonstrate this flexibility by training with up to $M=14$ predictive rewards. To avoid the combinatorial training and deployment burden of sampling preferences across a growing $M$-simplex, we learn a dominance-preserving, strictly positive, row-stochastic projection from the $M$ normalized rewards onto $D \ll M$ preference slots~\citep{park2025rdr}.

We evaluate reward performance, diversity, and structural quality on 45 held-out antibody complexes. In sum, we contribute:
\begin{enumerate}
  \item A preference-conditioned MOGFN fine-tuning pipeline for fixed-framework CDRH3 redesign, optionally incorporating a learned preference projection to support many rewards.
  \item A modular suite of fourteen antibody-design oracles covering inverse-folding backbone compatibility, general and antibody-specific sequence likelihood, immunogenicity risk, novelty, and developability liabilities.
  \item An autoregressive span-in-filling version of AMPLIFY-350M, serving as a fast and memory-efficient scaffold-conditioned CDRH3 generation prior.
  \item A structural quality-diversity evaluation pipeline designed to produce a Chai-1 \citep{chai2024chai1} and Rosetta \citep{Alford2017-nu} ranked shortlist of promising CDRH3 loop candidates using estimator scores weakly correlated with the training rewards.
\end{enumerate}

\section{Related Work}
\label{sec:related}
\vspace{-2mm}

\paragraph{Antibody and binder design.} Computational antibody design spans structure-based de novo design~\citep{Bennett2026-xw,stark2025boltzgen}, antigen-specific generators~\citep{Shanehsazzadeh2023.01.08.523187,Bio2025-zd,Levine2026.01.14.699389,Mille-Fragoso2026-fy,Wasdin2025-xz}, affinity prediction and optimization~\citep{Agarwal2025-bw}, sequence infilling~\citep{shuai2023iglm}, and antibody-specific foundation and inverse-folding models~\citep{Guloglu2025-xc,antibodympnn}. These systems typically optimize binding, sequence likelihood, or structure recovery and often rely on structural information, which covers only a small fraction of antibody sequence space: SAbDab contains $\approx10^4$ antibody--antigen structures~\citep{schneider2022sabdab}, compared with $>10^9$ observed antibody sequences in OAS~\citep{olsen2022oas}. In contrast, we train a single CDRH3 generator across many scaffolds using structural priors, sequence likelihood, developability, immunogenicity, and novelty as rewards. The resulting model generalizes to new scaffolds at test time, unlike antigen-specific generators that treat each target as a separate design problem~\citep{Shanehsazzadeh2023.01.08.523187,Bio2025-zd,Levine2026.01.14.699389,Mille-Fragoso2026-fy,Wasdin2025-xz}.
Previous work has also cast antibody design as a multiobjective problem. \citet{Makowski2022-ps} co-optimized affinity against non-specific binding across a CDR mutational library; \citet{Tagasovska2025-ie} affinity and expression from limited assay data; others decomposed interface-energy terms using diffusion-based sequence-structure co-design \citep{wen2026pareto} or sequential design \citep{Feng2025-dd}; and \citet{Ren2025-op} Rosetta binding energy and evolutionary plausibility while treating non-specific binding, self-association and stability as constraints. These approaches use at most a handful (2-5) and largely emphasize binding energy, whose predictive accuracy remains limited for mutated antibody-antigen interfaces \citep{sirin2016abbind}, and out-of-domain antigens~\citep{hummer2025graphinity}. We instead train a single multi-scaffold (623 bound SAbDab complexes), multi-task sampler to \emph{explore} the  sequence space of promising CDRH3 variants over a broader suite of up to fourteen objectives.

\paragraph{GFlowNets.} Generative Flow Networks learn policies whose terminal states are sampled proportional to a positive reward~\citep{NEURIPS2021_e614f646}, preserving diversity among high-reward candidates. Multi-objective GFNs~\citep{jain2023mogfn} and HN-GFN~\citep{zhu2023hngfn} extend this idea to Pareto-optimal generation under competing objectives. \citet{Yin2025-zz} leveraged a GFN to sample high likelihood sequences under a product of experts (ProGen2 from\citet{nijkamp2023progen2}, and a Potts model, respectively providing global protein sequence distribution and local antibody-domain distributions) to generate diverse CDR candidates. Instead, we condition on preferences over up to fourteen rewards to take advantage of the same diversity-seeking behaviour of GFNs.

\paragraph{Objective reduction.} Multi-objective optimization often relies on objective reduction to accommodate many rewards, via methods such as PCA~\citep{deb2006objred} or dominance-preserving reduction with an explicit $\delta$-error~\citep{brockhoff2006reduction,brockhoff2009theory}, motivated by the observation that Pareto dominance degenerates as the objective count grows. Recent work in multi-objective reinforcement learning (MORL; \citet{park2025rdr}) showed that learning to sample from the Pareto front of an affine, all-positive, row-stochastic reward reduction provably produces a policy that samples from regions of the non-reduced objective set Pareto front. We adopt this method and transfer it from value-based MORL to a preference-conditioned MOGFN which targets the whole reward distribution, and reward compression therefore trades against both diversity and optimality.

\paragraph{Structural filtering of designs.} Forward-folding self-consistency and co-folding are standard practices for validating designed sequences; we rely on both an antibody-specific structure predictor  ABodyBuilder3~\citep{abodybuilder3} and the general structure prediction model Chai-1~\citep{chai2024chai1}, which does not require multiple sequence alignment. Fold confidence and  empirical, energy-based binding estimators like FoldX~\citep{Delgado2025FoldX} and Rosetta~\citep{Alford2017-nu} are useful as binary gates for selecting   promising binders for experimental validation, although their ranking performance remains limited \citep{pacesa2025bindcraft,bennett2023binder,sirin2016abbind,hummer2025graphinity}.

\section{Methods}
\label{sec:methods}
\vspace{-2mm}

\subsection{Task Overview}
\label{sec:methods-task-overview}
\vspace{-2mm}

We denote a candidate CDRH3 amino acid sequence by $\bx = (x_1, \dots, x_L)$, where $L$ is the CDRH3 loop length. For a fixed antibody scaffold $s$, we redesign only the CDRH3 loop while fixing the surrounding framework context. Each design task is a multi-objective optimization problem which seeks feasible candidates $\bx^\star \in \cX$ that jointly maximize $M$ objectives $R(\bx) = [R_1(\bx), \ldots, R_M(\bx)]$ i.e., $\max_{\bx \in \cX} R(\bx)$. When objectives conflict, no single candidate simultaneously maximizes all objectives, so we seek Pareto-optimal candidates where improving one objective dimension would compromise another one.

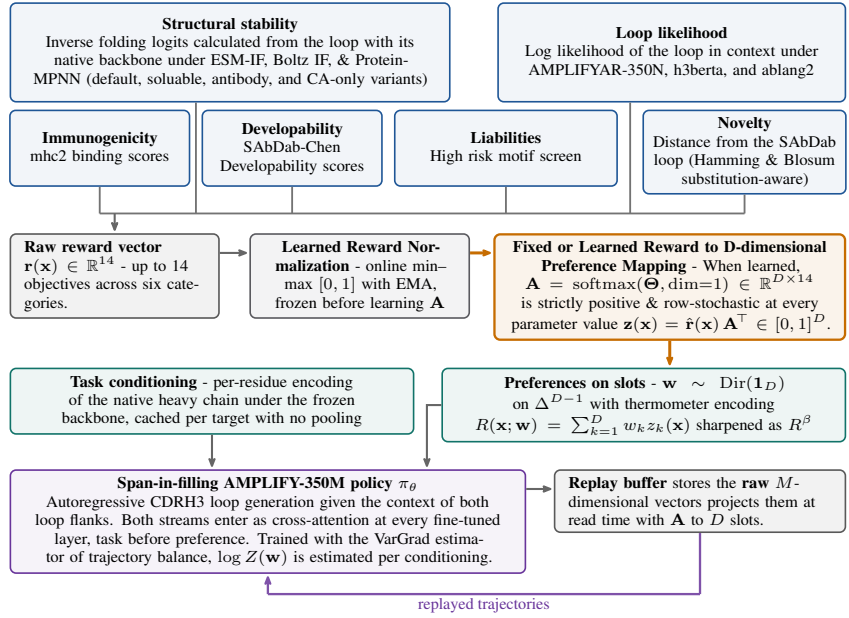
\begin{figure}[!t]
\centering
\scalebox{0.87}{\definecolor{tpOra}{RGB}{31,86,150}      
\definecolor{tpNorm}{RGB}{90,90,95}      
\definecolor{tpProj}{RGB}{198,108,0}     
\definecolor{tpPref}{RGB}{17,115,105}    
\definecolor{tpPol}{RGB}{104,52,154}     
\definecolor{tpOff}{RGB}{130,130,130}    

\pgfmathsetlengthmacro{\tpFull}{12.3cm}                
\pgfmathsetlengthmacro{\tpHead}{\tpFull + 0.1cm}       
\pgfmathsetlengthmacro{\tpSpine}{0.764*\tpFull}        
\pgfmathsetlengthmacro{\tpHalf}{0.5*\tpFull + 6pt}     
\pgfmathsetlengthmacro{\tpGut}{\tpHalf + 0.10cm - 0.5*\tpSpine - 4pt - 0.06cm}

\begin{tikzpicture}[
  font=\scriptsize,
  >={Stealth[round,length=2.0mm]},
  box/.style={draw, rounded corners=2.5pt, align=center,
              inner xsep=4pt, inner ysep=3pt, line width=0.7pt},
  oracle/.style={box, draw=tpOra, fill=tpOra!6},
  norm/.style={box, draw=tpNorm, fill=tpNorm!7},
  wire/.style={box, draw=black!65, fill=black!5},
  prose/.style={wire, align=left, inner xsep=6pt},
  proj/.style={box, draw=tpProj, fill=tpProj!8, line width=1.0pt},
  projsub/.style={box, draw=tpProj, fill=tpProj!4},
  pref/.style={box, draw=tpPref, fill=tpPref!7},
  pol/.style={box, draw=tpPol, fill=tpPol!6},
  off/.style={box, draw=tpOff, fill=black!2, dashed,
              dash pattern=on 2pt off 1.6pt, text=black!55},
  flow/.style={->, line width=0.9pt, draw=black!60, shorten >=1pt},
  pflow/.style={->, line width=1.3pt, draw=tpProj, shorten >=1pt},
  bus/.style={-, line width=0.8pt, draw=black!55},
  side/.style={->, line width=0.6pt, draw=black!35,
               dash pattern=on 1.5pt off 1.5pt},
  back/.style={-, line width=1.0pt, draw=tpPol!85},
  backa/.style={->, line width=1.0pt, draw=tpPol!85, shorten >=1pt},
  samp/.style={-, line width=1.0pt, draw=tpOra!85},
  sampa/.style={->, line width=1.0pt, draw=tpOra!85, shorten >=1pt},
  gut/.style={font=\scriptsize, text=black!60, text width=\tpGut},
  gutr/.style={gut, anchor=west, align=left},
  gutl/.style={gut, anchor=east, align=right},
  note/.style={font=\scriptsize, text=black!65, align=center},
  elab/.style={font=\scriptsize, align=center, inner sep=1.5pt},
]

\coordinate (L) at (-\tpHalf,0pt);   
\coordinate (R) at ( \tpHalf,0pt);   

\node[oracle, text width=6.4cm, minimum height=1.5cm, anchor=north west]
  (stab) at (L)
  {\textbf{Structural stability}\\
   Inverse folding logits calculated from the loop with its native backbone under ESM-IF, Boltz IF, \& ProteinMPNN (default, soluable, antibody, and CA-only variants)};

\node[oracle, text width=5.0cm, minimum height=1.5cm, anchor=north east]
  (nat) at (R)
  {\textbf{Loop likelihood}\\
   Log likelihood of the loop in context under AMPLIFYAR-350N, h3berta, and ablang2};

\node[fit=(stab)(nat), draw=none, inner sep=0pt] (rowa) {};

\node[oracle, text width=2.45cm, minimum height=1.15cm, anchor=north west]
  (mhc) at ($(rowa.south west) + (0,-0.10)$)
  {\textbf{Immunogenicity}\\ mhc2 binding scores};
\node[oracle, text width=2.45cm, minimum height=1.15cm, anchor=north west]
  (dev) at ($(mhc.north east) + (0.17,0)$)
  {\textbf{Developability}\\ SAbDab-Chen Developability scores};
\node[oracle, text width=3.10cm, minimum height=1.15cm, anchor=north west]
  (liab) at ($(dev.north east) + (0.17,0)$)
  {\textbf{Liabilities}\\ High risk motif screen};
\node[oracle, text width=2.80cm, minimum height=1.15cm, anchor=north east]
  (nov) at ($(rowa.south east) + (0,-0.10)$)
  {\textbf{Novelty}\\ Distance from the SAbDab loop (Hamming \& Blosum substitution-aware)};

\node[fit=(mhc)(dev)(liab)(nov), draw=none, inner sep=0pt] (rowb) {};

\coordinate (gapL) at ($(mhc.north east)!0.5!(dev.north west)$);
\coordinate (gapR) at ($(liab.north east)!0.5!(nov.north west)$);
\coordinate (busy) at ($(rowb.south) + (0,-0.28)$);

\node[prose, text width=2.75cm, anchor=north west]
  (raw) at ($(L |- busy) + (0,-0.30)$)
  {\textbf{Raw reward vector} $\br(\bx) \in \R^{14}$ - up to 14 objectives across six categories.};

\node[norm, text width=3.05cm, anchor=north west]
  (nrm) at ($(raw.north east) + (0.45,0)$)
  {\textbf{Learned Reward Normalization} - online min--max $[0,1]$ with EMA, frozen before learning $\bA$};

\node[proj, text width=5.05cm, anchor=north east]
  (mlearn) at ($(R |- busy) + (0,-0.30)$)
  {\textbf{Fixed or Learned Reward to D-dimensional Preference Mapping} - When learned, $\bA = \softmax(\bTheta, \mathrm{dim}{=}1) \in \R^{D \times 14}$ is strictly positive \& row-stochastic at every parameter value $\bz(\bx) = \hat{\br}(\bx)\,\bA^{\!\top} \in [0,1]^{D}$.
  };

\node[fit=(raw)(nrm)(mlearn), draw=none, inner sep=0pt] (rowc) {};

\node[pref, text width=5.85cm, anchor=north west]
  (tsk) at ($(L |- rowc.south) + (0,-0.40)$)
  {\textbf{Task conditioning} - per-residue encoding of the native heavy chain
   under the frozen backbone, cached per target with no pooling};

\node[pref, text width=5.85cm, anchor=north east]
  (prf) at ($(R |- rowc.south) + (0,-0.40)$)
  {\textbf{Preferences on slots} - $\bw \sim \mathrm{Dir}(\mathbf{1}_D)$
   on $\Delta^{D-1}$ with thermometer encoding
   $R(\bx;\bw) = \sum_{k=1}^{D} w_k z_k(\bx)$ sharpened as
   $R^\beta$};

\node[fit=(tsk)(prf), draw=none, inner sep=0pt] (rowd) {};

\node[pol, text width=7.55cm, anchor=north west]
  (pol) at ($(L |- rowd.south) + (0,-0.40)$)
  {\textbf{Span-in-filling AMPLIFY-350M policy $\pi_\theta$}\\
  Autoregressive CDRH3 loop generation given the context of both loop flanks. Both streams enter as cross-attention at every fine-tuned layer, task before preference. Trained with the VarGrad estimator of trajectory balance, $\log Z(\bw)$ is estimated per conditioning.};

\node[prose, text width=4.00cm, anchor=north east]
  (buf) at ($(R |- rowd.south) + (0,-0.40)$)
  {\textbf{Replay buffer} stores the \textbf{raw} $M$-dimensional vectors projects them at read time with $\bA$ to $D$ slots.};

\node[fit=(pol)(buf), draw=none, inner sep=0pt] (rowe) {};

\draw[bus] (mhc.south |- busy) -- (nov.south |- busy);
\draw[bus] (mhc.south)  -- (mhc.south  |- busy);
\draw[bus] (dev.south)  -- (dev.south  |- busy);
\draw[bus] (liab.south) -- (liab.south |- busy);
\draw[bus] (nov.south)  -- (nov.south  |- busy);
\draw[bus] (gapL |- stab.south) -- (gapL |- busy);
\draw[bus] (gapR |- nat.south)  -- (gapR |- busy);
\draw[flow] (raw.north |- busy) -- (raw.north);

\draw[flow]  ($(raw.north east) + (0,-0.30)$) -- ($(nrm.north west) + (0,-0.30)$);
\draw[pflow] ($(nrm.north east) + (0,-0.30)$) -- ($(mlearn.north west) + (0,-0.30)$);
\draw[pflow] (mlearn.south) -- (mlearn.south |- prf.north);
\coordinate (px) at ($(prf.west) + (-0.22,0)$);
\draw[flow]  (prf.west) -- (px) -- (px |- pol.north);
\draw[flow]  (tsk.south) -- (tsk.south |- pol.north);
\draw[flow]  ($(pol.north east) + (0,-0.30)$) -- ($(buf.north west) + (0,-0.30)$);

\coordinate (fby) at ($(rowe.south) + (0,-0.30)$);
\draw[back]  (buf.south) -- (buf.south |- fby);
\draw[back]  (buf.south |- fby) -- (pol.south |- fby);
\draw[backa] (pol.south |- fby) -- (pol.south);
\node[elab, text=tpPol, anchor=north]
  at ($(buf.south |- fby)!0.5!(pol.south |- fby)$)
  {replayed trajectories};

\end{tikzpicture}}
\caption{\textbf{Training pipeline.} The \textbf{14 rewards} in six categories feed one reward vector. Blue: oracles, orange: reward dimension reduction, teal: conditioning, purple: policy. \Cref{tab:objectives} lists every reward and \cref{sec:projection} derives the projection.}
\vspace{-4mm}
\label{fig:pipeline}
\end{figure}

Our method introduces four components: 1) a modular reward suite for antibody design (\cref{sec:obj}), 2) a learned reward projection onto a fixed-size preference simplex (\cref{sec:projection}), 3) a scaffold-conditioned span-in-filling policy that proposes CDRH3 sequences in framework context (\cref{sec:policy-architecture}), and 4) a preference-conditioned GFN objective that samples diverse high-reward sequences under different tradeoffs (\cref{sec:methods_mogfn}). Generated pools are then sharpened by a structure-based filtering pipeline (\cref{sec:postfilter-benchmark-evaluation}). \Cref{fig:pipeline} shows the training framework.

\subsection{Rewards}
\label{sec:obj}
\vspace{-2mm}

We optimize up to $M = 14$ independent reward terms in six categories: structural stability, immunogenicity, developability, liability, likelihood, and novelty (see \cref{tab:objectives}). All variance in the rewards is driven by changes to the CDRH3 loop, regardless of whether the oracles required additional framework-specific context. \textbf{Structural stability} guides the sampler to produce loops with high likelihood under the common inverse-folding models ESM-IF, Boltz-IF, and ProteinMPNN \citep{hsu2022esmif, dauparas2022proteinmpnn, stark2025boltzgen}. \textbf{Immunogenicty} is measured by an MHC-II binding predictor built on a frozen ESM-2 8M backbone~\citep{lin2023esm} to detect antibody designs that could trigger immune responses. \textbf{Developability} captures whether an antibody can be practically manufactured, and is measured using a fine-tuned ESM-2 650M sequence classifier~\citep{lin2023esm} on SAbDab-derived binary labels~\citep{chen2020developability}. \textbf{Sequence liability} (chemical and post-translational) was measured using a motif scan adapted from the Liability Antibody Profiler~\citep{satlawa2024lap}. \textbf{Sequence likelihood} was estimated under our AMPLIFY-AR model, a CDRH3-specific masked language model \texttt{h3berta}~\citep{rodella2026h3berta}, and \texttt{ablang2}~\citep{olsen2024ablang2}, which scores the loop inside its target's paired variable domain and is less germline-biased. \textbf{Novelty} was incentivized using the minimum normalized Levenshtein distance and BLOSUM62~\citep{henikoff1992blosum} substitution score to the training CDRH3 set, minimized over training loops of the same length. All rewards were normalized using an online min-max normalizer to facilitate multiobjective training. See \cref{apx:rewards} for details for more details.

\paragraph{Reduction to Slots.}
\label{sec:projection}

Covering a $M$-simplex to resolution $\epsilon$ requires $\binom{1/\epsilon + M - 1}{M-1}$ cells, so conditioning the policy on the native preference simplex is not tractable with larger values of $M$. We therefore use a learned reward projection with a fixed number of \emph{preference slots} $D \ll M$~\citep{park2025rdr}. Let $\hat{\br}(\bx) \in [0,1]^M$ be the normalized reward vector. The free parameter matrix $\bTheta \in \R^{D \times M}$ induces $\bA = \softmax(\bTheta, \, \mathrm{dim}=1) \in \R^{D \times M}$, which allows us to calculate the reduced $\bz(\bx) = \hat{\br}(\bx)\, \bA^\top \in [0,1]^D$, ensuring $\bA$ is strictly positive and row-stochastic regardless of convergence\footnote{Since every entry of $\bA$ is strictly positive, $\hat{\br}' \succeq \hat{\br}$ with $\hat{\br}' \neq \hat{\br}$ implies $\bA(\hat{\br}' - \hat{\br}) > 0$ in every coordinate, so $\bz'$ strictly dominates $\bz$. This preserves dominance and ensures that no $\bz$-Pareto-optimal sample is $\hat{\br}$-dominated.}. The influence of this reduction on the Pareto front is analyzed in \cref{sec:results-projection}. Preferences are then drawn on the $D$-simplex, $\bw \sim \mathrm{Dir}(\balpha)$ with $\balpha = (1,\dots,1) \in \R^D$, and the scalarized reward is a convex combination of the slot rewards $R(\bx; \bw) = \sum_{k=1}^{D} w_k \, z_k(\bx)$. $\bA$ is trained by reconstruction on a decoder $h_\psi$ reward batches\footnote{A stop gradient prevents the projection from being trained on the trajectory-balance loss to avoid a constant reward degenerate optimum.}:
$\Ls_{\mathrm{rec}} = \frac{1}{B M}\sum_{b,j} \frac{1}{\sigma_j^2}\Big( h_\psi(\bz_b)_j - \big(\hat{r}_{bj} - \mu_j\big) \Big)^2$, where $\bA$ is frozen before MOGFN training terminates so the the GFN can train against a static reward during the final stages of training (see \cref{apx:projection} for details).

\paragraph{Reward sharpening.} The scalarized reward is sharpened as $R(\bx;\bw)^\beta$ before entering the GFN objective to set the exploitation--controllability trade-off. Reported experiments use $\beta = 8$, see \cref{apx:beta} for details.

\subsection{AMPLIFY-AR Span-In-Filling Policy}
\label{sec:policy-architecture}
\vspace{-2mm}

We trained an autoregressive span-in-filling variant of AMPLIFY-350M~\citep{fournier2024amplify} from scratch on the original UniRef100~\citep{uniref}, OAS, and SCOP~\citep{chandonia2022scope} data (1{,}050{,}000 steps; $1.77\times10^{11}$ in-filling tokens; \cref{apx:amplify-infill}) to initialize $\pi_\theta$. Sequences were rearranged in FIM format:
\[
  \texttt{<bos> prefix <mask> suffix <sep> infill <eos>},
\]
and the model was trained with causal cross-entropy on the missing span. Span lengths were sampled uniformly from 1--50 residues, covering the 5--24-residue CDRH3 range. 
This enables teacher-forced log-probability calculation in one forward pass rather than the $L$ passes required for masked-model pseudo-likelihood, yielding a \bbSpeedup{}$\times$ throughput improvement on held-out CDRH3 loops (see~\cref{apx:amplify-infill}).
\textbf{Conditioning} used sequential task and preference cross-attention after self-attention and before each conditioned layer's feed-forward network. Task conditioning used unpooled per-residue native heavy-chain representations from the frozen backbone, computed once per target. Preference weights $w_d$ were thermometer-encoded, combined with learned slot embeddings, and projected to the AMPLIFY hidden dimension. This ordering allows the same preference trade-off to act differently across target-conditioned hidden states. Each cross-attention output projection was initialized to zero. During sampling, the model predicts each CDRH3 residue conditioned on the framework prefix and suffix, previously generated CDRH3 residues, and scaffold/preference conditioning using two sequential cross-attention modules per conditioned layer. The resulting trajectory log-probability is: $\log P_F(\tau \mid \bw) = \sum_{t=1}^{L} \log \pi_\theta(x_t \mid \bx_{<t}, \bx_{\mathrm{frame}}, \bw)$. Full conditioning and cross-attention details are provided in~\cref{apx:policy-details}.


\subsection{GFN Training Objective}
\label{sec:methods_mogfn}
\label{sec:methods_loss}
\vspace{-2mm}

We train a preference-conditioned MOGFN~\citep{jain2023mogfn} to sample CDRH3 sequences along the Pareto frontier of the design objectives. For a fixed preference vector $\bw$, the GFN target distribution is proportional to the scalarized reward, $R(\bx; \bw) / Z(\bw)$, where $Z(\bw)$ is the preference and context-dependent normalizing partition function. The FIM policy above defines the forward transition probability $P_F$: we use $\pi_\theta$ when emphasizing the per-step action distribution and $P_F$ when referring to trajectory-level probabilities.

We train with the trajectory balance (TB) objective~\citep{malkin2022tb}. Let $P_F(\tau)$ and $P_B(\tau)$ denote the forward and backward probabilities of a complete trajectory $\tau$ that generates sequence $\bx$. TB requires $\log Z(\bw) + \log P_F(\tau \mid \bw) = \log R(\bx; \bw) + \log P_B(\tau)$, but as each autoregressive trajectory has a unique backward decomposition, $P_B(\tau) = 1$ and can be eliminated. Rather than learning a separate $\log Z(\bw)$ for each conditioning vector, we use a VarGrad-style centered-residual estimator~\citep{richter2020vargrad,malkin2023gfnvi,zhang2023robust}. With the per-sample residual $\delta_n = \log P_F(\tau_n \mid \bw) - \log R(\bx_n; \bw)$, the batch mean estimates $-\log Z(\bw)$ and the VarGrad loss minimizes residual variance over a batch of $N$ trajectories:
\begin{equation}
  \Ls_{\mathrm{VarGrad}} = \frac{1}{N} \sum_{n=1}^{N} \ell\!\left( \delta_n - \bar{\delta} \right), \quad \bar{\delta} = \frac{1}{N} \sum_{n=1}^{N} \delta_n.
\end{equation}

At the fixed point all residuals are equal and the loss is zero\footnote{The resulting gradient can be viewed as a REINFORCE estimator~\citep{williams1992reinforce} with the batch mean residual as a control variate~\citep{malkin2023gfnvi}.}. We take $\ell$ to be a log-cosh penalty, $\ell(\delta) = \log\cosh(\beta_{\mathrm{loss}}\delta)/\beta_{\mathrm{loss}}$, rather than a squared error, which bounds the influence of the large residuals that a sharpened ($\beta = 8$) reward and structure-to-structure variance produces during training. To further reduce structure-to-structure gradient variance, each optimizer step averages $K = 4$ independent single-structure updates. Training details are provided in~\cref{apx:implementation}.

\subsection{Post-Filter Cascade, Benchmark, \& Evaluation}
\label{sec:postfilter-benchmark-evaluation}
\vspace{-2mm}

\paragraph{Post-Filter Cascade.}
\label{sec:postfilter}

We calculate scores on the distribution of designs from each sampler with oracles of interest to the drug development pipeline, including liability, developability, structural plausibility, and structural compatibility. Each stage of the funnel is increasingly more computationally expensive to run, and is performed on a smaller subset of the previous stage's designs. Briefly the sampled pool undergoes a Stage 0 sequence liability motif scan, followed by a Stage 1 forward fold under ABodyBuilder3 (collecting the CDRH3 pLDDT confidence scores and scRMSD structural fidelity scores), a Stage 2 measure of self-aggregation propensity, a Stage 3 forward fold under Chai-1 (pLDDT \& RMSD), followed by a Stage 4 estimation of relative binding  and stability energies  as measured by FoldX and Rosetta. Stage 3's budget is 1000 per arm due to computational constraints, and Stage 4's is 250 (budgets referred to as B3 and B4 in text). See \cref{apx:postfilter} for all details of the evaluation.

\paragraph{Dataset.}
\label{sec:benchmark-pdb-selection}

Antibody-antigen complexes were curated from the Structural Antibody Database (SAbDab; ~\citet{dunbar2014sabdab}) and split by heavy-chain scaffold to evaluate generalization of CDRH3 to novel antibody frameworks, yielding 623 training, 10 validation, and 55 held-out test targets. Antigen chains were used only to confirm that held-out test targets represent distinct binding problems. Each arm was evaluated at a matched sampling budget (32{,}768 designs per arm per target for the ranked pipeline, 100 designs per arm for the uniform pipeline). We compare three GFlowNet arms (a pre-defined 9/14-reward mapping and two learned objective-slot models with $D = 4$ and $D = 5$), the AMPLIFY-AR policy provided the loop design performance of the prior. We also evaluated a uniform random policy, and compared against two structural inverse-folding baselines: ProteinMPNN \& ESM-IF. Full details  are provided in~\cref{apx:data-pipeline}.

\paragraph{Evaluation.}
\label{sec:evaluation-metrics}

We evaluate all arms both for their ability to produce high quality antibodies on the test scaffolds (as defined by both the reward set and the held out postfilter pipeline) conditions and sample diversity. On the held out scaffolds, we report oracle scores, postfilter score distribtions, distinct clusters among the acceptable designs (50\% loop identity), pareto front enrichment by the postfilter pipeline, and sequence diversity in both sequence and embedding space.
\section{Results}
\label{sec:results}

\subsection{Learned Projection}
\label{sec:results-projection}
\vspace{-2mm}

An analysis of the reward landscape can be found in ~\cref{sec:reward-landscape-analysis}. Briefly, our learned projections recover expected groupings in the data. Fitted on the native-PDB reward matrix at $D = 4$, the slots resolve to an inverse-folding stability cluster, a novelty cluster, the prior likelihood clusters with the Antibody ProteinMPNN and BoltzIF inverse fold logits, and developability \& immunogenicity comprise the final axis. These groupings were not found to vary with other training hyperparameters. See \Cref{apx:lrsweep} for more details. We also report results for a GFN trained on 9 objectives with a fixed reduction mapping: all stability oracles collapsed into one, developability, AMPLIFY-AR likelihood, and immunogenicity.

\subsection{Quality-Diversity: More Distinct Acceptable Candidates per Design}
\label{sec:results-qd}
\vspace{-2mm}

An analysis of pool-wise rewards across all arms (\Cref{apx:rewards}) reveals that the GFNs sample designs far from the AMPLIFY-AR prior, and score higher on novelty, immunogeniciy, and developability than any trained model. The liability filter removes $\approx 30\%$ of sampled designs from the $D=4$ and $D=5$ GFNs, as well as the inverse folding baselines, compared to $\approx 50\%$ for the random baseline, AMPLIFY-AR prior, and Fixed GFN (which did not train on the liability reward). An analysis of sequence diversity at this stage of the filter reveals that the GFNs, AMPLIFY-AR prior, and random baseline retain substantially more sequence diversity as we sweep cluster thresholds until $0.8-0.9$. In addition, the Vendi score \citep{friedman2023vendi} --- the effective number of distinct designs in a set, which equals 1 when every design is identical and $n$ when all $n$ are mutually dissimilar --- shows that these methods maintain substantially more diversity in the tails of the distribution, where the strongest final designs reside. \Cref{fig:postfilter-identity} reports all three measures --- acceptability, the cluster sweep and the Vendi score --- with bootstrap intervals over targets. Each sampler's pre-and-post filter scores can be found in \cref{fig:postfilter-cascade-quality}.

\begin{figure}[!htb]
\centering
\includegraphics[width=\textwidth]{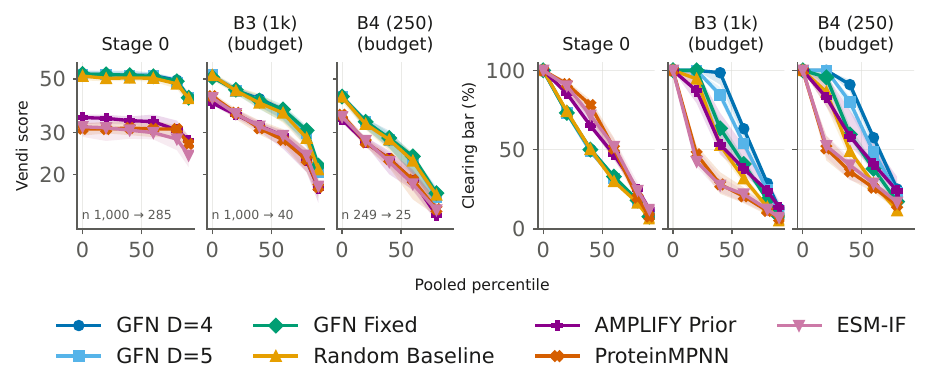}
\vspace{-7mm}
\caption{Diversity at matched quality per arm throughout the postfilter. Left: The Vendi score over ESM-2 650M embeddings of the loop in its own heavy-chain context, after subtracting each target's mean embedding, computed over percentile quality thresholds. Right: the fraction of each arm's designs over a percentile threshold. "Bar" represents the percentile of ABodyBuilder3 CDR-H3 pLDDT pooled over the arms within a target, Bands represent 95\% bootstrap CIs resampling targets. The GFN models show increased diversity relative to the other trained baselines across thresholds.}
\label{fig:postfilter-quality-diversity}
\vspace{-3mm}
\end{figure}

\begin{figure}[!htb]
\centering

\includegraphics[width=\textwidth]{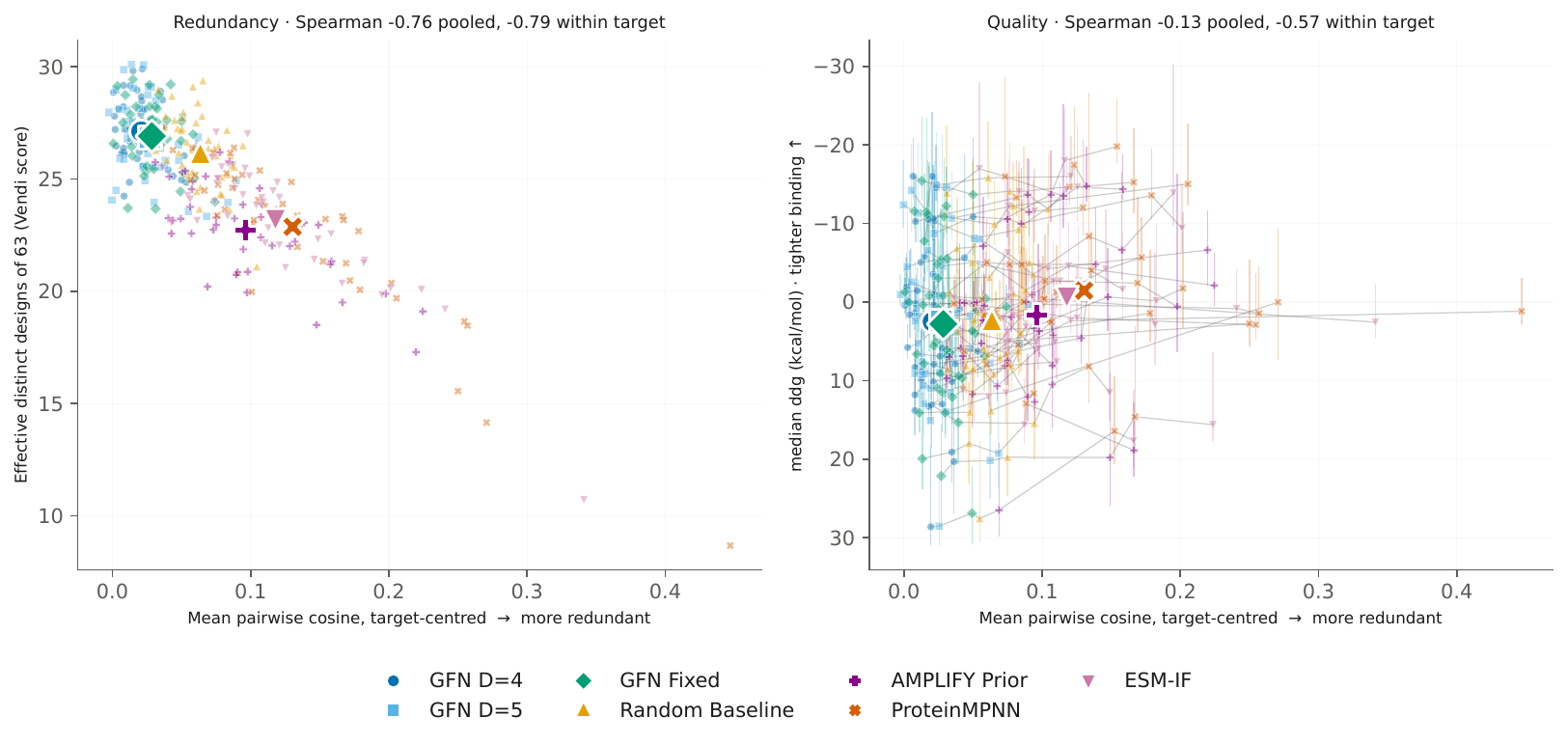}
\vspace{-5mm}
\caption{Sample redundancy (left) and final quality (right) among postfilter survivors ($n=250$ per arm). Sequences embedded in ESM-2 650M. Cosine distances taken after removing each target's mean. One dot per test-target. Large markers denote the arm's mean (right). Grey lines connect the designs across all arms for a single target, and whiskers denote within-arm 5-95\% CI. GFN-based models produce substantially more unique designs than all baselines in exchange for a reduction in median binding scores.}
\vspace{-5mm}
\label{fig:postfilter-cascade}
\end{figure}

\Cref{fig:postfilter-quality-diversity} shows how the diversity of our samples, measured by the Vendi score, varies with increasing the quality threshold (see also \Cref{fig:diversity} for baseline diversity measures at the beginning of the postfilter funnel). A clear pattern emerges: while the GFN models have lower quality scores and lower survival rates, the  sequences which do survive are still more diverse than the higher quality samples from our baseline methods. This relationship between quality and diversity is made most clear in \Cref{fig:postfilter-cascade}, which shows not only that the Vendi scores are highest for the GFlowNets at the end of the postfilter funnel (even higher than the random baseline), but the distribution of survivors from the GFNs contains very promising binders. \Cref{fig:landscape-rarefaction} demonstrates that the diversity gap between the GFN and structural approaches grows as more designs are drawn after the Chai-1 and Rosetta stages. \Cref{fig:landscape-umap} shows where each arm's surviving designs sit on a shared landscape, showing the GFN samples overlapping the design space occupied by structual models, our random baseline, and the AMPLIFY-AR prior.

\subsection{Reward Dominance}
\label{sec:results-pareto}
\vspace{-2mm}

\paragraph{Pooled Pareto front.}

\begin{table}[!htb]
\centering
\caption{Joint Pareto front pooled over all seven arms, per target, summed over \ParetoPooledFourTargets{} held-out targets. }
\label{tab:pareto-spaces}
\scriptsize
\setlength{\tabcolsep}{4pt}%
\begin{tabular}{@{}lrrccccccc@{}}
\toprule
Objectives & Front & \% of pool & ESM-IF & ProteinMPNN & GFN D=4 & GFN D=5 & GFN Fixed & AMPLIFY & Random \\
\midrule
4 (collapsed) & 19{,}501 & 0.19\% & 27.4\% & 12.4\% & 3.8\% & 2.8\% & 6.7\% & 46.5\% & 0.4\% \\
9 (common) & 86{,}972 & 0.85\% & 40.6\% & 25.0\% & 2.0\% & 1.5\% & 3.5\% & 27.1\% & 0.2\% \\
14 (full) & 2{,}115{,}506 & 20.71\% & 21.9\% & 22.1\% & 9.8\% & 8.2\% & 9.7\% & 24.3\% & 4.0\% \\
\bottomrule
\end{tabular}

\end{table}

The number of non-dominated samples naturally grows as $M$ grows, so we report three reductions of the Pareto front over rewards in \cref{tab:pareto-spaces}. The GFN models claim less of the front than the prior. The prior and inverse folding models hold roughly similar proportions of it. Preference controllability --- whether a design's realized rewards actually track the preference vector it was conditioned on --- is reported separately in \Cref{apx:controllability}.

\subsection{Structural Validation Through the Post-Filter}
\label{sec:results-postfilter}
\vspace{-2mm}

\begin{figure}[!htb]
\centering
\includegraphics[width=\textwidth]{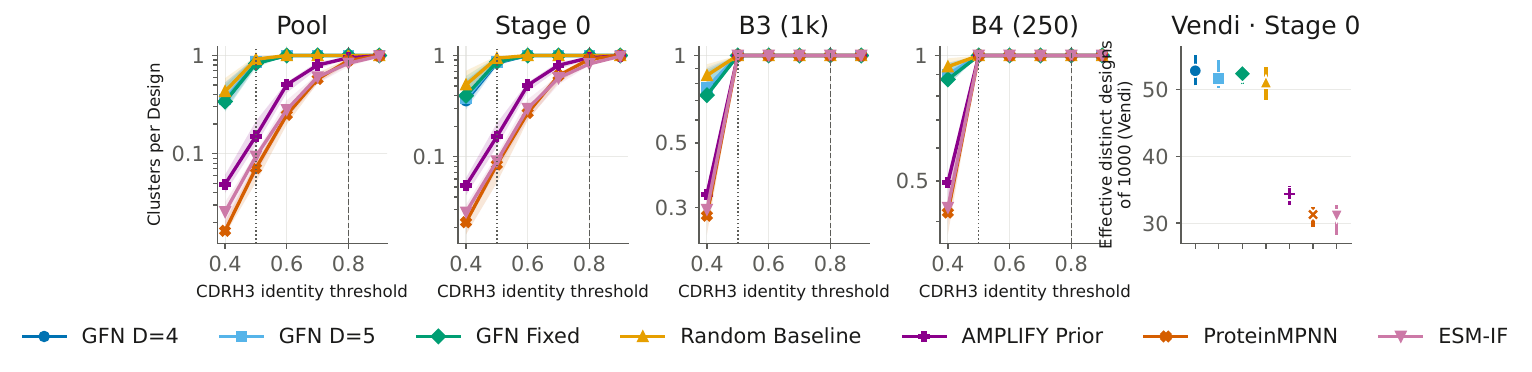}
\vspace{-7mm}
\caption{Design diversity throughout the post-filter pipeline across clustering thresholds. Dotted line at 0.50 marks the Stage-3 ranking threshold, \& dashed line at $\geq$0.80 marks the clonotyping reference for shared B-cell lineage. At earlier funnel stages, the three GFN arms and the random baseline retain more clusters at permissive thresholds up to the clonotyping threshold, while the AMPLIFY-AR prior tracks the inverse-folding baselines. Rightmost panel: Vendi score over ESM-2 650M embeddings of CDRH3 in heavy-chain context after subtracting each target's mean embedding. Bands show 95\% bootstrap CIs over resampling targets.}
\vspace{-5mm}
\label{fig:postfilter-identity}
\end{figure}

The GFNs match or modestly outperform the random baseline, while trailing structural models on end-funnel quality metrics. Recall that the GFNs outperform random baselines on the other outcome metrics and pipeline diversity (\Cref{apx:rewards}). Among non-dominated designs, stability oracles correlate most strongly with structural (as expected) and likelihood endpoints; novelty is most strongly anticorrelated, immunogenicity mildly so, while developability and liability are largely uncorrelated (\Cref{{fig:front-bias}}). CDRH3 spatial aggregation propensity (SAP)~\citep{chennamsetty2009sap} across Stage-1 folds is lower for three GFN arms than all baselines in both the distribution bulk and tail (\cref{tab:postfilter-summary}). \Cref{fig:postfilter-identity} further shows that structural models perform best on training-aligned rewards, whereas GFNs match or outperform baselines on novelty, immunogenicity, developability, and sequence liability.


\begin{table}[!htb]
\centering

\caption{Design performance through the funnel across 45 held-out targets and seven arms. Stage 2 is CDRH3 spatial aggregation propensity~\citep{chennamsetty2009sap} over the full pre-funnel Stage-1 population (ABodyBuilder3), where lower is better. Stage 3 reports forward fold RMSD to the crystal structure, and the CDRH3 agreement RMSD in \AA{} from ABodyBuilder3$\leftrightarrow$ and Chai-1 (lower is beter). CDRH3 pLDDT  \emph{Retained} is the fraction of each target's pLDDT top \acTopK{} surviving the best-\acKeepPercent{} agreement threshold. \emph{Stage 4} reports Top-1 $\Delta\Delta G$ (kcal/mol for FoldX, REU for Rosetta) and IMP, the fraction with $\Delta\Delta G < 0$.}
\label{tab:postfilter-summary}
\scriptsize
\vspace{-3mm}
\resizebox{\textwidth}{!}{%
\begin{tabular}{@{}lr rrr rrrr rrrrr@{}}
\toprule
& Stage 0 (45) & \multicolumn{3}{c}{Stage 2 --- CDR-H3 SAP (45)} & \multicolumn{4}{c}{Stage 3 --- fold agreement (45)} & \multicolumn{5}{c}{Stage 4 --- binding $\Delta\Delta G$ (45)} \\
\cmidrule(lr){2-2}\cmidrule(lr){3-5}\cmidrule(lr){6-9}\cmidrule(lr){10-14}
Arm & Surv.\ & Median & Best 10\% & Won & RMSD (\AA) & pLDDT & $\rho$ & Retained & FoldX & Rosetta & FoldX IMP & Rosetta IMP & $\rho$ \\
\midrule
GFN $D{=}4$ & 71.1\% & $-0.0088$ & $-0.0877$ & 0/45 & $3.54$ & $77.7$ & $-0.351$ & 40.6\% & $-4.96$ & $-8.88$ & 23.1\% & 17.2\% & $+0.47$ \\
GFN $D{=}5$ & 68.3\% & $-0.0116$ & $-0.0886$ & 0/45 & $3.65$ & $77.1$ & $-0.340$ & 40.2\% & $-5.50$ & $-9.73$ & 24.0\% & 17.5\% & $+0.46$ \\
GFN Fixed & 47.1\% & $-0.0200$ & $-0.0959$ & 40/45 & $3.55$ & $77.6$ & $-0.359$ & 40.1\% & $-4.78$ & $-9.31$ & 22.7\% & 16.8\% & $+0.46$ \\
\midrule
Random baseline & 46.0\% & $+0.0177$ & $-0.0619$ & 0/45 & $3.87$ & $76.0$ & $-0.283$ & 39.6\% & $-5.48$ & $-9.34$ & 24.0\% & 18.0\% & $+0.47$ \\
AMPLIFY prior & 80.1\% & $+0.0271$ & $-0.0443$ & 0/45 & $2.94$ & $80.4$ & $-0.337$ & 36.5\% & $-5.09$ & $-9.34$ & 24.2\% & 19.6\% & $+0.47$ \\
ProteinMPNN & 68.7\% & $+0.0138$ & $-0.0514$ & 2/45 & $3.02$ & $82.0$ & $-0.331$ & 35.0\% & $-6.20$ & $-14.52$ & 30.4\% & 24.6\% & $+0.52$ \\
ESM-IF & 72.1\% & $+0.0096$ & $-0.0587$ & 3/45 & $3.18$ & $80.6$ & $-0.297$ & 35.9\% & $-6.09$ & $-13.07$ & 27.7\% & 23.3\% & $+0.50$ \\
\bottomrule
\end{tabular}%
}

\vspace{-2mm}
\end{table}

\begin{figure}[!htb]
\centering
\includegraphics[width=\textwidth]{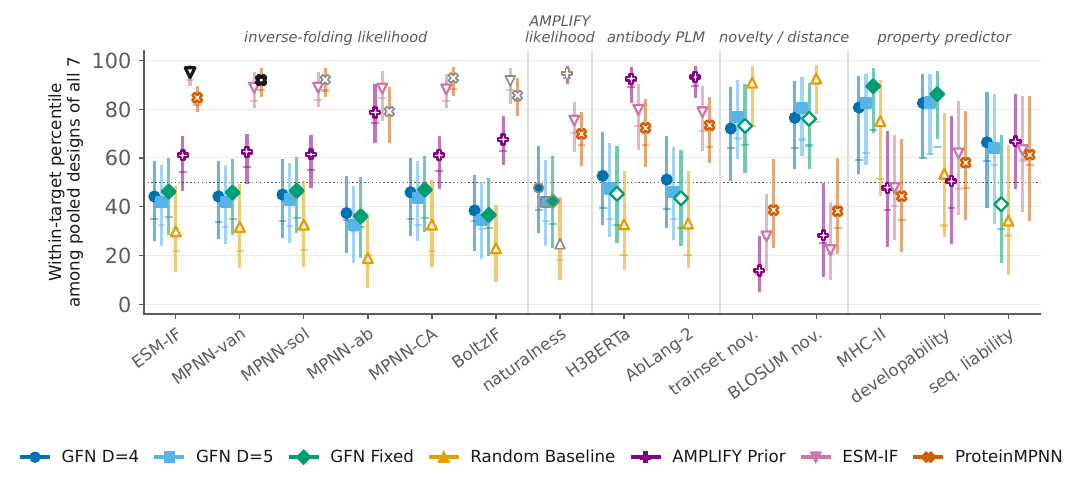}
\vspace{-7mm}
\caption{Non-dominated designs per generator. Markers show the median within-target percentile of each arm's front members per objective, bars span the interquartile range. Inverse-folding models favor non-dominated structural quality, whereas GFNs favor novelty while satisfying liability, developability, and immunogenicity objectives. Filled markers denote objectives used to train that arm; black rings indicate evaluation on a method's own outputs. Ticks denote whole-pool medians, with the tick-to-marker gap indicating distance to the front.}
\vspace{-5mm}
\label{fig:front-bias}
\end{figure}

Both inverse-folding baselines and the AMPLIFY prior exceed all three GFN arms in forward-fold loop confidence, while all arms exceed the random baseline (\cref{fig:funnel-forest}). It is important to note that forward-fold confidence is useful as a binary binder-rejection gate but not as a binding ranker~\citep{Bennett2026-xw}. Inverse-folding baselines also lead on Rosetta and FoldX endpoints (\cref{tab:postfilter-summary}). Rosetta Top-1 $\Delta\Delta G$ is $\sfourproteinmpnnvanillaRosettaTopOne$ for ProteinMPNN and $\sfouresmifRosettaTopOne$ for ESM-IF, versus $\sfourexpandedFourteenDFiveRosettaTopOne$ to $\sfourexpandedFourteenDFourRosettaTopOne$ for the three GFN arms. The FoldX margin is narrower ($\sfourproteinmpnnvanillaFoldxTopOne$ and $\sfouresmifFoldxTopOne$ versus $\sfourgflownetFoldxTopOne$ to $\sfourexpandedFourteenDFiveFoldxTopOne$) but consistent; the GFN arms do not outperform random on this measure. This advantage is expected because the energy functions assess sequence--backbone compatibility, closely aligned with the inverse-folding training objective, whereas the GFNs' inverse-folding logits provide only a rough proxy. The GFNs instead distinguish themselves in diversity at matched acceptability (\cref{sec:results-qd}). The Rosetta and FoldX pipelines largely corroborate each other.

\section{Limitations \& Conclusions}
\label{sec:limitations-and-conclusions}
\vspace{-2mm}

All experiments are computational and rely on learned or sequence and structure-derived oracles which may not reflect experimental outcomes. Generated candidates therefore require experimental validation before their potential therapeutic relevance can be assessed. Our sampler is also restricted to a fixed CDRH3 length per scaffold to support the inverse-folding oracles, substantially reducing the accessible sequence space. Future work could enable variable-length generation using alternative structural priors and potentially more accurate, albeit considerably more expensive, oracle evaluations.

Our method frames fixed-framework CDRH3 redesign as preference-conditioned multi-objective generation followed by structural filtering. By fine-tuning a span-in-filling AMPLIFY backbone against fourteen modular oracles, we generate substantially more distinct acceptable candidates than inverse-folding baselines and explore regions of protein design space not recovered by structure-based approaches, at the cost of lower average per-sample quality. The practical value of this quality-diversity trade-off requires experimental validation. However, given that many functional proteins exist outside the sequence space represented in the UniRef90 and MGnify protein sequence databases~\citep{diplock2026unobserved}, sequence-based generators may provide a complementary approach for exploring novel functional protein sequence space. Our framework can incorporate additional objectives, subject to oracle and computational cost, while the projection mechanism prevents the cost of covering preference space from growing with the number of objectives.

\newpage

\section{Ethics Statement}
\label{sec:ethics}

This work uses public protein sequence and structure data and does not involve human subjects, private personal data, or clinical deployment. The remaining risk is dual-use: generated binders should not be synthesized, tested, or deployed without appropriate biosafety review, target-risk assessment, experimental validation, and regulatory oversight. Our models inherit biases from SAbDab, UniRef100, and the learned reward models, including biases toward well-studied antibody families, antigens, and assay conditions. This work should not be considered safe to use in any autonomous discovery program.

\section{Reproducibility Statement}
\label{sec:reproducibility}

We will release all code, including dataset curation scripts, model training code, oracle training code, and the postfilter pipeline scripts, as a open-source monorepo upon publication. We will additionally release all oracles, the FIM AMPLIFY-AR model, and fine-tuned GFNs as open source checkpoints. The release will also include the SAbDab train and held-out splits, oracle preprocessing code, and method for generating all cached training artifacts (e.g., inverse-folding logits) in a reproducible manner. We will additionally release training configurations, random seeds, evaluation scripts, and figure-generation scripts. These artifacts are intended to allow the community to inspect the work, improve the framework within the domain of antibodies, as well as adapt it to other biological applications.

\section{AI Assistance Disclosure}
\label{sec:ai-assistance}

Large language models (LLMs) were used to assist with manuscript drafting (primarily consistency checks, accuracy of claims validation, and LaTeX automation), and for code generation for all model evaluations. Many elements of this codebase predate the advent of human-level coding agents, so most core methods and oracles were implemented entirely by hand, with some assistance on individual methodological improvements. LLMs were used under heavy supervision for figure and table generation. They were not used for research ideation, method design, or experiment design. The authors reviewed all code underlying reported results and re-wrote any generated manuscript text by hand.

\newpage
\bibliography{bibliography}
\bibliographystyle{iclr2027_conference}

\clearpage
\appendix
\etocdepthtag.toc{appendix}
\etocsettagdepth{mainmatter}{none}
\etocsettagdepth{appendix}{subsection}

\tableofcontents

\newpage

\section{Implementation Details}
\label[appendix]{apx:rewards}

\paragraph{Structural stability.}
\label{sec:stability-reward}

We guide our sequence sampler to produce loops with high likelihood under common inverse-folding models conditioned on the native backbone coordinates. For a candidate sequence $\bx$ and structure $s$, the single-predictor stability score is the mean conditional log-probability $\widetilde{R}_{\mathrm{stab}}^{(m)}(\bx, s) = \frac{1}{L} \sum_{i=1}^{L} \log p_m(x_i \mid \mathbf{X}_s^{\mathrm{bb}})$, where $\mathbf{X}_s^{\mathrm{bb}}$ denotes the backbone coordinates of structure $s$ and $p_m$ is the amino acid distribution predicted by a single inverse folding predictor. These are pre-computed from six inverse-folding model families as independent objectives and used as lookup tables during training \ref{apx:data-pipeline}). See \cref{apx:stability} for model details and validation.

\paragraph{Immunogenicity.}
\label{sec:mhc2-reward}

We train an MHC-II binding predictor built on frozen ESM-2 8M backbone~\citep{lin2023esm} to detect antibody designs that could trigger immune responses. Five models were trained via cross-validation, which we ensembled as our oracle. Each candidate CDRH3 is scored against a reference panel of 27 human leukocyte antigen (HLA) class~II alleles~\citep{greenbaum2011functional}, and the reward penalizes the maximum predicted immunogenicity against any target $R_{\mathrm{mhc2}}(\bx) = 1 - \max_{h \in \cH} \, f_{\mathrm{mhc2}}(\bx, h)$, where $f_{\mathrm{mhc2}}$ is the ensemble predictor, $\cH$ is the 27-allele HLA-II panel. Architecture, training, and evaluation details are in \cref{apx:mhc2}.

\paragraph{Developability.}
\label{sec:developability-reward}

Developability captures whether an antibody can be practically manufactured. We fine-tune an ESM-2 650M sequence classifier~\citep{lin2023esm} on SAbDab-derived binary labels~\citep{chen2020developability} based on the Developability Index~\citep{lauer2012di}, which combines CDR spatial aggregation propensity with variable-region net charge, computed by \citet{chen2020developability} over the variable region. The heavy chain with the candidate loop spliced in is scored as the developable-class probability $R_{\mathrm{dev}}(\bx) = \softmax(f_{\mathrm{dev}}(\bx^{-}_{\mathrm{frame}} \oplus \bx \oplus \bx^{+}_{\mathrm{frame}}))_1$, where $\bx^{\pm}_{\mathrm{frame}}$ are the framework prefix and suffix and index 1 is that class. Training and cross-validation details are in \cref{apx:developability}.

\paragraph{Sequence liabilities.}
\label{sec:liability-reward}

Chemical and post-translational liabilities are a distinct failure mode from aggregation-driven developability, and can be estimated directly from sequence. We score the designed loop with a motif scan adapted from the Liability Antibody Profiler~\citep{satlawa2024lap}, restricted to the designed span and covering deamidation (\texttt{NG}), isomerization (\texttt{DG}), N-linked glycosylation sequons (\texttt{N-x-[S/T]}), unpaired cysteine, oxidation-prone Met/Trp, and \texttt{DP} fragmentation $R_{\mathrm{liab}}(\bx) = - \sum_{u \in \mathcal{U}} \kappa_u \, n_u(\bx)$, where $n_u$ counts occurrences of motif $u$ and $\kappa_u$ is its severity weight (hard-flagged motifs carry larger weight). See~\cref{apx:liability} for motif definitions and the full derivation.

\paragraph{Sequence likelihood.}
\label{sec:naturalness-reward}

Three likelihood oracles score each candidate loop at distinct scopes. Our AMPLIFY-AR model is used to measure how likely a generated loop is under the pretrained span-in-filling AMPLIFY prior (\cref{sec:policy-architecture}). The candidate loop is teacher-forced after \texttt{<sep>}, and the reward is the length-normalized conditional log-likelihood $R_{\mathrm{nat}}(\bx) = \frac{1}{L} \sum_{i=1}^{L} \log p_{\mathrm{FIM}}(x_i \mid \bx_{<i}, \bx_{\mathrm{frame}})$, where $p_{\mathrm{FIM}}$ is the frozen span-in-filling model, $\bx_{\mathrm{frame}}$ denotes the framework prefix and suffix, and $\bx_{<i}$ are previously generated CDRH3 residues. \texttt{h3berta}~\citep{rodella2026h3berta} is a RoBERTa-base masked language model pretrained on over 18M isolated CDRH3 strings from healthy-donor repertoires context-free, serving as a generic CDRH3 prior. \texttt{ablang2}~\citep{olsen2024ablang2} scores the loop inside its target's paired variable domain. Most antibody protein language models (PLMs) predict the germline residue at $>80\%$ of somatically-mutated positions~\citep{olsen2024ablang2}, in conflict with our novelty and exploration objectives -- \texttt{ablang-2} is trained with a focal loss~\citep{lin2017focal} on rare non-germline predictions to avoid this. Validation details for all three are in \cref{apx:naturalness}.

\paragraph{Novelty.}
\label{sec:novelty-reward}

Two rewards incentivize moving away from the SAbDab distribution. Training set novelty is the minimum normalized Levenshtein distance to the training CDRH3 set, Blosum novelty is the analogous distance under a BLOSUM62~\citep{henikoff1992blosum} substitution score, minimized over training loops of the same length. PLM embeddings are intentionally not used for this reward so that the downstream diversity analysis which depends them is not a training signal.

\paragraph{Normalization.}
\label{sec:reward_norm}

As rewards span heterogeneous scales\footnote{Log-probabilities for stability and likelihood oracles, probability-derived scores for immunogenicity risk and developability, and counts for liabilities.}, we normalize each reward with a learned min-max bounds that adapt by EMA before freezing during the final training phase. The clipped normalized reward is $\hat{R}_k(\bx) = \mathrm{clip}\!\left(\frac{R_k(\bx) - b_k^-}{b_k^+ - b_k^-},\, 0,\, 1\right)$, where $b_k^- = r_k^{\min} - \epsilon_{\mathrm{m}} (r_k^{\max} - r_k^{\min})$ and $b_k^+ = r_k^{\max} + \epsilon_{\mathrm{m}} (r_k^{\max} - r_k^{\min})$\footnote{Margin $\epsilon_{\mathrm{m}} = 0.05$ to prevent score pileup at the boundaries.}. The warmup, EMA, and freeze updates are specified in \cref{apx:policy-details}.

\label[appendix]{apx:implementation}

This appendix records implementation settings for the AMPLIFY span-in-filling backbone and AbGFN fine-tuning.

\subsection{AMPLIFY-350M Autoregressive Span-In-Filling}
\label[appendix]{apx:amplify-infill}
\vspace{-2mm}

The GFN backbone is an AMPLIFY-350M-scale transformer~\citep{fournier2024amplify} with hidden size $d_h = 960$ and 32 transformer layers, trained on the same data as the AMPLIFY-350M checkpoint: UniRef100~\citep{uniref}, Observed Antibody Space (OAS), and SCOP~\citep{chandonia2022scope}, with the original AMPLIFY data versions and filtering rules, from the identical preprocessed files (2024-09-25 snapshot; 386 shards; 3{,}611{,}708 OAS and 24{,}495 SCOP training sequences, and $\approx$394M UniRef100 sequences per epoch). Every training sequence contained one interior span with a length uniformly sampled between 1-50, clamped by sequence length.This range covers the 5--24-residue CDRH3 lengths used in our experiments. The span start location was sampled uniformly from valid positions. In contrast to IgLM, no CDR-annotated or otherwise guided spans were used. The input format was \texttt{<bos> prefix <mask> suffix <sep> infill <eos>} over raw amino-acid tokens, the objective was causal shifted-token cross-entropy over the missing span. It trained for 1{,}050{,}000 optimizer steps over 17 epochs in two stages that share an effective batch of $1{,}998{,}848$ tokens per optimizer step: the first 1{,}000{,}000 steps at context length 512 (32 $\times$ 62{,}464 tokens per device, no accumulation), then the next 50{,}000 steps at context length 2048 (32 $\times$ 15{,}616 tokens per device, 4 accumulation steps). Sequences were packed with FlashAttention~\citep{dao2022flashattention} variable-length attention. Training comprised $1.77 \times 10^{11}$ loss-bearing in-filling tokens ($2.10 \times 10^{12}$ tokens including padding and unmasked context). Training used the AdamW optimizer~\citep{loshchilov2019adamw} (fused) at peak learning rate $10^{-3}$, $\beta = (0.9, 0.95)$, $\epsilon = 10^{-8}$, weight decay $0.01$, and gradient-norm clip $1.0$, with cosine decay to a floor of $10\%$ of peak ($10^{-4}$) over a 900{,}000-step horizon after 1{,}000 warmup steps. The final 150{,}000 steps were run at the clamped floor. Training used bf16 mixed precision with TF32 matmuls. The run used 32 $\times$ NVIDIA H100 80GB (4 nodes $\times$ 8) for 130.6 hours wall-clock ($\approx$ 4{,}180 GPU-hours). Three held-out sets are evaluated every 1{,}000 steps: \texttt{uniprot\_dev} (10{,}000 sequences), \texttt{oas\_dev} (20{,}000 chains from 10{,}000 paired antibodies), and \texttt{scop\_dev} (10{,}000 domains), with a final in-filling loss of $1.430$ (\texttt{uniprot\_dev}) and $0.238$ (\texttt{oas\_dev}). 

For CDRH3 generation, the native CDRH3 is treated as the missing FIM span, such that the model autoregressively generates the loop following the \texttt{<sep>} token while conditioning on the framework prefix and suffix. This formulation also permits the conditional log-probability of a complete CDRH3 sequence to be calculated in a single teacher-forced forward pass, rather than the $L$ forward passes required for leave-one-out pseudo-likelihood evaluation with a masked language model for a loop of length $L$. On \bbThroughputN{} held-out CDRH3 loops, the autoregressive model evaluates \bbArLoopsPersec{} loops/s compared with \bbMlmLoopsPersec{} loops/s for masked-model leave-one-out pseudo-likelihood, a \bbSpeedup{}$\times$ throughput improvement. The observed speedup approximately tracks the median loop length of \bbMedianSpan{} residues, reflecting the reduction from $L$ model evaluations to one while accounting for fixed computational overhead.

\subsection{MOGFN Fine-Tuning}
\label[appendix]{apx:mogfn-finetuning}
\vspace{-2mm}

The policy was initialized with all but the final 4 of its 32 transformer layers frozen, and a freshly initialized output head (MLP $d_h \to 2 d_h \to 2 d_h \to 20$) over the 20 amino-acid action space. Each fine-tuned transformer layer was wrapped in two sequential cross-attentions (scaffold, then preference), with 8 heads, dropout 0.1, independent projections, and zero-initialized output projections. Preferences were encoded per slot by a thermometer encoding with $B = 50$ bins, joined with a learned 64-dimensional slot-identity embedding and passed through a shared MLP $B + 64 \to 2 d_h \to d_h$ (SiLU, dropout 0.1) and RMSNorm, which gave $D$ key-value tokens. The task embedding was the per-residue hidden states of the native heavy-chain scaffold with its natural CDRH3 sequence, taken after the frozen layers of the span-in-filling backbone and cached per target; it gave $L_{\mathrm{task}}$ key-value tokens with a key-padding mask, unpooled. Trajectories were per-amino acid autoregressive CDRH3 generation after \texttt{<sep>} employing KV-cache sampling. The optimizer was AdamW at $\mathrm{lr}=10^{-4}$ with 200 warmup steps, cosine decay to $0.1\times$, and gradient-norm clipping, over $K = 4$ target structures per gradient step. The log-cosh scale is per-structure and annealed, $\beta_{\mathrm{loss}}(t) = c(t)/(\hat{\sigma}_k + \epsilon)$ with $c(t) = c_{\max} e^{-t/\tau_\beta} + c_{\min}(1 - e^{-t/\tau_\beta})$, where $\hat{\sigma}_k$ is a per-structure EMA of the residual standard deviation for structure $k$ (see \cref{apx:logcosh}). Each scaffold update drew one distinct target structure and one preference vector $\bw \sim \mathrm{Dir}(\alpha)$, and sampled $N = 256$ completions under that fixed scaffold and preference\footnote{Holding both fixed across a sub-batch allows the FIM prefix be shared through a KV cache (\cref{apx:kvcache}) and gives a valid $\log Z(\bw)$ estimate for that one condition.}. The distinct $K$ structures are drawn uniformly before drawing from the replay buffer, and afterwards with a staleness priority. All $KB = 1024$ candidates were scored in one batched oracle call and each sub-task contributes proportionally to the gradient. Training trajectories were sampled using temperature scaling on the policy per-step logits, $q(\tau) = \prod_t \mathrm{softmax}(z_t/T)_{a_t}$, with $T$ annealed linearly from $2$ to $1.5$ over training. The trajectory-balance loss used the untempered $\log P_F$ throughout, and all reported samples are drawn from the untempered policy (see \cref{apx:policy-details} for details). The replay buffer stores raw $M$-dimensional reward vectors, which are re-projected using $\bA$ per step.

\subsection{Log-Cosh Residual Penalty}
\label[appendix]{apx:logcosh}
\vspace{-2mm}

The log-cosh training schedule used $c_{\max}=10$, $c_{\min} = 2$, $\sigma_0=10$, EMA decay $\gamma=0.999$ and $\epsilon$ of $10^{-6}$, and $\tau_\beta = 875$. Training proceeded for either 1750 train steps (9 objective GFN) or 3000 steps for the 14-objective arms. The schedule results in the loss quadratic-to-linear knee moving from $\lvert\delta\rvert = 1.0$ out to 3.2 and 4.4, respectively, over training. Replay had capacity 20{,}000, staleness 3.0, and warmup 262 steps. Reward normalization was online min--max with warmup, EMA adaptation ($\gamma_{\mathrm{ema}}=0.998$), a freeze at step 350, and margin $\epsilon_{\mathrm{m}}=0.05$. The projection used either $D = 4$ or $D = 5$, $\lambda_{\mathrm{dec}} = 0$, EMA $\tau = 0.99$, a freeze at $0.60\times$ train steps, a warm start of 2000 steps, and reference fraction 0.5. Evaluation during training happens every 250 steps on 32 Dirichlet-sampled preferences over 5 validation structures.

\citet{hu2025lossdesign} show that the regression loss $g$ applied to trajectory-balance residuals determines an $f$-divergence between the target and the learned distribution;\citet{fawkes2026ftb} extend the correspondence to translation-invariant losses and off-policy data. They classify losses by two properties of that divergence: zero-forcing ($f(0) = \infty$, which penalises putting mass outside the target distribution, i.e., favouring exploitation) and zero-avoiding ($f'(\infty) = \infty$, which penalises missing mass, favouring exploration). The squared loss $g(t) = \tfrac12 t^2$ is zero-forcing and not zero-avoiding. The three losses they propose all grow faster than quadratic as $t \to +\infty$, but only Shifted-Cosh does so in both tails: Linex(1) and Linex(1/2) are linear as $t \to -\infty$, which is why neither is zero-forcing. Log-cosh sits at the opposite end from the squared loss, so it is neither zero-forcing nor zero-avoiding, and does not contain the squared loss exploitation pressure. Our annealing of $c(t)$ moves the loss back towards quadratic over the course of training, growing progressively biased toward exploitation, in an attempt to reduce the magnitude of tail residuals without discarding them; \citet{chen2026agfn} anneal the policy mixing weight instead. The gradient is $\tanh(\beta_{\mathrm{loss}}\delta)$ --- $0.81$, $0.55$ and $0.38$ times the squared-error gradient at $1\sigma_\delta$, $2\sigma_\delta$ and $3\sigma_\delta$ --- so the penalty acts as a smooth per-sample gradient clip that relaxes as $c(t)$ decreases.

\subsection{Policy Conditioning and Sampling}
\label[appendix]{apx:policy-details}
\vspace{-2mm}

For AMPLIFY-AR span-in-filling pretraining, we sample a single interior span by first drawing length uniformly from 1 to 50 residues, clamped by the sequence length, and then drawing a valid start position uniformly at random. The tokenized sequence is rearranged as
\[
    \texttt{<bos> prefix <mask> suffix <sep> infill <eos>}.
\]

The preference vector $\bw \in \Delta^{D-1}$ is encoded via thermometer encoding into $\R^{BD}$ ($B = 50$ bins per slot). Given $B$ linearly spaced thresholds $\xi_1, \dots, \xi_B$ over $[0, 1]$ with gap $\Delta\xi = \xi_{j+1} - \xi_j$, the thermometer encoding of a scalar $c$ is:

\begin{equation}
  \mathrm{therm}(c)_j = \frac{\mathrm{clamp}(c - \xi_j,\, 0,\, \Delta\xi)}{\Delta\xi}, \quad j = 1, \dots, B.
\end{equation}

Each slot is tokenized separately. We joint the thermometer slice $\mathrm{therm}(w_d) \in \R^{B}$ with a learned slot-identity embedding $\mathbf{a}_d \in \R^{64}$ and passed through a shared two-layer MLP ($B + 64 \to 2 d_h \to d_h$, with SiLU and dropout 0.1) and a final RMSNorm, where $d_h = 960$ is the AMPLIFY hidden dimension. This gives $D$ preference key-value tokens $\Epref \in \R^{D \times d_h}$. The MLP is shared across slots, followed by a RMSNorm.

Scaffold (or task) conditioning is handled by the key-value set $\Etask \in \R^{L_{\mathrm{task}} \times d_h}$, i.e., the full sequence of per-residue hidden states of the native heavy chain $\bx_{\mathrm{frame}}$ (including its natural CDRH3) after the frozen layers of the span-in-filling backbone, with a key-padding mask over padded positions. It is computed once per target and cached, with no pooling applied. The cross attention conditioning mechanism rests between the self-attention and the feed-forward network of the transformer backbone, first applying scaffold cross-attention before preference cross-attention. This ordering enabled the preferences to modulate generation according to preferences from hidden states informed by the task-scaffold, i.e., the same objective trade-off can be differently applied for different targets. Each cross-attention output projection was initialized to zero. Given input token representations $\mathbf{H}^{(\ell)} \in \R^{L_{\mathrm{seq}} \times d_h}$ at layer $\ell$, and writing $\mathrm{RMS}$ for RMS normalization:

\begin{align}
  \mathbf{H}' &= \mathbf{H}^{(\ell)} + \mathrm{SelfAttn}\!\left(\mathrm{RMS}_1(\mathbf{H}^{(\ell)})\right) \\
  \mathbf{H}'' &= \mathbf{H}' + \mathrm{Dropout}\!\left(\mathrm{CrossAttn}_{\mathrm{task}}\!\left(\mathrm{RMS}_2(\mathbf{H}'),\, \Etask,\, \Etask\right)\right) \\
  \mathbf{H}''' &= \mathbf{H}'' + \mathrm{Dropout}\!\left(\mathrm{CrossAttn}_{\mathrm{pref}}\!\left(\mathrm{RMS}_3(\mathbf{H}''),\, \Epref,\, \Epref\right)\right) \\
  \mathbf{H}^{(\ell+1)} &= \mathbf{H}''' + \mathrm{FFN}\!\left(\mathrm{RMS}_4(\mathbf{H}''')\right).
\end{align}

Each $\mathrm{CrossAttn}(\mathbf{Q}, \mathbf{K}, \mathbf{V})$ is multi-head cross-attention with 8 heads and dropout 0.1. The two cross-attention mechanisms have independent query, key, value, and output projections, with no weight sharing, and both output projections are initialized to zero.




\paragraph{Replay buffer.} The buffer is partitioned by target structure. Each entry holds one trajectory's actions, its per-step log-probabilities as written, and the raw $M$-dimensional reward vector. When a partition is consumed, its entries are re-scored under the current policy and filtered on the log importance ratio $\rho = \log P_F^{\mathrm{now}}(\tau) - \log P_F^{\mathrm{stored}}(\tau)$. Entries with $\lvert\rho\rvert > c_\rho$ are discarded, with $c_\rho = 3$. The $K$ structures per-step are drawn from the buffer without replacement with probability proportional to $t - t_s + 1$, where $t_s$ is the last step for structure $s$.

\paragraph{Reward normalization.} Three phases: 1) During warmup, a running min and max for each reward expands monotonically using an expanding window: $r_k^{\min} \leftarrow \min(r_k^{\min},\, \min_{\mathrm{batch}} R_k)$ and $r_k^{\max} \leftarrow \max(r_k^{\max},\, \max_{\mathrm{batch}} R_k)$. 2) During EMA adaptation, these bounds update with decay $\gamma_{\mathrm{ema}}$: $r_k^{\min} \leftarrow \gamma_{\mathrm{ema}} \, r_k^{\min} + (1 - \gamma_{\mathrm{ema}}) \min_{\mathrm{batch}} R_k$ (likewise for the max). 3) During the frozen phase, these bounds lock so the multi-objective GFN can learn from a stationary reward distribution.

\subsection{KV-Cache Sampling}
\label[appendix]{apx:kvcache}

To avoid sampling $t$ times during the FIM pass $[B, P + t]$ where $P \approx 120$ is the fill-in-the-middle prompt and $L \approx 5$--$25$, we use a standard KV-cache prefill-plus-incremental-decode under \texttt{no\_grad}. Prefill runs the full prompt through the frozen and conditioned layers capturing each layer's K and V, and each decode step embeds one new token, applies rotary embeddings at the token's absolute position, appends to the cache, and attends over it. Conditioned layers insert the same two cross-attentions against $\Etask$ and $\Epref$, which are fixed for the whole step and never cached: $\Etask$ is computed once per target and $\Epref$ once per preference vector. Within a sub-batch every sequence shares the same $P$ and $L$, and the $K$ sub-batches are sampled separately.

\section{Data Pipeline Details}
\label[appendix]{apx:data-pipeline}

\subsection{SAbDab Curation}
\label[appendix]{apx:sabdab-curation}

The SAbDab~\citep{dunbar2014sabdab} dataset comprises paired heavy/light chain sequences with antigen co-crystal structures, associated metadata (resolution, species, affinity where available), and PDB coordinate files. We removed entries missing heavy-chain, light-chain, antigen-chain, or antigen-type annotations, excluded multi-chain antigens, and retained only protein and peptide antigens. Entries were deduplicated by PDB code and heavy-chain identifier. Heavy-chain, light-chain, and antigen sequences were parsed from PDB ATOM records using Biopython~\citep{cock2009biopython}, with non-standard residues mapped to canonical one-letter amino acid codes where possible. CDRH3 boundaries were identified using ANARCI~\citep{dunbar2016anarci} under the IMGT numbering scheme~\citep{lefranc2003imgt} (positions 105--117), and gap characters at insertion positions were removed. Structures with CDRH3 length $<5$ were excluded because their sequence space is sufficiently small to enumerate exhaustively.

Structures resolved better than \curationResCut~\AA{} were retained, leaving \curationAfterRes{}. The retained set spans \curationResLowest--\curationResHighest~\AA{} (median \curationResMedian) over \curationProtein{} protein and \curationPeptide{} peptide antigens. We additionally remove \curationParserDropped{} samples due to parser disagreements between the structures and inverse-folding models: \curationBlacklist{} entries are excluded by a maintained list, and the rest fail the per-structure consistency check of the loop logits.

Train, validation, and test sets were defined based on the heavy-chain scaffold to evaluate whether the loop-design policy generalizes to novel antibody frameworks rather than memorizing CDRH3 sequences associated with nearly identical scaffolds. Because antibody frameworks are germline-conserved (median scaffold sequence identity $\approx0.85$ across unrelated antibodies in our dataset), we used a near-identity threshold of $\geq0.95$ to identify potential leakage. This removes exact and closely related affinity-matured scaffold duplicates across splits while retaining members of the same germline family as distinct design problems. Antigen information was not used to define the primary split, but was used to verify that held-out targets represented distinct antigen-binding problems. After curation, filtering, and leakage control, the final dataset comprised 623 training, 10 validation, and 55 held-out test targets from an initial universe of 784 structures, with 161 structures excluded during processing.

\begin{table}[!htb]
\centering
{\small
\begin{tabular}{@{}p{0.23\textwidth}p{0.40\textwidth}rr@{}}
\toprule
Step & Rule & Removed & Retained \\
\midrule
SAbDab entries after annotation cleaning & complete H/L/antigen annotation; single-chain protein or peptide antigen; one entry per PDB & --- & 1{,}905 \\
Resolution & $0 <$ resolution $\leq 2.3$~\AA{} & 1{,}032 & 873 \\
Parser blacklist & 17 PDB entries with known structure/inverse-folding parser mismatch & 17 & 856 \\
\midrule
\multicolumn{4}{@{}l}{\emph{Per-structure checks inside the inverse-folding loop}} \\
Chain residue-count check & parsed residue count must equal the heavy-chain sequence length & 67 &  \\
BoltzIF resolved-residue check & BoltzIF resolved residues must equal the sequence length & 4 &  \\
ProteinMPNN parse & structure must parse under every ProteinMPNN variant & 1 & 784 \\
Cross-model CDRH3 length agreement & all 24 models must agree on CDRH3 length, under Chothia and under IMGT & 0 & 784 \\
\midrule
\multicolumn{4}{@{}l}{\emph{Recorded per structure; never a cull}} \\
Sequence recovery & argmax agreement with the native CDRH3 & 0 & 784 \\
Boundary-shift test & native CDRH3 must score above every $\pm1..\pm5$ shift and above chance & 0 & 784 \\
\bottomrule
\end{tabular}
}

\caption{SAbDab curation, from the cleaned metadata table to the structures carrying precomputed inverse-folding logits. The first three steps act on the metadata; the next four are per-structure checks inside the inverse-folding loop, which drop a structure when the parsers disagree with each other or with the deposited sequence. Sequence recovery and the boundary-shift test are recorded per (structure, model) and never remove a structure, so a structure may be admitted carrying a failing flag. The per-structure drops are recovered from the run log, since the stored artifact records only the admitted set.}
\label{tab:curation-funnel}
\end{table}

\subsection{Inverse-Folding Logit Alignment}
\label[appendix]{apx:inverse-folding-logits}

We computed inverse folding logit matrices across the full curated SAbDab dataset for the stability oracles as follows:
\begin{enumerate}
  \item Each PDB structure is processed independently by ESM-IF, four ProteinMPNN variant families, and BoltzIF.
  \item For each model family $m$, stored logits are indexed as $\mathbf{g}_{v,e}^{(m)}(s) \in \R^{|\cA| \times L_{\mathrm{chain}}}$, where $v \in \cV_m$ indexes weight variants, $e \in \{1,\dots,E\}$ indexes ensemble members, $|\cA|$ is the vocabulary size, and $L_{\mathrm{chain}}$ is the chain length.
  \item Because models parse PDB files differently, each model records per-position PDB residue numbers. Cross-model alignment is performed at inference time via these residue identifiers.
  \item Full-chain logits are sliced to the CDRH3 region using both Chothia~\citep{allazikani1997standard,chothia1987canonical} and IMGT boundary definitions.
  \item The pipeline performs sequence-recovery, boundary-shift, and cross-model CDRH3 length-agreement checks (\cref{tab:curation-funnel}). Separately, a maintained list excludes \curationBlacklist{} PDB entries whose parsers are known to disagree.
\end{enumerate}

Sequence recovery (fraction of CDRH3 positions where the argmax prediction matches the native residue) was performed as a correctness evaluation for each structure, variant, and ensemble member:
\begin{equation}
  \mathrm{Recovery}_{v,e}^{(m)}(s) = \frac{1}{L} \sum_{i=1}^{L} \mathbb{1}\!\left[\argmax_a \, \mathbf{g}_{v,e}^{(m)}(s)_{a,i} = x_i^{\mathrm{native}}\right].
\end{equation}

A boundary-shift test was also performed to ensure correct loop placement: the mean log-probability of the native CDRH3 sequence is computed at the stated logit columns and at $\pm 1, \dots, \pm 5$ shifted positions. The stated position must yield a higher score than all shifted positions and must exceed uniform chance ($\log(1/20) \approx -3.0$). For cross-model agreement, structures where the \curationNModels{} models disagree on CDRH3 length under either the Chothia or the IMGT scheme are excluded, though no structure in the current set triggers this. The shift test is recorded rather than enforced: all \curationNModels{} models pass it on \curationShiftChothia{} of \curationUniverse{} structures under Chothia and on \curationShiftImgt{} under IMGT, and the per-model flags are stored so that a consumer can require agreement where it matters.

\subsection{Train, Valid, and Test Splits}
\label[appendix]{apx:split}

The generator conditions only on the heavy-chain scaffold; the light chain and antigen never enter generation, and the training reward is antibody-intrinsic (binding is computed only in the post-filter). The CDRH3 loop is the model's output. We therefore generate splits using a clustering of the framework region.

\begin{table}[!htb]
\centering
{\small
\begin{tabular}{@{}lrr@{}}
\toprule
Exclude scaffold identity $\geq T$ & training kept & excluded \\
\midrule
0.50 & 1 & 707 \\
0.70 & 43 & 665 \\
0.80 & 136 & 572 \\
0.90 & 529 & 179 \\
\textbf{0.95} & \textbf{660} & \textbf{48} \\
0.97 & 669 & 39 \\
1.00 & 691 & 17 \\
\bottomrule
\end{tabular}
}

\caption{Leakage menu on the heavy scaffold (heavy chain minus CDRH3) over the \curationUniverse-antibody pool: normalized identity of each training scaffold to its nearest held-out test scaffold. The $T = 1.00$ row excludes exact duplicates only.}
\label{tab:leakage-menu}
\end{table}

Since antibody VH frameworks all descend from a small set of germline genes, their scaffolds are similar even between unrelated antibodies: measured percentiles of nearest-test-scaffold identity are p50 $= \scafIdentityPfifty$, p75 $= \scafIdentityPseventyfive$, p90 $= \scafIdentityPninety$, p95 $= \scafIdentityPninetyfive$, p99 $= \scafIdentityPninetynine$.

Since the population sits at $\approx$0.85 median scaffold identity, ``zero leakage'' is not achievable in an absolute sense,  leakage on the scaffold axis is necessarily a near-identity notion we opt for 0.95 clustering to retain more training examples.

The minimum CDRH3 length of 5 removes \splitShortDropped{} antibodies: Loops of length $\leq 4$ span at most $20^4 = 160$k sequences which can be searched exhaustively, $L = 5$ jumps to 3.2M and serves as the lower length bound.

\begin{table}[!htb]
\centering
{\small
\begin{tabular}{@{}lr@{}}
\toprule
 & count \\
\midrule
Universe (structures with precomputed inverse-folding logits) & 784 \\
After minimum CDRH3 length $\geq 5$ & 753 \\
TRAIN & \textbf{623} \\
VALID (CDRH3-length-stratified) & \textbf{10} \\
TEST (RAbD-derived, reserved) & \textbf{55} \\
\texttt{exclude\_from\_training} & 161 \\
\bottomrule
\end{tabular}
}

\caption{Final split. The exclusion set (\splitExclude) is much larger than TEST $+$ VALID because leakage is defined on the scaffold: training antibodies whose scaffold is $\geq$0.95 identical to a held-out target are removed as well (\splitTestLeaks{} test-leaks, \splitValidLeaks{} valid-leaks), on top of the \splitShortDropped{} dropped by the length filter. Validation CDRH3 lengths are \{\splitValidHthrees\}.}
\label{tab:split}
\end{table}

The scaffold split above only controls what the GFN is fine-tuned on. The backbone was pretrained on UniRef100, OAS and SCOP (\cref{apx:amplify-infill}). No held-out antibody is in the the SCOP training set (24{,}495 sequences), the SCOP development set (10{,}000), or the OAS training set (3{,}611{,}708). Four test and one validation sequence to have a matched PDB entries on the antigen side: HER2 (\texttt{1N8Z\_C}), the cytochrome $bc_1$ complex (\texttt{3CX5}), and the TSH receptor (\texttt{2XWT\_C}). The generator never conditions on these. Additionally, only five test and one valid CDRH3 loops are present in the OAS, which is a chance occurance rate. Held-out loops are not enriched relative to the rest of the 784-structure pool.

\begin{table}[!htb]
\centering
\caption{Verbatim CDRH3 recurrence in the OAS pretraining fill with held-out targets against the rest of the structure pool, by loop length. A short loop appearing somewhere in 3.6M repertoire sequences means nothing on its own, so the interpretable quantity is the rate relative to non-held-out loops at the same length.}
\label{tab:oas-loop-recurrence}
\small
\begin{tabular}{@{}lrrrr@{}}
\toprule
& \multicolumn{2}{c}{held-out (TEST $+$ VALID)} & \multicolumn{2}{c}{rest of pool} \\
\cmidrule(lr){2-3}\cmidrule(lr){4-5}
CDRH3 length & entries & recur & entries & recur \\
\midrule
$\leq$5 & 2 & 2 (100\%) & 76 & 76 (100\%) \\
6--7 & 3 & 1 (33\%) & 53 & 29 (55\%) \\
8--9 & 12 & 2 (17\%) & 130 & 21 (16\%) \\
10--12 & 23 & 1 (4\%) & 204 & 2 (1\%) \\
\midrule
\textbf{total ($\leq$12)} & \textbf{40} & \textbf{6 (15.0\%)} & \textbf{463} & \textbf{128 (27.6\%)} \\
\bottomrule
\end{tabular}
\end{table}

The test set is the RAbD benchmark~\citep{adolfbryfogle2018rabd} targets for which scaffolds are available and which clear the length filter (\splitRabdAvail{} of \splitRabdTotal), deduplicated on scaffold $\geq$0.90 and antigen $\geq$0.90 so every test entry is a distinct binding problem; at these thresholds \splitRabdCollapsed{} targets collapse. Generation covered 45 of the these 55 test targets as they are members of the original 784-structure universe.

\subsection{Reward Metrics Through the Funnel}
\label[appendix]{apx:rewards}

\begin{figure}[!htb]
\centering
\includegraphics[width=\textwidth]{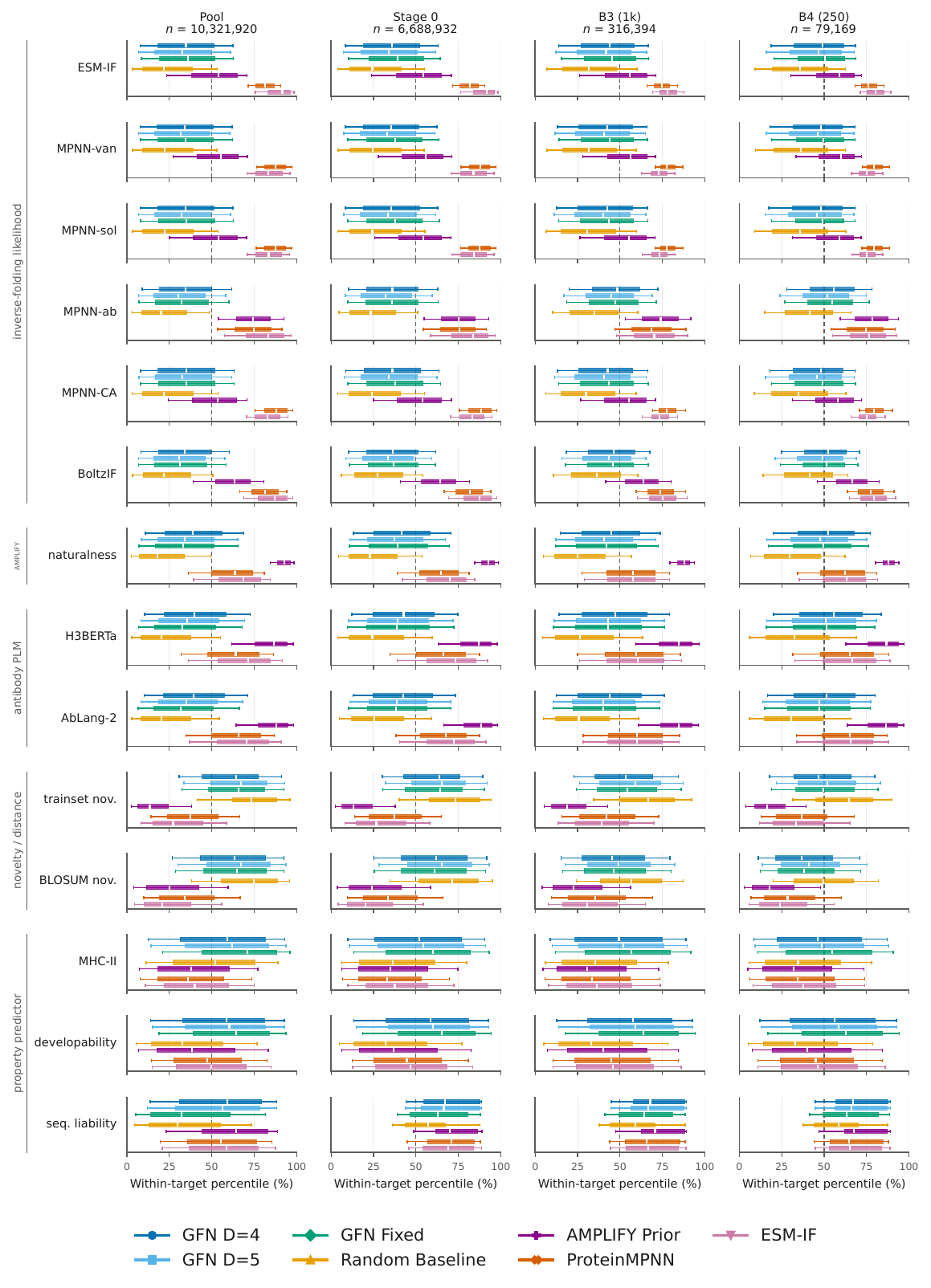}
\caption{Distribution of per-objective scores for each sampler at each level of the postfilter funnel, relative to the pool-level statistics. Within-target percentile is calculated against the pooled pool of all arms. 50\% is calculated as the pooled-pool medium, which explains why values sum to $>100\%$ at the end of the funnel. Boxplots denote the inter-quartile range, with a tick at the medium and whiskers denoting 10\% and 90\%.}
\label{fig:rewards}
\end{figure}

In \Cref{fig:rewards} we show the the rewards from all arms, shown as a percentile against the full pool. The learned GFNs sample designs with substantially different properties than the AMPLIFY-AR prior. Two broad trends can be seen in this analyses: 1) the structural models and AMPLIFY-AR prior outperform the learned GFNs and random sampler at the outset of the pool -- this gap is closed via the application of the postfilter pipeline, not completely. On these metrics, the learned GFNs outperform a random sampler. 2) The learned GFNs outperform all baselines on novelty, developability, MHC-II binding, and perform at parity with respect to liability. The delta in performance between the AMPLIFY-AR prior and the various trained objectives suggests that our training recipe was not optimal across all objectives, and a more precise training procedure which prioritizes particular reward dimensions adaptively during training might produce better tradeoffs in the final sampler.

\subsection{Diversity Metrics}
\label[appendix]{apx:diversity-metrics}

Pool diversity is measured over ESM-2 650M sequence embeddings, with the Vendi score \citep{friedman2023vendi} (the effective number of distinct modes; the exponential of the Shannon entropy of the eigenvalues of $\mathbf{K}/n$, where $\mathbf{K}_{ij}$ is the embedding similarity between designs $i$ and $j$ and $n$ is the pool size, which is scale-free and mode-count-aware) and mean pairwise $L_2$ distance. We do not report 2D UMAP convex-hull area as a diversity metric due to volumes in this space being a projection artifact. UMAP is used for visualization only. Vendi counts effective modes and therefore grows with sample size, so every arm is subsampled to the same $n$ before scoring. Embeddings are calculated with the loop in context of it's heavy chain. The GFN methods are close to the untrained control on both diversity measures while adding objective controllability:

\begin{figure}[!htb]
\centering
\includegraphics[width=\textwidth]{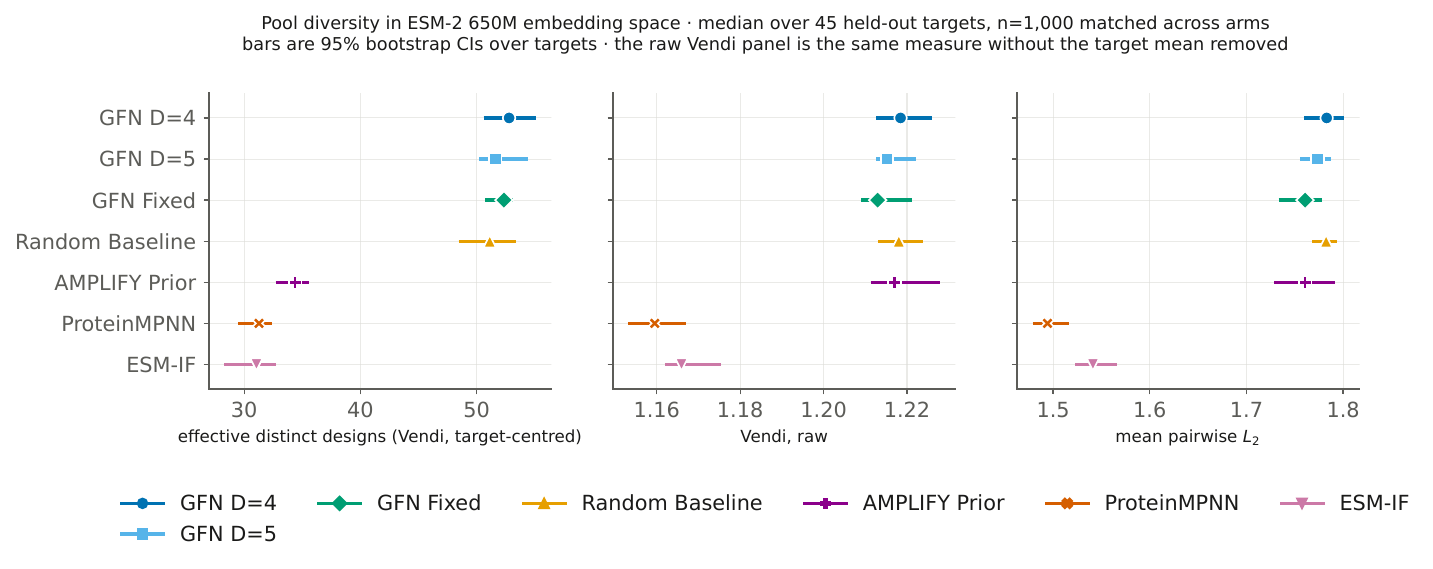}
\caption{Sequence diversity over ESM-2 650M embeddings of the CDRH3 loop in its own heavy-chain context, on the post-filter design pools. Median over held-out targets with 95\% bootstrap CIs resampling targets.}
\label{fig:diversity}
\end{figure}

A Vendi score at one sample size cannot separate many modes from a dense sample of few modes, because it is a single number at a single $n$. We therefore also resample each arm's pool at increasing draw sizes and report how the effective count grows.

\begin{figure}[!htb]
\centering
\includegraphics[width=\textwidth]{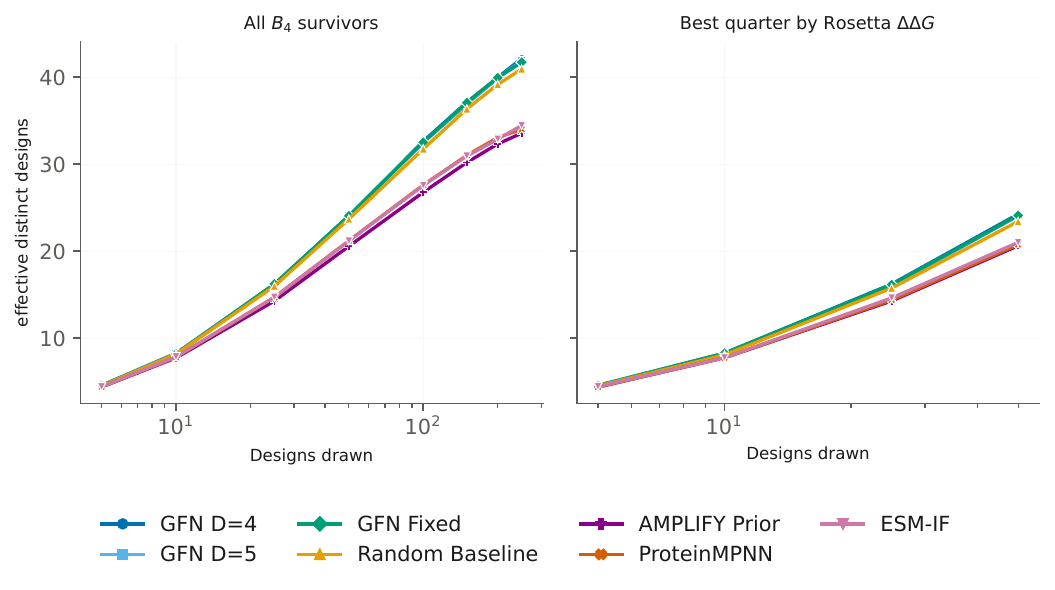}
\caption{Rarefaction of the $B_4$ survivor pools, which measures the effective number of distinct designs against how many designs are drawn per pool / target, averaged over \pfparetotargets{} held-out targets. Draws come from the same pool and are repeated at each size. The Vendi score is taken over ESM-2 650M embeddings of the loop in its own heavy-chain context, after removing each target's mean embedding, so an arm is measured against the other arms at that target rather than against the target. \emph{Left:} all $B_4$ survivors. \emph{Right:} the same measurement over the best quarter of those survivors by Rosetta $\Delta\Delta G$. Both curves clearly show that the GFN arms draw more unique designs per draw than the structural baselines.}
\label{fig:landscape-rarefaction}
\end{figure}

The same populations can also be shown in the space itself rather than as a count, which says \emph{where} in the landscape an arm still has designs once the cascade has run.

\begin{figure}[!htb]
\centering
\includegraphics[width=\textwidth]{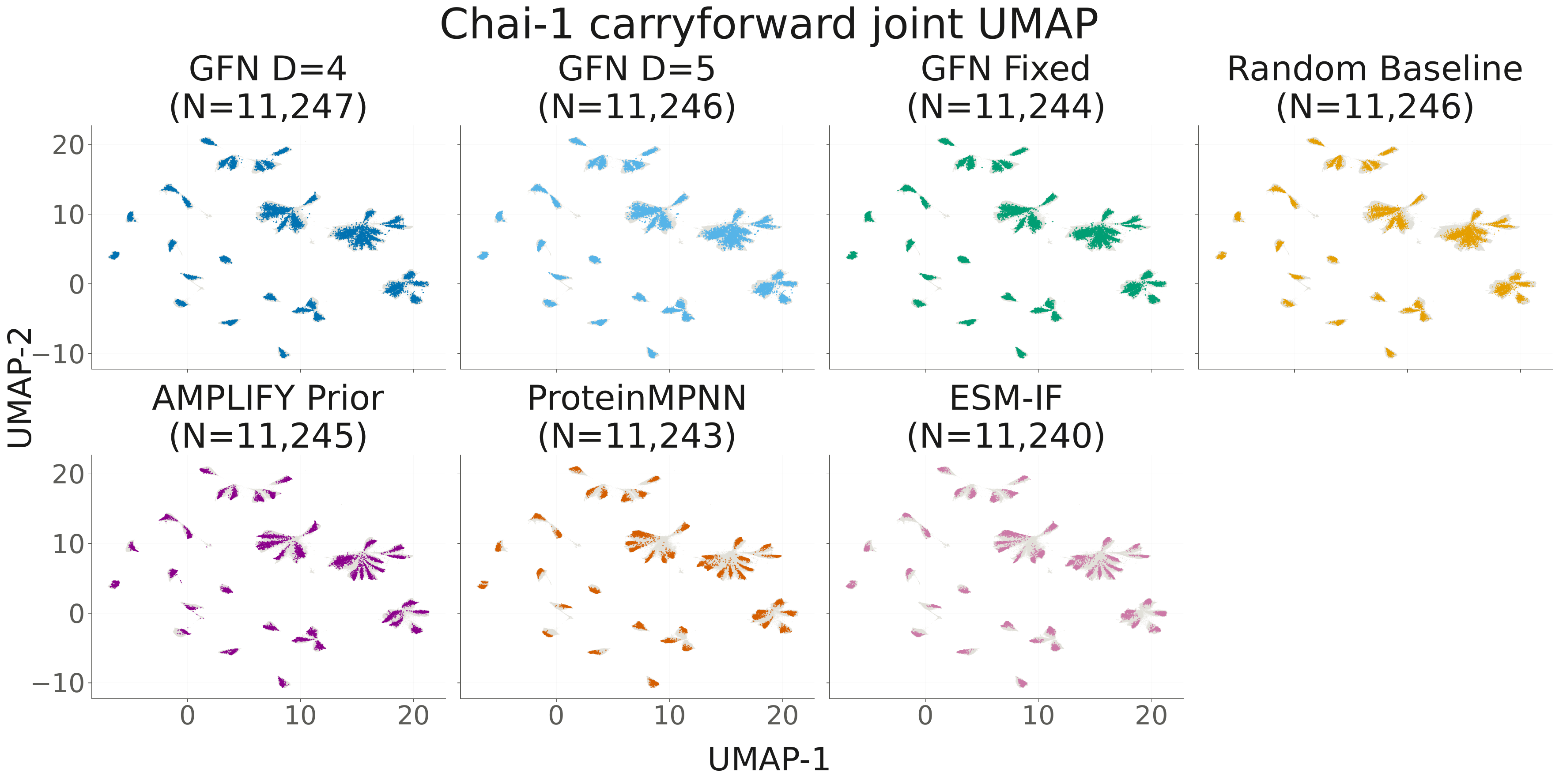}\\[6pt]
\includegraphics[width=\textwidth]{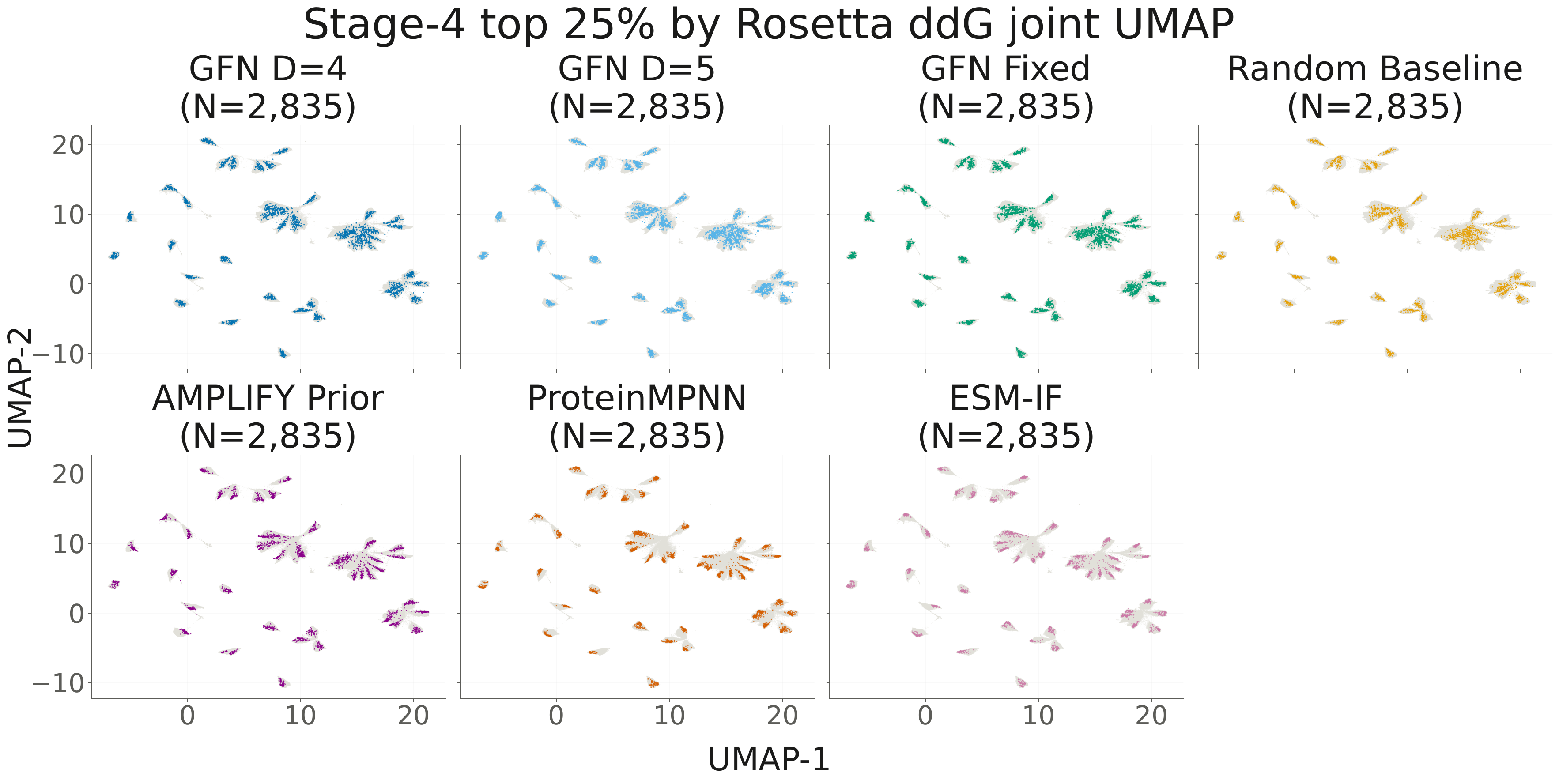}
\caption{Where each arm's surviving designs sit in the shared sequence landscape. The UMAP is fit once over the background distribution of every embedded design of every arm and target, and each panel then plots that background distribution in grey. The coloured points are the designs of that arm which passed the post-filter thresholds. \emph{Top:} the designs carried forward to the Chai-1 stage, $B_4$. \emph{Bottom:} the best quarter of those by Rosetta $\Delta\Delta G$. Embeddings are ESM-2 650M over the CDRH3 loop in its own heavy-chain context under cosine distance. The islands largely represent individual test-set targets. The plots demonstrate qualitatively how the different methods coexist in design space -- in particular, how the GFN arms reach across the boundaries of both the structural models, the prior, and the random baselines.}
\label{fig:landscape-umap}
\end{figure}




\section{Conditioning, Controllability, and Learned Preference Slots}
\label[appendix]{apx:conditioning}
\label[appendix]{apx:projection}
\label[appendix]{apx:controllability}

Since the learned projection centres the target ($\hat{r}_{bj} - \mu_j$) and weights the reconstruction loss by $1/\sigma_j^2$, a reward with a large dynamic range cannot win a slot on scale alone. Reconstruction quality is measured as $R^2_j = 1 - \mathrm{MSE}_j/\mathrm{Var}(\hat{r}_j)$. $\bA$ is warm-started on 2{,}000 steps of offline rewards over native CDRH3 loops. $\bA$ mixes min-max normalized rewards. Following~\citet{park2025rdr} the learned reduction is affine, positive and row-stochastic, $f(r) = \bA r + b$ with $b = 0$, which keeps every slot a convex combination and therefore $\bz \in [0,1]^D$. An EMA target $\bar{\phi} \leftarrow \tau\bar{\phi} + (1-\tau)\phi$ supplies the rewards during training, decoupling how fast the projection learns from how much the reward can vary across scaffolds. \textsc{gfn}~$D{=}4$ and \textsc{gfn}~$D{=}5$, which learn their slots over all 14 objectives, and \textsc{gfn fixed}, whose four axes are fixed config groups over 9.

\begin{figure}[!htb]
\centering
\includegraphics[width=\textwidth]{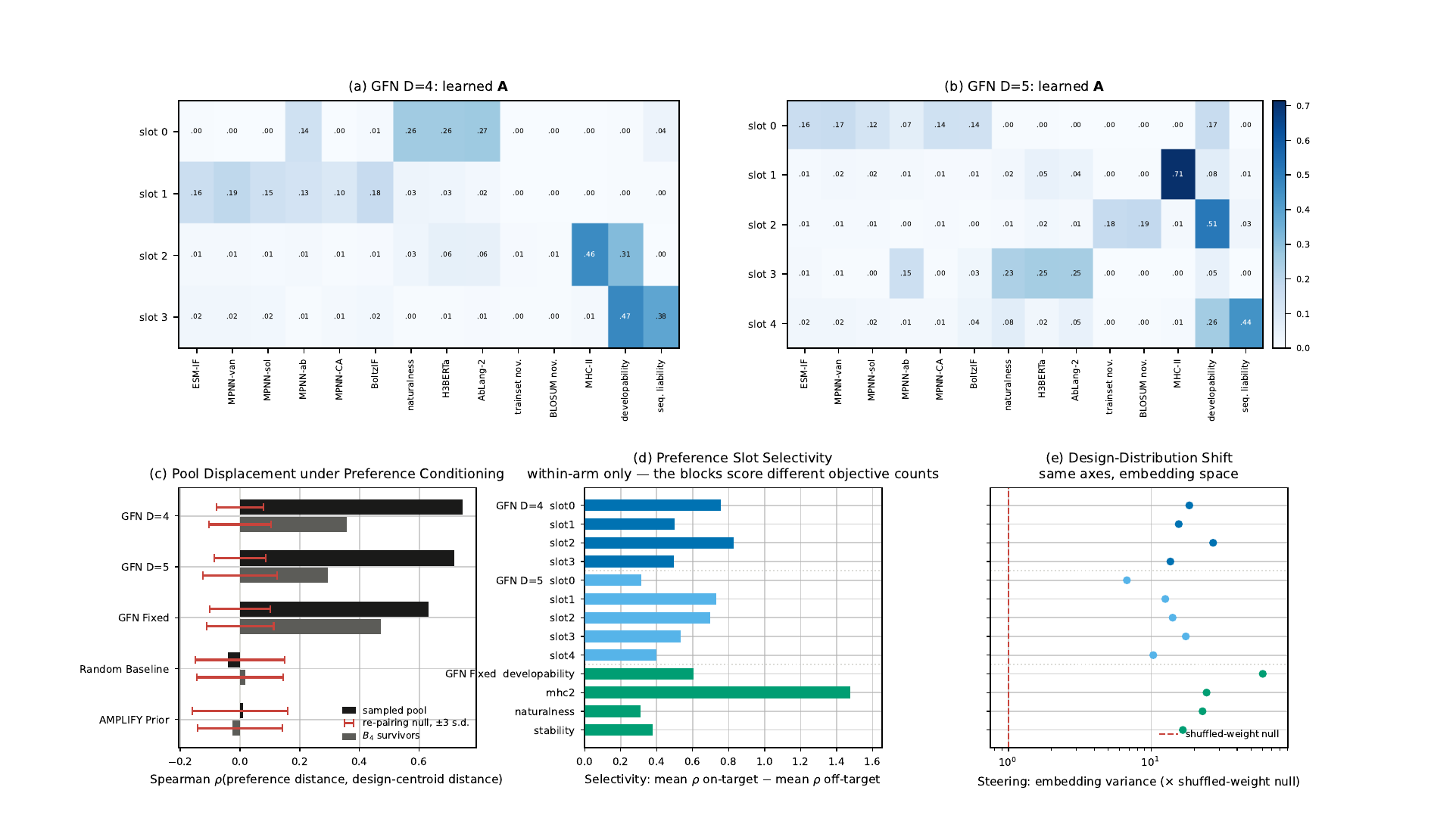}
\caption{\textbf{(a,b)} Learned $\bA$ weights over the 14 rewards for the $D=4$ and $D=5$ GFNs. Each row $d$ sums to $1$. While many groupings are broadly similar between the two slot counts, some parings change dramatically, such as the isolation of MHC-II in slot 1 for $D=5$. \textbf{(c)} Spearman $\rho$ of the Euclidean distances between every pair of an arm's preference vectors, and the distance between the matching design centroids (the mean target-centred ESM-2 embedding of a preference's designs) pooled over targets. Red bars span $\pm3$ s.d. Only conditioned models show design dispersion outside of this null region. \textbf{(d)} $R_{dj}$ is the Spearman correlation, across preferences, between the weight on axis $d$ and the per-preference mean of objective $j$, averaged over 45 targets (i.e., it is performed in reward space). Let $w_d$ be the group indicator of an axis's row of $\bA$. Selectivity is the difference between on target $\sum_j w_{dj}R_{dj}/\sum_j w_{dj}$ minus the off-target $\max_j w_{dj} - w_{dj}$. \textbf{(e)} The same axes as (d) in embedding space space which shows the degree to which each axis displaces the embedding.}
\label{fig:conditioning}
\end{figure}

An analyses of the learned preference slots (\cref{fig:conditioning}a,b) reveals that both learned groupings recover reasonable and repeatable structure: both isolate MHC-II, each gives the three sequence-likelihood oracles a single slot in near-equal thirds, and each keeps the six inverse-folding stability oracles (roughly) together. At $D{=}5$ the two novelty terms concentrate in one slot, carrying \condDFiveNoveltyMass{} of the total row mass; at $D{=}4$ they carry \condDFourNoveltyMass{} of 4. The $D{=}4$ reduction effectively discards novelty rather than compressing it. The conditioned arms return $\rho = \condFixedPoolRho{}$--$\condDFourPoolRho{}$ Spearman's correlations between the preferences and the rewards, while the unconditioned models are centred on zero (\condAmplifyPriorPoolRho{} and \condAmplifyFimPriorPoolRho{}); \cref{fig:conditioning}c). Interestingly, $\rho$ falls by roughly half among $B_4$ survivors, suggesting that the structural post-filter pipeline removes designs which are too extremely committed to any preference.

Selectivity (\cref{fig:conditioning}d) measures specific modulation of rewards over the full sampled pool. It measures whether an axis moves the objectives it with higher weight under $\bA$ than those with lower weight, and reports how reliably an
objective tracks the axis weight, rather than by how much. Zero means the axis moves everything similarly. Steering (\cref{fig:conditioning}e) is a magnitude in sequence space, and measures how much of the design distribution moves with the axis weight in general. It is a ratio against a shuffled null on that axes, so $1$ means indistinguishable from no conditioning.

Selectivity is positive on every axis of all three arms, and the higher weighted axes are the top-responding objectives. The MHC-II axis is the highest-selectivity axis in all three (\condDFourBestSel{}, \condDFiveBestSel{} and \condFixedBestSel{} for
\condDFourBestAxis{}, \condDFiveBestAxis{} and \condFixedBestAxis{}). The six-way stability mixture is in the bottom half of each, which is unsuprising as each inverse folding model was trained on different data distributions. Within each arm the two measurements order the axes similarly (Spearman \condDFourAgree{}, \condDFiveAgree{} and \condFixedAgree{}).

\section{Post-Filter Details}
\label[appendix]{apx:postfilter}

\begin{figure}[!htb]
\centering
\includegraphics[width=\textwidth]{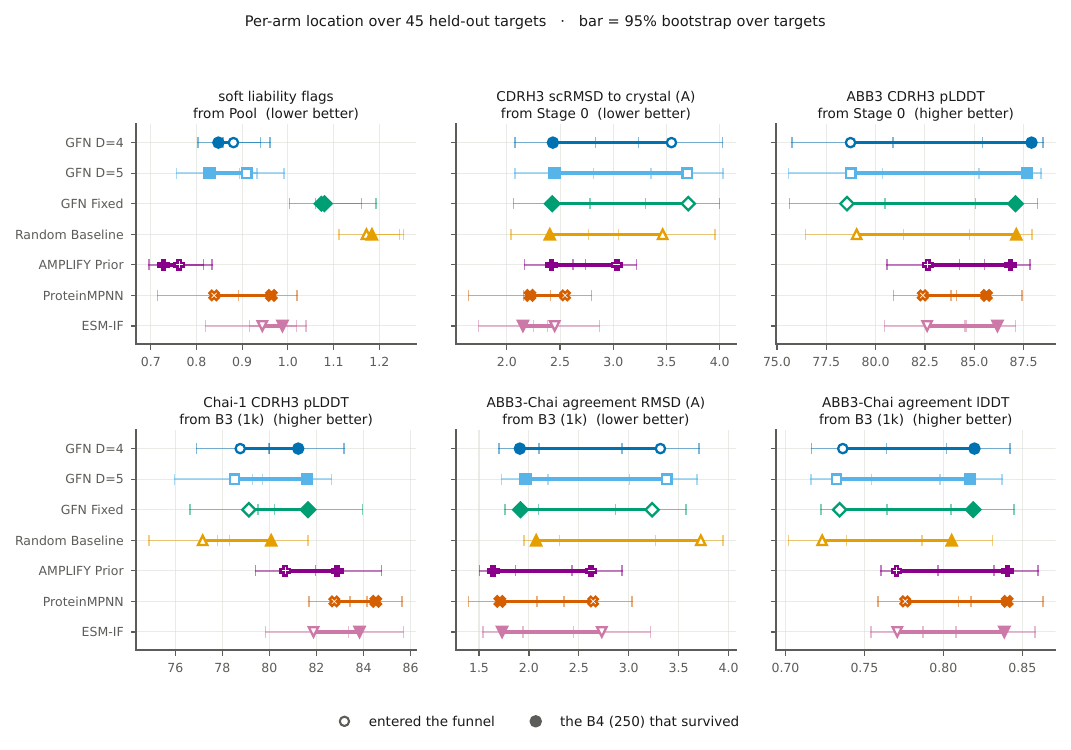}
\caption{Scores of each arm at the beginning and end of the funnel (B4). Whiskers represent 95\% bootstrap intervals over the held-out targets. Generally speaking, the improvement imposed by the funnel was greater for the random baselines and GFNs.}
\label{fig:postfilter-cascade-quality}
\end{figure}

This analysis' purpose is to refine the samples generated by each arm according to increasingly strict and expensive held-out oracles, in lieu of wet lab tests of sample quality. A liability gate first removes every design that fails sequence-level requirements. Afterwards, all removals are done according to a \emph{budget} informed by our computational constraints. These designs are ranked and the top performers are carried forward to the final analysis. Stage 0 gates on clear sequence liabilities. Stage 1 folds every Stage-0 survivor with ABodyBuilder3, clusters the survivors' CDRH3 at 50\% identity, and carries $B_3$ designs forward: a cluster of $n$ members earns $\lfloor \log_{10} n \rfloor + 1$ slots and spends them on its members of highest mean CDRH3 pLDDT. Stage 2 computes spatial aggregation propensity on those same
Stage-1 structures and records it for analysis only. Stage 3 forward folds the carried designs a second time under Chai-1, keeps the best $B_4$ designs of the target's own ABodyBuilder3--Chai agreement distribution. These are assembled onto the crystal antigen in Stage 3c and scored for interface energy by FoldX and Rosetta in Stage 4. The scores of each arm entering and exiting the post-filter funnel can be found in \cref{fig:postfilter-cascade-quality}.

\begin{table}[!htb]
\centering
{\small
\begin{tabular}{@{}llll@{}}
\toprule
Stage & Signal & Role & Cost \\
\midrule
0 & Sequence liability motif scan & Removes undesirable designs & $\mu$s, CPU \\
1 & ABodyBuilder3: H3 pLDDT, scRMSD & Cheap structural quality filter & 0.25~s/design \\
2 & Aggregation propensity & Developability score & 0.15~s/design \\
3 & Chai-1: RMSD, lDDT, pLDDTs & Expensive quality filter & $\leq$31~s/design \\
3c & Assembles predicted Fv w/ antigen crystal & Assemble structures & 0.27~s/design \\
4a & Rosetta:interface energy & Evaluate & 2.66--44~min/complex \\
4b & FoldX: interface energy & Evaluate & 1.05--2.3~min/complex \\
\bottomrule
\end{tabular}
}

\caption{This evaluation cascade is designed for reference-free mode recall. Stage 0 gates on clear sequence liabilities. Stage 1 folds all surviving sequences, clusters at $50\%$ similarity, and carries forward the top designs. Stage 2 calculates aggregation propensity on the Stage-1 structures. Stage 3 forward folds top designs under Chai-1, which are analyzed both for fold confidence and structural similarity to ABB3. The best sequences are then brought forward for binding estimation with the crystal scaffold under FoldX and Rosetta.}
\label{tab:postfilter-stages}
\end{table}

Stage 2 SAP multiplies solvent exposure by side-chain hydrophobicity and sums the product over a 5~\AA{} radius. It mostly measures composition and exposure. A design can score low either due to fewer hydrophobic residues, or buried residues. The \texttt{developability} objective is related to this due to training labels derived from the Developability Index~\citep{lauer2012di}, which is built on CDR aggregation propensity (\cref{sec:obj}). The six inverse-folding likelihoods also influence this outcome variable because they score burial on the native backbone.

Stage 3 uses Chai-1, a MSA-free folding model which is trained on the general PDB rather than on SAbDab alone. Agreement between Chai-1 and ABodyBuilder3 is measured either by  RMSD after superposition on the VH framework or by bidirectional H3 lDDT, which is superposition-free and bounded. These provide a measure of structural quality in the absence of a reference. All confidence scores are evaluated over the CDRH3 loop residues.

\begin{table}[!htb]
\centering
\caption{CDRH3 confidences over the \cpNTargets{} targets present in all \cpNArms{} arms (\cpNDesigns{} designs). Values are mean Chai pLDDT read from the Stage-3c complex: over CDRH3 residues within 5~\AA{} of the antigen, over all CDRH3 residues, and over the whole Fv. \emph{Contacting H3} is the fraction of loop residues meeting the 5~\AA{} criterion. The final row is the between-arm range: the whole-Fv column is nearly invariant, so an effect measured on it is compressed relative to the same effect measured on the loop. Budgets rank on CDRH3 pLDDT, and the whole-Fv column is reported only as the comparison scale.}
\label{tab:contact-plddt}
\small
\begin{tabular}{@{}lrrrrr@{}}
\toprule
Arm & Targets & H3 contact & H3 all & Fv & Contacting H3 \\
\midrule
GFN $D{=}4$ & 45 & $74.58$ & $80.53$ & $93.91$ & 47.1\% \\
GFN $D{=}5$ & 45 & $74.14$ & $80.10$ & $93.84$ & 47.0\% \\
GFN Fixed & 45 & $74.58$ & $80.51$ & $93.93$ & 47.2\% \\
\midrule
Random baseline & 45 & $73.12$ & $78.87$ & $93.58$ & 47.6\% \\
AMPLIFY prior & 45 & $76.01$ & $82.31$ & $94.23$ & 46.4\% \\
ProteinMPNN & 45 & $77.18$ & $83.55$ & $94.41$ & 45.1\% \\
ESM-IF & 45 & $75.98$ & $82.45$ & $94.24$ & 45.4\% \\
\midrule
\emph{between-arm range} & & $4.06$ & $4.68$ & $0.83$ & \\
\bottomrule
\end{tabular}

\end{table}

Stage 3c builds each scored complex from the Chai-1 predicted antibody structure and a crystal structure-derived antigen. Chai-1 Fv is rigid-body superimposed onto the crystal structure-derived antibody using the VH framework, after which the crystallographic antibody is removed while the antigen is retained in its experimental coordinate frame. The resulting complex therefore combines the predicted antibody structure with the experimentally determined antigen pose. This construction assumes that changes in the CDRH3 loop do not substantially alter  the orientation of the VH framework relative to the antigen.

Finally, Rosetta~\citep{Alford2017-nu} performs a constrained FastRelax of the Stage-3c complex, adapting the new CDRH3 loop to the complex by optimizing its backbone and side-chain conformations while relieving local structural strain and steric clashes,  and then scores the resulting interface. FoldX~\citep{Delgado2025FoldX} subsequently repairs the relaxed complex and independently scores the interface. Binding and stability energies are reported relative to a matched WT reference ($\Delta\Delta E = \Delta E_{\mathrm{design}} - \Delta E_{\mathrm{WT}}$).  Binding scores are additionally penalized for unfavorable steric interactions by adding the FoldX intraclash or Rosetta \texttt{fa\_rep} term, respectively, to the calculated binding energy.  Diversity is tracked throughout the funnel using the the number of distinct CDRH3 loop sequences, the mean per-position Shannon entropy of the CDRH3 (maximum $\log_2 20 = 4.32$ bits), and the number of clusters at 70\% identity.

\begin{figure}[!htb]
\centering
\includegraphics[width=\textwidth]{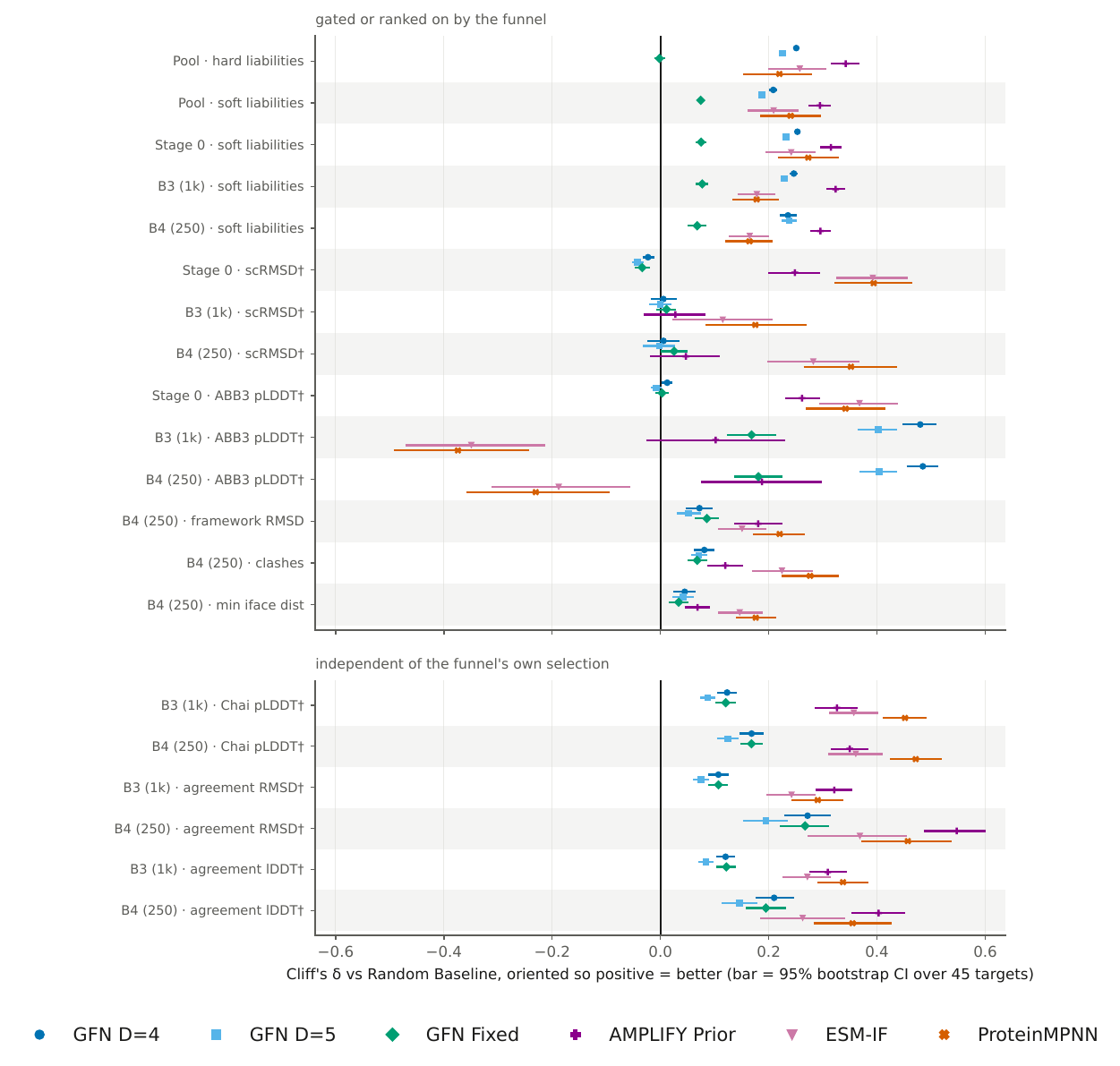}
\caption{Every arm against the random baseline across the postfilter. Cliff's $\delta$ is computed within each target and averaged over the 45 held-out targets. Bars are 95\% bootstrap CIs resampling targets. Signs are oriented so that positive is favourable. $\dagger$ marks metrics which reward proximity to the native structure. The lower panel holds the metrics nothing upstream in the cascade gates or ranks on. The size of the selection effect that separation exists for is visible in ABodyBuilder3 pLDDT, which moves from roughly zero at Stage 0 to $+0.5$ at B3 for the same arm.}
\label{fig:funnel-forest}
\end{figure}

\begin{table}[!htb]
\centering
\caption{Estimate of liability removal criteria impacts.}
\label{tab:culls}
\small
\begin{tabular}{@{}p{0.18\textwidth}p{0.2\textwidth}p{0.10\textwidth}p{0.41\textwidth}@{}}
\toprule
Criterion & Removes & Cull & Diversity cost \\
\midrule
Free cysteine & any loop containing C & $\approx$46\% est. & Deletes disulfide-stapled and cyclic CDRH3 entirely --- a real and productive antibody class. Pairing cannot be decided from sequence, so this rule cannot distinguish a reactive free thiol from a stabilizing intra-loop disulfide. \\
N-glyc sequon & N-x-[S/T], $x \neq$ P & $\approx$5\% est. & A real liability, but the motif is short and its occurrence is not correlated with function. \\
Deamidation NG & any loop containing NG & $\approx$3\% est. & NG occurs in the CDRs of approved antibodies; fast deamidation is a formulation problem, not a binding one. \\
Isomerisation DG & any loop containing DG & $\approx$3\% est. & As NG. \\
Soft flags & NS, NN, NT, NH; DS, DD, DT, DH; M, W; DP & none & Recorded only, no cull. \\
\bottomrule
\end{tabular}
\end{table}

\section{Span-In-Filling vs.\ Masked-Language-Model Backbone}
\label[appendix]{apx:backbone-comparison}



\subsection{Pretraining Validation}
\label[appendix]{apx:backbone-validation}

\begin{figure}[!htb]
\centering
\includegraphics[width=\textwidth]{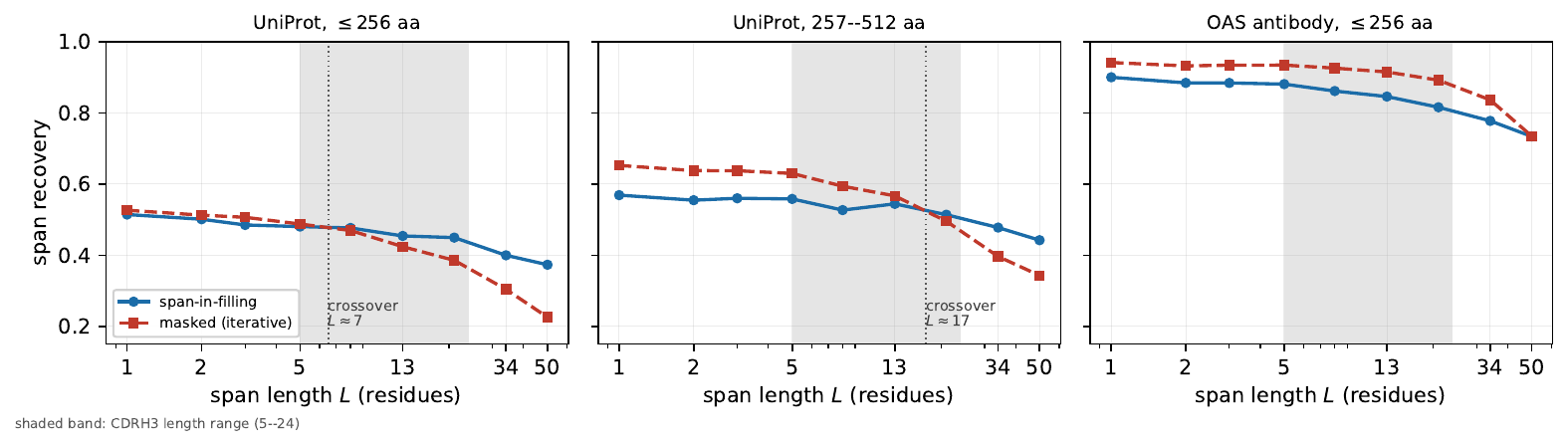}
\caption{Span recovery against span length on the in-filling and masked-language models, both producing the span conditioned on their own predictions. On general protein sequence the masked model leads at short spans and in-filling leads at long ones, with the crossover moving from $L\approx$8 on proteins of at most 256 residues to $L\approx$21 on longer ones. On antibody sequences there is no crossover in range: the masked model leads marginally at every span length tested. The shaded band is the CDRH3 length range our paper's tasks operate in.} 
\label{fig:backbone-recovery}
\end{figure}

Held-out in-filling loss, computed by separate code from the training loop, reproduces the pretraining log's own final evaluation to within \bbUniprotDevLogPct{}\% on \texttt{uniprot\_dev} (\bbUniprotDevNll{} against \bbUniprotDevLogNll{}) and \bbScopDevLogPct{}\% on \texttt{scop\_dev} (\bbScopDevNll{} against \bbScopDevLogNll{}). \texttt{oas\_dev} is the one set where the two disagree (\bbOasDevNll{} against \bbOasDevLogNll{}), because OAS loss is bimodal --- germline-conserved framework spans score near zero and CDR spans score high --- so its mean summarizes a mixture. Memorization is also well controlled: on OAS, held-out data scores \bbOasGap{} nats against training data (\bbOasDevNll{} against \bbOasTrainNll{}), while SCOP shows \bbScopGap{} (\bbScopDevNll{} against \bbScopTrainNll{}). That SCOP gap survives length-matching and is not explained by homology: \texttt{scop\_dev} entries that share a UniProt protein with \texttt{scop\_train} score worse. On matched span recovery, the MLM pulls ahead in performance (\cref{tab:backbone-validation-recovery}, \cref{fig:backbone-recovery}).

\begin{table}[!htb]
\centering
\caption{Span recovery against the original masked-language-model AMPLIFY-350M, at matched span length. Recovery is argmax accuracy over the 20 canonical residues. Note that the exact likelihood calculated under AMPLIFY-AR is not directly comparable to the pseudo-likelihood calculated under the original AMPLIFY. $\mathrm{AR}_{\mathrm{gen}}$ and $\mathrm{MLM}_{\mathrm{gen}}$ both generate the span conditioned only on their own predictions. $\mathrm{AR}_{\mathrm{nat}}$ and $\mathrm{MLM}_{\mathrm{nat}}$ see native context. $n$ falls with span length because the $L$-forward decodings cost $O(L)$ for MLM models. We use the same sequences for all model evaluations.}
\label{tab:backbone-validation-recovery}
\small
\begin{tabular}{@{}llrrrrrrr@{}}
\toprule
& & & \multicolumn{3}{c}{generation (self-conditioned)} & \multicolumn{2}{c}{native-conditioned} & \\
\cmidrule(lr){4-6}\cmidrule(lr){7-8}
Corpus & $L$ & $n$ & $\mathrm{AR}_{\mathrm{gen}}$ & $\mathrm{MLM}_{\mathrm{gen}}$ & $\Delta$ & $\mathrm{AR}_{\mathrm{nat}}$ & $\mathrm{MLM}_{\mathrm{nat}}$  \\
\midrule
UniProt, $\leq$256 aa & 1 & 2000 & 0.514 & 0.526 & -0.012 & 0.514 & 0.526 \\
 & 2 & 2000 & 0.501 & 0.513 & -0.012 & 0.510 & 0.526 \\
 & 3 & 2000 & 0.485 & 0.507 & -0.021 & 0.502 & 0.525 \\
 & 5 & 2000 & 0.481 & 0.488 & -0.007 & 0.519 & 0.519 \\
 & 8 & 2000 & 0.477 & 0.470 & \textbf{+0.007} & 0.528 & 0.533  \\
 & 13 & 1230 & 0.454 & 0.424 & \textbf{+0.030} & 0.532 & 0.528  \\
 & 21 & 761 & 0.450 & 0.386 & \textbf{+0.064} & 0.553 & 0.543  \\
 & 34 & 470 & 0.400 & 0.304 & \textbf{+0.096} & 0.552 & 0.535  \\
 & 50 & 400 & 0.373 & 0.226 & \textbf{+0.148} & 0.547 & 0.544  \\
\midrule
UniProt, 257--512 aa & 1 & 2000 & 0.569 & 0.652 & -0.084 & 0.569 & 0.652  \\
 & 2 & 2000 & 0.555 & 0.638 & -0.083 & 0.564 & 0.640  \\
 & 3 & 2000 & 0.560 & 0.638 & -0.078 & 0.576 & 0.652  \\
 & 5 & 2000 & 0.558 & 0.630 & -0.072 & 0.592 & 0.661  \\
 & 8 & 2000 & 0.527 & 0.594 & -0.067 & 0.579 & 0.650  \\
 & 13 & 1230 & 0.544 & 0.567 & -0.023 & 0.614 & 0.673  \\
 & 21 & 761 & 0.514 & 0.496 & \textbf{+0.018} & 0.612 & 0.663  \\
 & 34 & 470 & 0.478 & 0.396 & \textbf{+0.081} & 0.614 & 0.657  \\
 & 50 & 400 & 0.442 & 0.343 & \textbf{+0.099} & 0.618 & 0.672  \\
\midrule
OAS antibody, $\leq$256 aa & 1 & 2000 & 0.900 & 0.942 & -0.041 & 0.900 & 0.942 \\
 & 2 & 2000 & 0.885 & 0.932 & -0.048 & 0.888 & 0.936  \\
 & 3 & 2000 & 0.884 & 0.934 & -0.050 & 0.896 & 0.938  \\
 & 5 & 2000 & 0.881 & 0.935 & -0.053 & 0.902 & 0.942 \\
 & 8 & 2000 & 0.862 & 0.926 & -0.064 & 0.899 & 0.941  \\
 & 13 & 1230 & 0.846 & 0.915 & -0.069 & 0.903 & 0.943  \\
 & 21 & 761 & 0.816 & 0.892 & -0.076 & 0.901 & 0.941  \\
 & 34 & 470 & 0.778 & 0.837 & -0.059 & 0.902 & 0.942 \\
 & 50 & 400 & 0.734 & 0.735 & -0.001 & 0.901 & 0.945  \\
\midrule
native CDRH3 (held-out) & native & 55 & 0.121 & 0.314 & -0.193 & 0.343 & 0.416 \\
\bottomrule
\end{tabular}
\end{table}

\clearpage
\section{Oracle Details}
\label[appendix]{apx:oracles}

\begin{table}[!htbp]
\centering
\caption{The 14 reward objectives. Context denotes how much information in addition to the CDRH3 loop is presented to the oracle during scoring.}
\label{tab:objectives}
\small
\begin{tabular}{@{}llll@{}}
\toprule
Category & Objective & Context & Source \\
\midrule
Structural stability & ESM-IF & loop + backbone & ESM-IF~\citep{hsu2022esmif} \\
 & ProteinMPNN & loop + backbone & ProteinMPNN~\citep{dauparas2022proteinmpnn} \\
 & ProteinMPNN (Soluble) & loop + backbone & ProteinMPNN \\
 & ProteinMPNN (Antibody) & loop + backbone & AbMPNN~\citep{antibodympnn} \\
 & ProteinMPNN (Ca Only) & loop + backbone & ProteinMPNN \\
 & BoltzIF & loop + backbone & BoltzIF~\citep{stark2025boltzgen} \\
\midrule
Naturalness & AMPLIFY-AR & loop in HC & AMPLIFY-AR (\cref{sec:policy-architecture}) \\
 & h3berta & loop & H3BERTa (OAS CDRH3 MLM) \\
 & ablang2 & loop in Fv & AbLang-2~\citep{olsen2024ablang2} \\
\midrule
Immunogenicity & MHC-II Binding & loop & MHC-II ensemble (\cref{apx:mhc2}) \\
Developability & Developability & loop in HC & Fine-tuned ESM-2 650M classifier~\citep{lin2023esm} \\
Liabilities & Sequence Liability & loop & PTM/glycosylation motif scan~\citep{satlawa2024lap} \\
\midrule
Novelty & Hamming Distance & loop & edit distance to training loops \\
 & Blosum Novelty & loop & BLOSUM62 distance to training loops \\
\bottomrule
\end{tabular}
\end{table}

\begin{figure}[!htb]
\centering
\includegraphics[width=\textwidth]{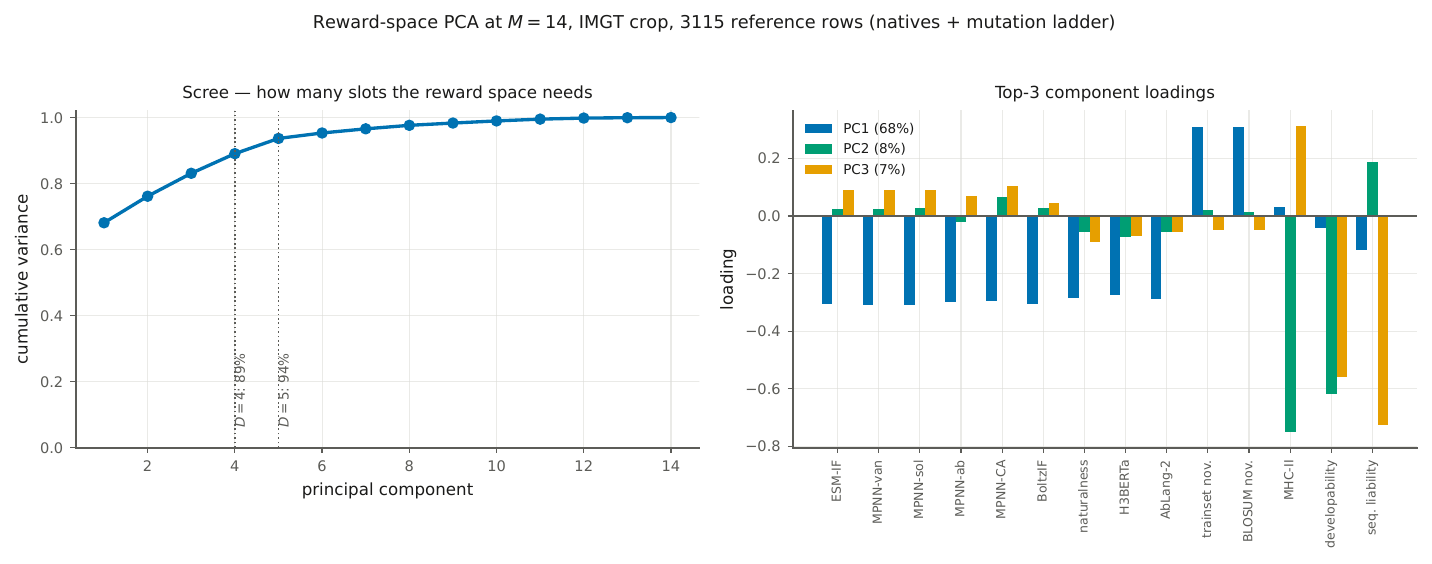}
\caption{PCA of the reward space at $M = 14$, IMGT crop, over the natives-plus-mutation-ladder reference set ($n = 3{,}115$). Left: cumulative variance explained, with the two production slot counts marked. Right: loadings of the top three principal components on each objective. The ladder rather than natives alone, for the reason given in \cref{fig:reward_correlation}: on natives both novelty terms are constant, and a PCA over them silently runs in 12 dimensions while reporting 14.}
\label{fig:reward_pca}
\end{figure}

\Cref{fig:reward_pca} shows the latent substructure of the reward landscape. The strongest negative pairwise correlation on designs is \tradeoffMostOpposedRho{}, between \tradeoffMostOpposed{}, which is modest, so this reward set does not contain severe \emph{pairwise} tradeoffs --- which is a caution about how much preference conditioning can be expected to do, not a defect: if objectives do not trade off, preferences have little to control. Reward sharpening does not manufacture the tensions this set lacks: it raises the response of realized rewards to preference weights without strengthening the objective-to-objective correlations (\cref{apx:beta}).

\subsection{The Reward Landscape}
\label{sec:reward-landscape-analysis}

\begin{figure}[!htb]
\centering
\includegraphics[width=0.75\textwidth]{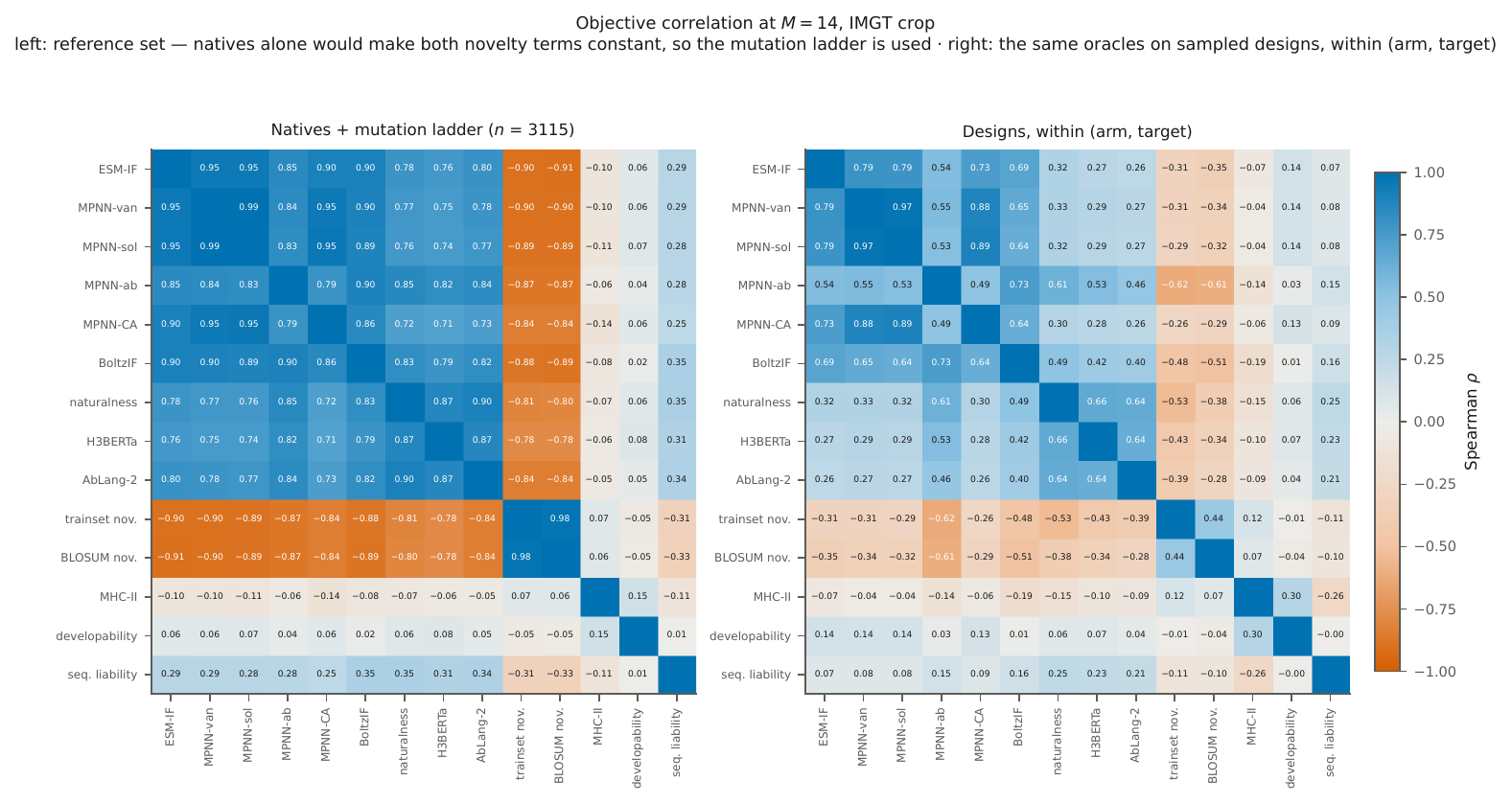}
\caption{Spearman rank correlation between all pairs of reward objectives at $M = 14$, IMGT crop. \textbf{Left:} the reference set --- natives plus a mutation ladder at rates 0.1--0.75 ($n = 3{,}115$). \textbf{Right:} the same oracles on sampled designs, computed within (arm, target) and averaged. Correlation on natural antibodies is inflated by every likelihood oracle liking real antibodies: \texttt{naturalness}, \texttt{H3BERTa} and \texttt{AbLang-2} agree at 0.87--0.90 on the reference set but only 0.64--0.66 on designs, and the two novelty terms fall from 0.98 to 0.44. On the other hand, \texttt{MPNN}/\texttt{MPNN-sol} stay at 0.97, which is genuine redundancy. The reduction in agreement of these oracles on generated designs motivates our use of this broader reward set.}
\label{fig:reward_correlation}
\end{figure}

The reward suite should provide complementary signals rather than redundant scores. Across the curated SAbDab set, the six stability rewards cluster together but retain some variation, while immunogenicity risk, developability, and naturalness show weak correlations with the other objectives (\cref{fig:reward_correlation}). Positive correlations are stronger than negative ones throughout: \tradeoffMostAlignedRho{} for the most aligned pair, \tradeoffMostAligned{}, against \tradeoffMostOpposedRho{} for the most opposed.

\begin{figure}[!htb]
\centering
\includegraphics[width=\textwidth]{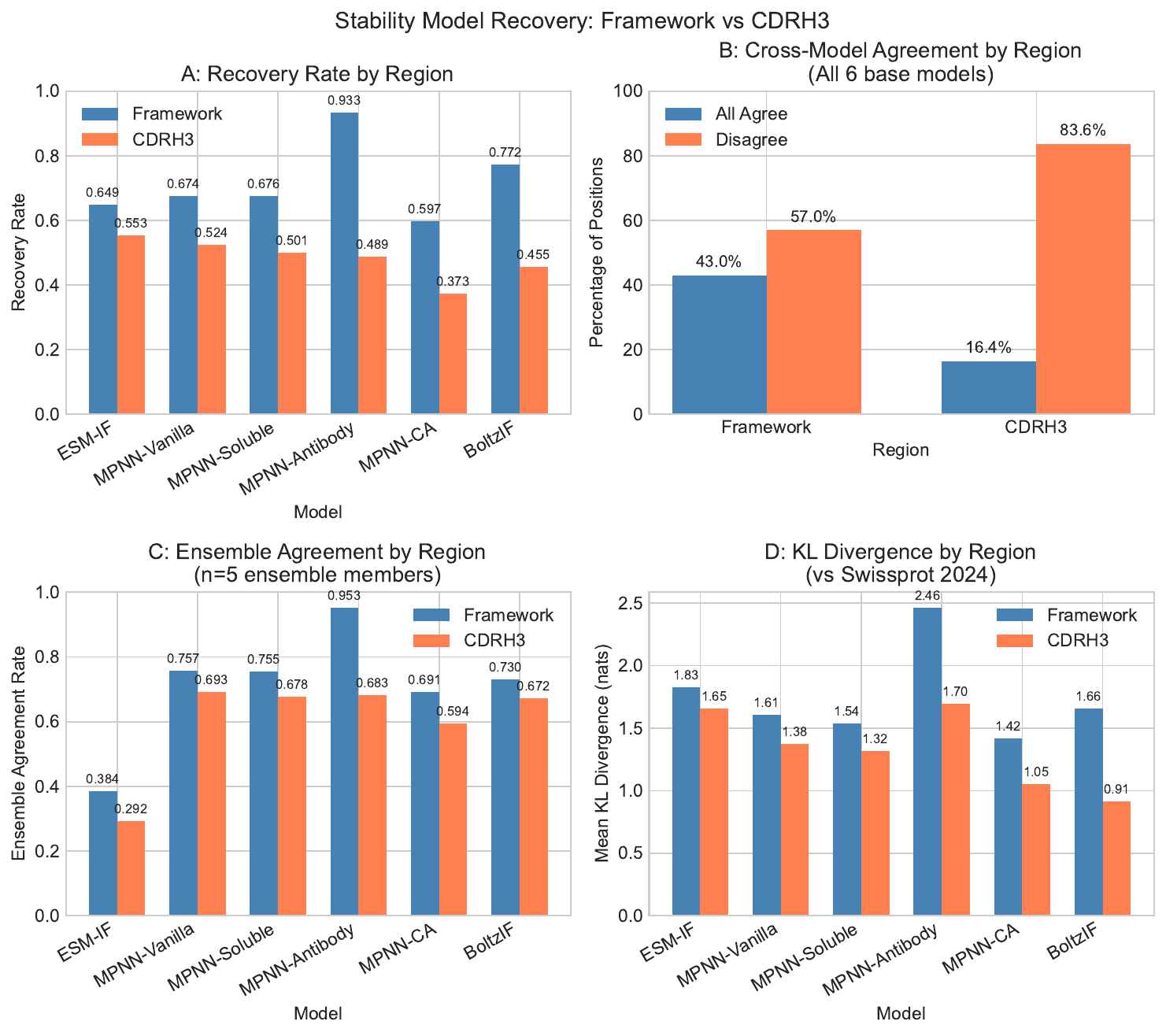}
\caption{Regional comparison of inverse folding predictions: framework versus CDRH3. Models show consistently lower recovery, higher entropy, and higher KL divergence in the CDRH3 region.}
\label{fig:logits_regional}
\end{figure}

\Cref{fig:logits_regional} further shows that inverse-folding recovery drops in CDRH3 relative to the framework, with higher predicted sequence variability and ensemble disagreement in the loop. This is the regime where a single stability model is least likely to be sufficient, motivating the use of multiple structure-conditioned rewards as separate objectives.

Over the reference set at $M = \pcaNobjectives$, $D = 4$ PCA components retains \pcaCumDfour\% of the variance and $D = 5$ retains \pcaCumDfive\%, so a fifth slot buys \pcaGainDfourToDfive{} points. $D$ is treated as an experimental variable in~\cref{sec:results-projection} rather than determined by a variance target.

\subsection{MHC-II Binding Oracle}
\label[appendix]{apx:mhc2}

The MHC-II binding oracle predicts peptide--allele binding affinity to estimate immunogenicity risk for candidate CDRH3 sequences. The model scores each candidate against a panel of 27 reference human leukocyte antigen (HLA) class~II alleles covering the major HLA-DP, HLA-DQ, and HLA-DR supertypes~\citep{greenbaum2011functional}. The model uses ESM-2 (8M parameters, 6 layers, 320-dimensional hidden states)~\citep{lin2023esm} as a frozen backbone encoder. Peptide and MHC-II pseudo-sequences are encoded independently through the backbone and mean-pooled across sequence positions (ignoring padding). The two 320-dimensional representations are concatenated into a 640-dimensional vector and passed through an MLP with two hidden layers ($640 \to 512 \to 512 \to 1$), GELU activations, dropout ($p = 0.25$), and a sigmoid output activation. The backbone is frozen throughout training. Only the MLP head (591K parameters) is optimized. Total model size including the frozen backbone is 8.4M parameters.

\begin{figure}[!htb]
\centering
\includegraphics[width=\textwidth]{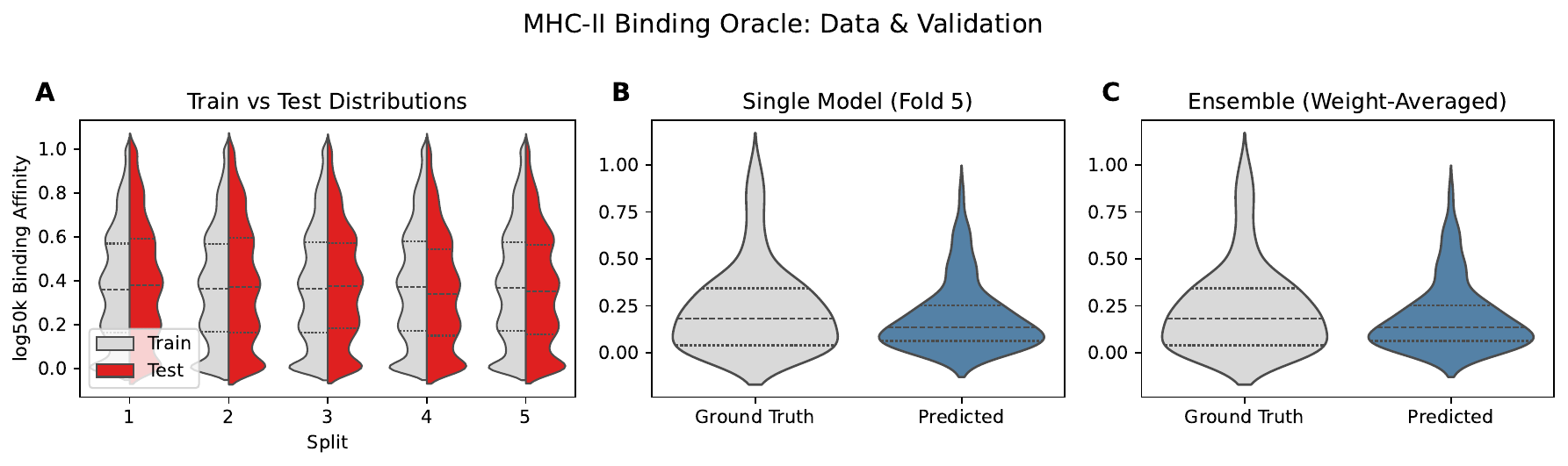}
\caption{MHC-II binding affinity distributions across the 5 cross-validation splits, comparing train (gray) and test (red) partitions (left). MHC-II oracle validation: predicted versus ground-truth binding affinity distributions for a single-fold model (middle) and the weight-averaged ensemble (right), showing little difference across folds. The shipped oracle averages output probabilities instead.}
\label{fig:mhc2_validation}
\end{figure}

We train on binding affinity data from NetMHCIIpan~4.0~\citep{reynisson2020netmhcpan}, which provides log$_{50k}$-transformed binding affinities ($y = 1 - \log_{50k}(\text{IC}_{50}\text{ nM})$, $y \in [0, 1]$) for peptide--allele pairs across 79 HLA class~II alleles. We perform 5-fold cross-validation using the provided BA1--BA5 partitions (${\sim}87$K training samples per fold). The model is trained for 40{,}000 steps with batch size 128 using AdamW ($\text{lr} = 10^{-3}$, no weight decay) and mean squared error (MSE) loss. A linear warmup schedule (500 steps) followed by linear decay is applied.

\begin{table}[!htb]
\centering
\caption{MHC-II binding oracle: 5-fold cross-validation results (ESM-2 8M backbone, frozen, MSE loss).}
\label{tab:mhc2_cv}
\begin{tabular}{lcc}
\toprule
Fold & AUROC & MSE \\
\midrule
1 & 0.833 & 0.039 \\
2 & 0.879 & 0.033 \\
3 & 0.862 & 0.031 \\
4 & 0.868 & 0.030 \\
5 & 0.866 & 0.033 \\
\midrule
Mean $\pm$ std & $0.862 \pm 0.015$ & $0.033 \pm 0.003$ \\
\bottomrule
\end{tabular}
\end{table}

At inference time, the five fold-specific heads share the frozen backbone and their output probabilities are averaged. Each candidate CDRH3 is scored against all 27 reference alleles in parallel. Higher rewards correspond to lower predicted immunogenicity. \Cref{tab:mhc2_cv} reports 5-fold cross-validation performance of the ensemble-averaged model evaluated on each fold's held-out test set. Binder classification uses the standard 500~nM threshold ($y > 0.426$). \Cref{fig:mhc2_validation} shows the label distribution across the 5 splits and final ensemble performance against the labeled ground truth. Our oracle achieves good median performance but under-predicts extreme positive binding affinity values.

For efficiency at training-loop scale, the 27 allele pseudo-sequence embeddings are cached once, reducing molecule-side forward passes per scoring call from $B \times 27$ to zero and peptide-side from $B \times 27$ to $B$. The cached path is verified exactly identical to the uncached path (maximum absolute difference $0.0$ on a 256-candidate batch).

\subsection{Developability Oracle}
\label[appendix]{apx:developability}

We use a SAbDab-derived dataset~\citep{chen2020developability} that assigns binary developability labels to 2,409 antibodies based on the Developability Index~\citep{lauer2012di}. The original definition of the index combines the spatial aggregation propensity of the CDRs with net charge computed on the full-length antibody structure. \citet{chen2020developability} computes it from variable-region structures alone, so the net-charge term in these labels is a variable-region quantity. Antibodies in the bottom 20\% of Developability Index scores, corresponding to the lowest aggregation risk, are labeled developable. This yields an imbalanced dataset with approximately 482 positives and 1,927 negatives.

The developability classifier fine-tunes ESM-2 650M~\citep{lin2023esm} with a BOS-token-pooled sequence classification head and 2-class output. The heavy-chain sequence is provided as input. Only the head and the layer-normalization parameters are fine-tuned, yielding an efficient adaptation of the pretrained backbone. Training uses 10-fold stratified cross-validation for 15 epochs with the Adam optimizer~\citep{kingma2015adam} (learning rate $5 \times 10^{-4}$), batch size 8 via gradient accumulation, and early stopping with patience 10.

Note that this oracle is a sequence classifier of \emph{ease of synthesis} and contains no spatial-aggregation or charge-patch descriptors. It is nontheless correlated with the the structural aggregation signal supplied in the postfilter analysis.

\subsection{Naturalness and Antibody-Repertoire Oracles}
\label[appendix]{apx:naturalness}

The naturalness oracle is a frozen copy of the AMPLIFY-350M span-in-filling model used to initialize our GFN. A candidate CDRH3 loop is scored by rendering the heavy chain as \texttt{<bos> prefix <mask> suffix <sep> CDRH3 <eos>} and computing the length-normalized log-likelihood of the CDRH3 tokens under teacher forcing. This gives a direct estimate of $p(\mathrm{CDRH3}\mid\mathrm{framework})$ under the same factorization used by the generator. As masked models score loops through leave-one-out pseudo-likelihood~\citep{salazar2020pll}, we measured a \bbSpeedup{}$\times$ throughput penalty at the median CDRH3 length compared to the autoregressive variant (\cref{sec:policy-architecture}). The two antibody-specific likelihood oracles are described in~\cref{sec:naturalness-reward}.

\subsection{Sequence Liability Motifs}
\label[appendix]{apx:liability}

The motif catalog is adapted from the Liability Antibody Profiler (LAP)~\citep{satlawa2024lap}, which catalogs 70 motifs and retains 30 that either recur across sources or carry strong evidence. Hard-flagged (gate-rejecting in the post-filter, maximally weighted in the reward): deamidation \texttt{NG}, isomerization \texttt{DG}, N-linked glycosylation sequon \texttt{N-x-[S/T]}, and free or unpaired cysteine, which LAP likewise treats as severe. The cysteine rule is a deliberately conservative sequence-only heuristic --- every cysteine in the designed loop is hard-flagged, since pairing cannot be determined from sequence. Soft-flagged (recorded and weighted, not gating): slow deamidation and isomerization variants, methionine and tryptophan oxidation, and \texttt{DP} fragmentation. The scanner emits per-motif occurrence counts, so the same code serves as a training reward, a post-filter gate, and an evaluation statistic. LAP is a screening tool which consumes a paired variable domain, folds it with ABodyBuilder2, and returns a flag list.

Our sequence-based scan is unable to identify solvent-based scans as it flags buried liability counts identically to exposed ones. Second, we do not include the "germline" or "therapeutic" exclusions to prevent our sampler from being biased towards those distributions. Third, we only evaluate the designed region and treat this score as a binary accept/reject criteria.

\subsection{Novelty Oracles}
\label[appendix]{apx:novelty}

Both novelty oracles precompute their reference set once at task setup from the \emph{resolved training split}, never from the full universe --- assembling the reference by hand is only correct when training on all targets, and is silently wrong for any subset. The reference is bucketed by loop length, both for speed (distance is at least $|\Delta L|$, so most buckets prune immediately) and, for the BLOSUM variant, because distance is minimized over training loops of the same length only.

CDRH3 length is a primary determinant of loop conformation, so comparing within a length class avoids introducing gap-open and gap-extend penalties as two free parameters that would then drive most of the length-dependent behaviour. The training set spans every length from 5 to 24 with no gaps. Note that both novelty rewards are \emph{constant on the native reference set by construction} (a native loop scored against a reference containing itself is exactly zero), so they must be scored leave-one-out when building the offline reference matrix for the projection warm start, or they vanish from the spectrum entirely.

\subsection{Stability Oracle}
\label[appendix]{apx:stability}

For a model family $m$ with weight variants $\cV_m$ and $E = 5$ ensemble members per variant, the operational stability reward averages over both levels:
\begin{equation}
  R_{\mathrm{stab}}^{(m)}(\bx, s) = \frac{1}{|\cV_m|} \sum_{v \in \cV_m} \frac{1}{L} \sum_{i=1}^{L} \log \softmax\!\left(\frac{1}{E} \sum_{e=1}^{E} \mathbf{g}_{v,e}^{(m)}(s)\right)_{x_i, i}
  \label{eq:rstab}
\end{equation}
where $\mathbf{g}_{v,e}^{(m)}(s) \in \R^{|\cA| \times L}$ are the precomputed logits for model family $m$, variant $v$, ensemble member $e$, and structure $s$ (here $L$ is the CDRH3 length after slicing from full-chain logits). The mean over ensemble members is taken before the log-softmax.

The six model families are correlated but not interchangeable: ProteinMPNN-Vanilla and ProteinMPNN-Soluble measure $+0.97$, but ESM-IF$\leftrightarrow$Vanilla measures $0.76$ while Vanilla$\leftrightarrow$CA-only measures $0.73$, so no correlation threshold separates them cleanly, and ProteinMPNN-Antibody tracks \emph{naturalness} ($+0.60$) rather than the other structural models. Any merge is therefore an arbitrary decision about which biological signals are ``the same''. This is precisely the decision the learned projection in~\cref{sec:projection} makes soft, data-driven, and reversible instead.

Structural stability of possible loop changes was assessed as the likelihood of loop sequences under structure-based inverse-folding models: ESM-IF~\citep{hsu2022esmif}, ProteinMPNN~\citep{dauparas2022proteinmpnn} variants, antibody ProteinMPNN~\citep{antibodympnn}, and BoltzIF~\citep{stark2025boltzgen}. ESM-IF is a 142M-parameter GVP-Transformer inverse-folding model. Predictions use all-atom coordinates from ATOM records, and ensembles are generated by adding Gaussian coordinate noise with $\sigma_{\mathrm{noise}} \in \{0, 0.0375, 0.075, 0.1125, 0.15\}$~\AA{}. BoltzIF is the inverse-folding model from the BoltzGen framework; ensembles are generated with 5 forward passes using different random seeds.

ProteinMPNN was used with four weight variants trained on different data distributions. \textbf{Vanilla} was trained on all PDB assemblies resolved to better than 3.5~\AA{} (clustered at 30\% sequence identity), using full backbone atom coordinates (N, C$_\alpha$, C, O, and a virtual C$_\beta$). \textbf{Soluble} uses the same architecture and featurization as Vanilla but excludes trans-membrane proteins from the training set, biasing surface residue preferences toward soluble compositions. \textbf{Antibody} (AbMPNN)~\citep{antibodympnn} is fine-tuned from Vanilla weights in two stages: first on ${\approx}108$k paired variable region clusters from the Observed Antibody Space predicted by ABodyBuilder2, then on ${\approx}1{,}700$ experimentally resolved antigen-binding clusters from SAbDab. \textbf{CA-only} uses only C$_\alpha$ coordinates as input, replacing inter-atom distances with orientation features derived from consecutive C$_\alpha$ displacement vectors. This variant is applicable to low-resolution or coarse-grained structures where full backbone coordinates are unavailable. Each variant was trained with multiple levels of Gaussian backbone noise (0.02--0.30~\AA{} for Vanilla and Soluble, 0.02--0.20~\AA{} for CA-only, and 0.0--1.0~\AA{} for Antibody), producing multiple sets of weights per base model.

\begin{figure}[!htb]
\centering
\includegraphics[width=\textwidth]{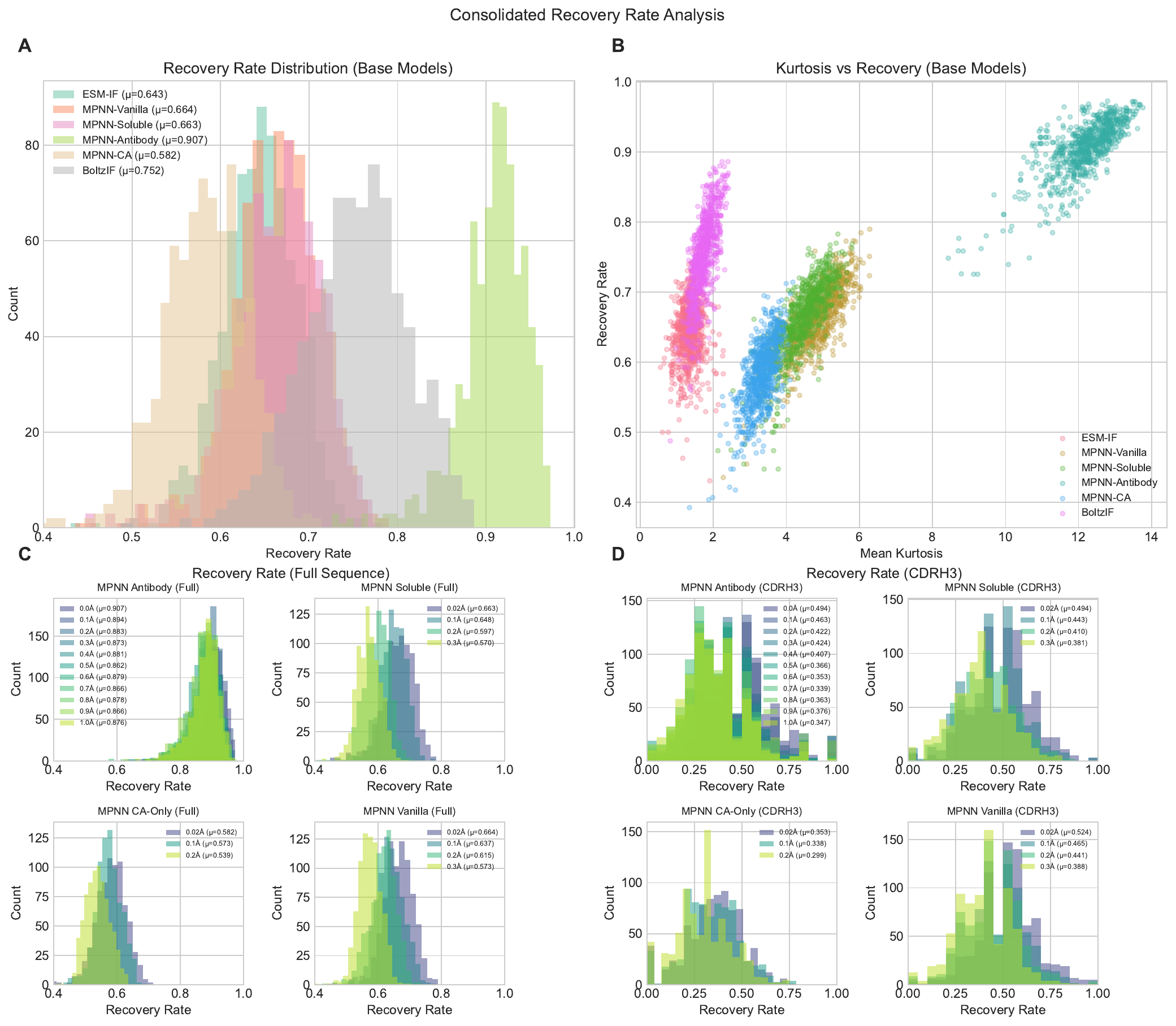}
\caption{Native sequence recovery rates across inverse folding models. Recovery is the fraction of
positions where the model's top prediction matches the native amino acid. A shows the distribution of
recovery rates across all base inverse folding models (no extra noise added) on all SAbDab entries.
ProteinMPNN-Antibody clearly has the highest recovery rate (this data is in-distribution for
the model, so this is expected). Panel B shows the model confidence (measures via logit kurtosis)
against recovery rate, which generally show strong positive correlations. Panel C shows the distribution
of recovery rates for each ProteinMPNN model with varying levels of structural noise added across the full
sequence. Additional noise produces a leftward shift in the distribution of recovery rates. Panel D
shows the same analysis in just the CDRH3 loop region, showing a large and expected drop in recovery rate
due to the inherent variability of this structure.}
\label{fig:logits_recovery}
\end{figure}

\begin{figure}[!htb]
\centering
\includegraphics[width=\textwidth]{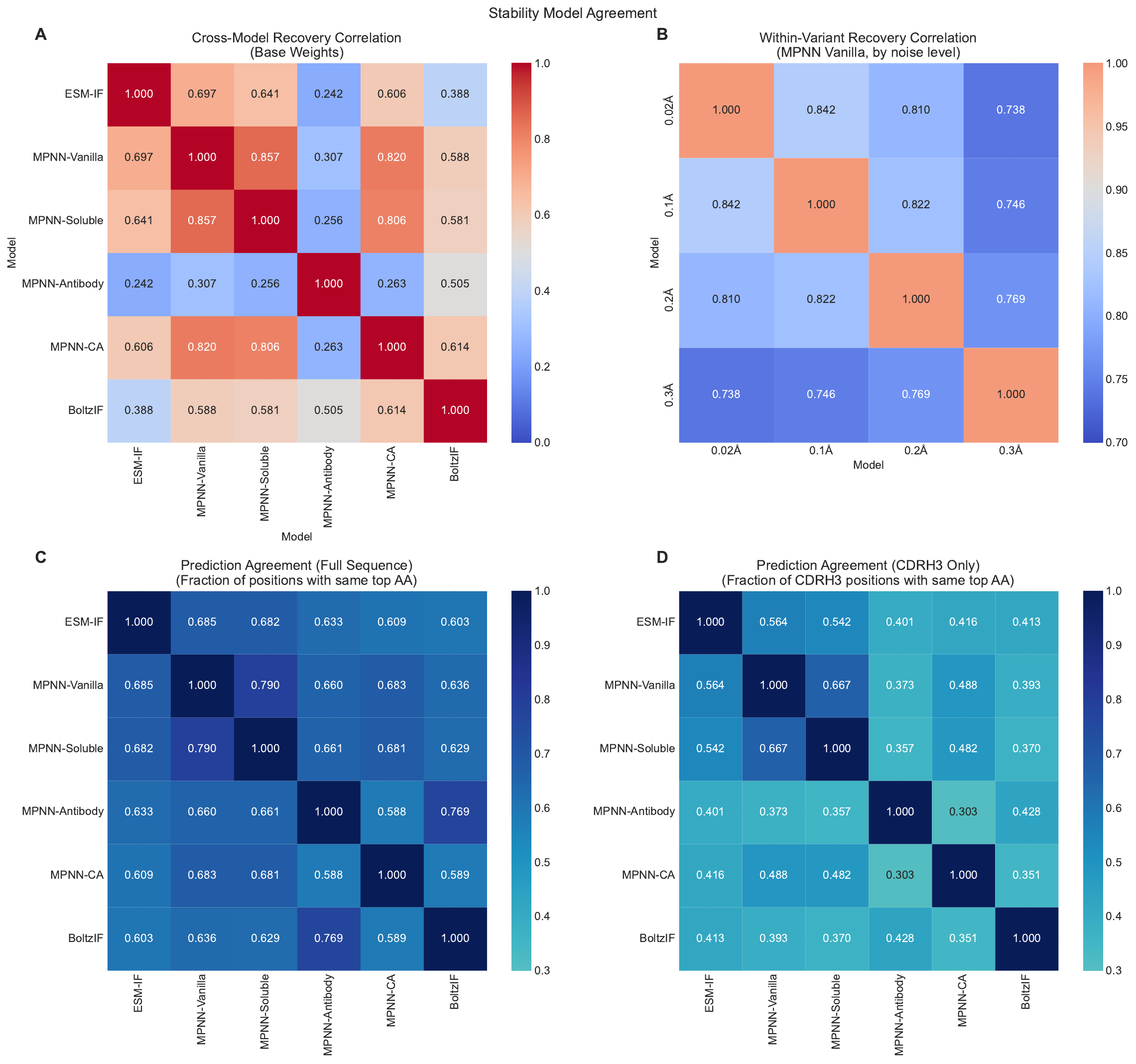}
\caption{Cross-stability model correlation and agreement. Pairwise Spearman correlations of
per-structure recovery rates and per-position prediction agreement across base models. Panel A
compares the base model of each major class of inverse folding model, showing their general
disagreement (while correlations are modest, each model clearly captures different signal). Panel B
shows the gradual drop in recovery agreement with additional structural noise added (this pattern
is consistent across ProteinMPNN models, data not shown). Panel C shows prediction agreement (\% of
positions with the same top amino acid predicted) across all base models for the entire sequence,
where agreement is generally high. Panel D shows the same analysis restricted to the CDRH3 loop,
showing these models capture different statistics in this functionally more variable region.}
\label{fig:logits_agreement}
\end{figure}

\begin{figure}[!htb]
\centering
\includegraphics[width=\textwidth]{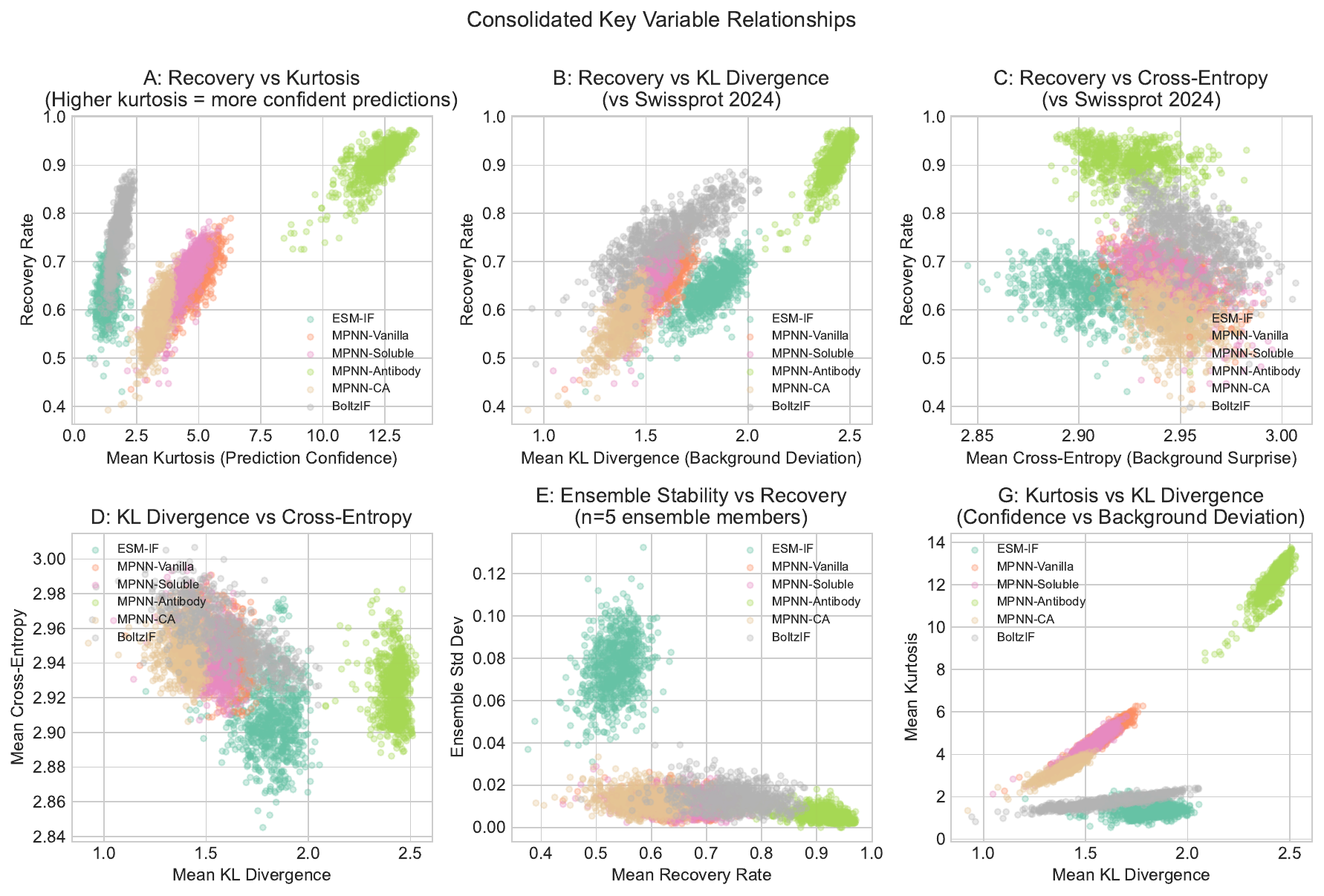}
\caption{Key variable relationships across inverse folding models: recovery versus KL divergence, entropy, kurtosis, and ensemble variability.}
\label{fig:logits_key_relationships}
\end{figure}

We validate these models in~\cref{fig:logits_recovery} by assessing sequence recovery for the CDRH3 loop and the surrounding framework under each for the full curated SAbDab set we used for our experiments. Panel A of~\cref{fig:logits_recovery} shows the distribution of recovery rates per model (where ``Base Model'' indicates the model with no extra noise added during training, where applicable). Panel B shows the relationship between logit kurtosis (a proxy for model confidence) and recovery. We find a strong correlation here for all models. Panel C shows the progressive reduction in recovery seen for ProteinMPNN models trained with increasing levels of additive noise, and panel D shows the expected drop in recovery for these models in the more variable CDRH3 region. \Cref{fig:logits_agreement} shows cross-model correlations of the recovery statistics (A), the effect of added structural noise (B), and cross-model prediction agreement between inverse folding models on the full sequence (C) and the CDRH3 (D), showing a substantial drop in agreement between models in the more variable CDRH3 region. Taken together, these results show that these inverse folding models all capture different statistics of the structural properties of the CDRH3 region, supporting our decision to use them all as unique reward models in the training of our GFN model. 
See ~\cref{fig:logits_key_relationships} for a more detailed view of the relationship between measures of the predicted logit distribution, recovery rate, and ensemble stability. In general, model confidence measures (KL Divergence, kurtosis) correlated strongly with the recovery rate while cross entropy did not. Ensemble stability was uncorrelated with recovery aside from the ESM-IF model.

\subsection{Length Dependence of Rewards}
\label[appendix]{apx:reward-length-dependence}

\begin{figure}[!htb]
\centering
\includegraphics[width=\textwidth]{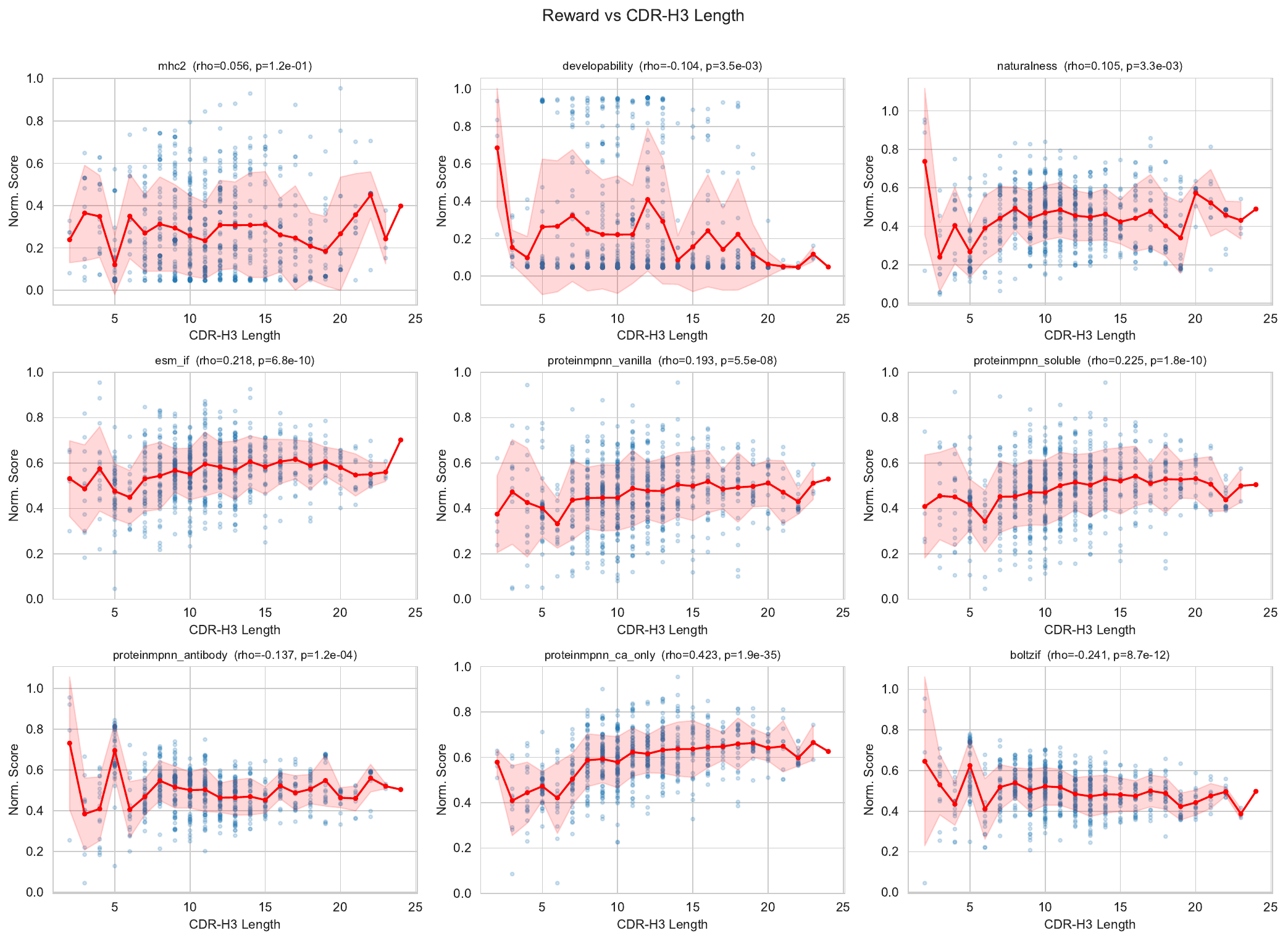}
\caption{Normalized reward scores as a function of CDRH3 loop length. Red lines show mean per length with $\pm 1$ standard deviation bands. Spearman $\rho$ and $p$-values are shown per objective.}
\label{fig:reward_vs_length}
\end{figure}

In general, rewards did not correlate with sequence length, with some exceptions at extremely short loop lengths. See~\cref{fig:reward_vs_length} for a per-reward breakdown. Length dependence nonetheless dominates several \emph{pipeline} quantities --- Stage-0 liability survival spans 19--61\% across targets and tracks CDRH3 length at $r = -0.97$ --- which is why every comparison in this work is paired per target.

\section{Hyperparameter Tuning}
\label[appendix]{apx:sweeps}

\subsection{Learning Rate, on the Objective-Slots Arms}
\label[appendix]{apx:lrsweep}

\lrnarms{} arms, identical except for the policy learning rate, at each of $D = 4$ and $D = 5$ on the $M = 14$ objective set: \lrsteps{} steps, $\beta = 8$, the scaffold-leakage split, and every other setting held at the values in \Cref{apx:implementation}. The grid brackets the learning rate $\pi_{\mathrm{lr}} = 1\times10^{-4}$ on both sides as well.
\begin{table}[!htb]
\centering
\begin{tabular}{llrrrr}
\toprule
$D$ & $\pi_{\mathrm{lr}}$ & Mean reward & Block SE & TB loss & $\Delta$ vs.\ best ($\sigma$) \\
\midrule
$4$ & $1\times 10^{-5}$ & -0.8027 & 0.0037 & 0.1155 & 0.0224 (4.8) \\
 & $3\times 10^{-5}$ & -0.7950 & 0.0030 & 0.0930 & 0.0146 (3.5) \\
 & $\mathbf{1\times 10^{-4}}$ & \textbf{-0.7803} & 0.0028 & \textbf{0.0792} & --- \\
 & $3\times 10^{-4}$ & -0.8115 & 0.0040 & 0.1330 & 0.0312 (6.3) \\
 & $1\times 10^{-3}$ & -0.8623 & 0.0039 & 0.1906 & 0.0820 (16.9) \\
\midrule
$5$ & $1\times 10^{-5}$ & -0.7608 & 0.0028 & 0.0954 & 0.0232 (5.1) \\
 & $3\times 10^{-5}$ & -0.7545 & 0.0035 & 0.0826 & 0.0169 (3.4) \\
 & $\mathbf{1\times 10^{-4}}$ & \textbf{-0.7376} & 0.0036 & \textbf{0.0727} & --- \\
 & $3\times 10^{-4}$ & -0.7593 & 0.0032 & 0.0908 & 0.0217 (4.5) \\
 & $1\times 10^{-3}$ & -0.7956 & 0.0040 & 0.1454 & 0.0580 (10.8) \\
\bottomrule
\end{tabular}

\caption{Learning-rate sweep on the $M = 14$ objective-slots arms. Mean reward and trajectory-balance loss over the final \lrwindow{} training steps; $\Delta$ vs.\ best is the reward gap to that $D$'s selected arm, in units of the pooled block standard error.}
\label{tab:lr-sweep}
\end{table}

$\pi_{\mathrm{lr}} = 1\times10^{-4}$ is selected at both $D$, by at least \lrminzDfour{} block standard errors at $D = 4$ and \lrminzDfive{} at $D = 5$, and the ordering is a clean single optimum degrading in both directions. Both production arms therefore use the anchor value: the sweep confirmed it rather than displacing it. Note that the two $D$ columns are not comparable to each other --- reward is measured in each arm's own slot space, and $D$ changes that space.

\subsection{Reward Sharpening: The $\beta$ Sweep}
\label[appendix]{apx:beta}

Three arms identical except for the reward-sharpening exponent $\beta$, one seed, \betanprefs{} Dirichlet preferences $\times$ \betansamplesperpref{} samples $\times$ \betantargets{} evaluation targets $=$ \betansamples{} designs per arm. Controllability is the per-axis diagonal Spearman correlation between preference weight and realized reward.

\begin{table}[!htb]
\centering
\begin{tabular}{@{}lccc@{}}
\toprule
axis / statistic & $\beta = 1$ & $\beta = 8$ & $\beta = 32$ \\
\midrule
immunogenicity & 0.021 & 0.127 & 0.247 \\
developability & 0.019 & 0.176 & 0.266 \\
naturalness & 0.004 & 0.061 & 0.200 \\
stability & 0.009 & 0.076 & 0.103 \\
\textbf{mean $|$diagonal$|$} & \textbf{0.013} & \textbf{0.110} & \textbf{0.204} \\
signal-to-noise (diagonal / off-diagonal) & 1.49 & 1.40 & 1.57 \\
mean pool reward & 0.435 & 0.521 & 0.656 \\
top-decile pool reward & 0.569 & 0.669 & 0.764 \\
\midrule
unique-sequence fraction & 1.000 & 1.000 & 0.997 \\
within-preference pairwise Hamming & 12.1 & 12.0 & 11.0 \\
distinct-mode fraction @70\% identity & 0.999 & 0.996 & 0.935 \\
global objective coverage (mean $\sigma$) & 0.179 & 0.201 & 0.207 \\
within-preference local spread & 0.146 & 0.160 & 0.145 \\
\midrule
hypervolume (training, final) & 0.503 & 0.875 & 0.904 \\
trajectory-balance residual spread (final) & 0.13 & 0.86 & 4.72 \\
\quad relative change, last 40\% of training & -4.5\% & -10.9\% & -11.5\% \\
\bottomrule
\end{tabular}

\caption{Effect of reward sharpening $\beta$ on controllability, pool quality and diversity.}
\label{tab:beta-controllability}
\end{table}

Controllability rises monotonically with $\beta$ on every axis. Mean $|$diagonal$|$ \betameandiagOne{} $\to$ \betameandiagEight{} $\to$ \betameandiagThirtyTwo{} --- so near-zero controllability at $\beta = 1$ is substantially a $\beta$ effect rather than a conditioning-architecture limit. The effect is large relative to the split difference between the arms, but it is not independent of it, and the measurement deserves repeating on the leakage-free split.

There is no mode collapse at high $\beta$, though we began to observe some diversity loss. At $\beta = 32$, \betauniqueThirtyTwo{} of designs are unique and \betamodefracThirtyTwo{} are distinct modes at \betaidentity{} CDRH3 identity, against \betauniqueOne{} and \betamodefracOne{} at $\beta = 1$; global objective coverage \emph{rises} (\betacoverageOne{} $\to$ \betacoverageThirtyTwo{}) while within-preference local spread falls \betalocalspreaddrop{}. Increasing $\beta$ did not have an adverse effect on training dynamics. While $\beta = 32$'s trajectory-balance residual spread is far larger in absolute terms (\betadeltastdThirtyTwo{} against \betadeltastdEight{}), in its scale-free form, the fitted trend over the last \betawindowfrac{} of training was roughly the same between $\beta = 8$ (\betatrendEight{}) and $\beta = 32$ (\betatrendThirtyTwo{}). 


Our data does not show that objectives develop genuine tensions at high $\beta$. Sharpening does raise the off-diagonals of the preference-to-reward matrix: at $\beta = 32$, preference on immunogenicity against realized naturalness is \betaprefoffnatThirtyTwo{} and against realized stability \betaprefoffstabThirtyTwo{}, from \betaprefoffnatOne{} and \betaprefoffstabOne{} at $\beta = 1$. But those off-diagonals carry the simplex confound of \cref{apx:controllability} --- raising one preference weight necessarily lowers the others --- so any policy that responds to preferences at all must produce them, and they say nothing about the objectives. The objective-to-objective correlations, where a real trade-off would appear, move the other way: immunogenicity against naturalness reads \betaobjcorrnatOne{} at $\beta = 1$ and \betaobjcorrnatThirtyTwo{} at $\beta = 32$, weaker rather than stronger, and immunogenicity against stability \betaobjcorrstabOne{} $\to$ \betaobjcorrstabThirtyTwo{}. Sharpening makes the policy \emph{steerable}; it does not make the objectives \emph{conflict}.

On trained-objective metrics alone, this table selects $\beta = 32$. However, as our goal was to emphasize the value of diversity in building loop samplers, we opted to use $\beta = 8$. Future work should explore the performance benefits of using sharpened rewards in this setting for improved steerability and, with the right optimization methods, coverage of the reward landscape. For this work, it was not selected largely because the generate-then-filter workflow's value depends on \emph{pool breadth reaching the post-filter}, and $\beta = 8$ was seen to be on the operating point of the controllability$\leftrightarrow$exploration trade-off for this workflow. In addition, due to compute constraints, many promising avenues were not able to be explored in this work - the reader should not interpret the lack of these experiments as the authors believing they are not worth performing. 

Absolute controllability remains modest even at $\beta = 32$ (mean \betameandiagThirtyTwo{}; the stability axis alone \betactrlstabThirtyTwo{}), so while these analyses shows \emph{$\beta$-dependence of controllability}, it does not show strict controlability over the generations.  All metrics are on trained objectives only. The lack of training collapse is conditional on the dual-cross-attention span-in-filling architecture adopted here -- earlier architectures \emph{did} collapse at $\beta = 32$ during training.

\end{document}